\documentclass[prd,twocolumn,floatfix]{revtex4-1}

\usepackage{graphicx,color}
\usepackage{amsmath,amssymb,amsfonts}
\newcommand{\be}{\begin{equation}}
\newcommand{\ee}{\end{equation}}
\newcommand{\ba}{\begin{eqnarray}}
\newcommand{\ea}{\end{eqnarray}}
\newcommand \nn {\nonumber}

\begin{document}

\title{Quark-Hadron Transition and Entanglement}

\author{Berndt M\"uller}
\affiliation{Department of Physics, Duke University, Durham, NC 27708-0305, USA}
\author{Andreas Sch\"afer}
\affiliation{Institut f\"ur Theoretische Physik, Universit\"at Regensburg, D-Regensburg, Germany}

\begin{abstract}
The dual holographic description has enjoyed many successes in explaining fundamental properties of the early stages of relativistic heavy ion collisions up to the formation of a minimal-viscosity quark-gluon fluid. However, there have been few attempts to extend its application beyond this stage. Here we explore the prospects for such an extension beyond the time of hadronization. Our discussion makes use of recent insights into the duality of entanglement properties of field theory states in the edge of Anti-de Sitter space and non-trivial topologies of horizons in the bulk, often referred to as ER = EPR duality. We discuss this topic from the point of view of heavy-ion phenomenology, review several relevant concepts, and map out a path toward combining them into a comprehensive, at least semiquantitative description of relativistic heavy ion collisions. We outline possible next steps in this direction.   
\end{abstract}

\maketitle

\section{Introduction}
\label{sec:Intro}
One of the fundamental questions of high energy physics is how apparently thermal behavior emerges in relativistic heavy ion collisions. A large number of measurements show such thermal behavior and are best described when one assumes that the produced transient Quark-Gluon Plasma (QGP) is in a fully thermalized state. Yet the nuclear reaction, which occurs in isolation, cannot produce a state of high von Neumann entropy because the reaction is governed by the laws of quantum chromodynamics (QCD) which assures unitarity of the S-matrix. The transient QGP and all the hadrons measured in the final state must exist in a highly entangled quantum state that mimics the properties of a thermal ensemble that is in interaction with an external heat bath. 

One is immediately led to the question whether there are any observable differences between a thermal ensemble and such a highly entangled excited state. The answer is not as simple as one might think as local observables and, more generally, observables that depend on only a fraction of the complete many-particle final state have difficulty differentiating between these two alternatives.

Whether entropy can be generated in an isolated system in a process that is governed by time reversal invariant laws or not, depends on the definition of generalized, non-equilibrium ``entropy''. Several proposals for such a notion exist and, unsurprisingly, the answer depends on the entropy definition used. In addition, as the deviation from equilibrium is time dependent, the most suitable notion of entropy may yield an answer that depends on the time elapsed after the start of the collision.

The complexity of this issue is illustrated by the fact that the Anti-de Sitter space-conformal field theory (AdS/CFT) duality allows to map the production of a QGP to the formation of a black hole in five-dimensional Anti-de Sitter (AdS$_5$) space. Any description of entropy production will share the subtleties of the black hole information paradox expressed most succinctly by the Page curve \cite{Page:1993wv,Page:2013dx}. For any pure quantum state of an isolated many-particle system the von Neumann entropy is zero and remains zero under unitary time evolution. If, however, only part of the system is observed, i.e. if the information residing in the rest of the system is discarded, the reduced system appears thermal with the corresponding thermal entropy. Although the Page curve was specifically introduced to resolve the information problem of black holes, it is understood to be the characteristic property of any generic many-body quantum system \cite{Page:1993df}. Its existence has been confirmed experimentally in various experiments, see, e.g., Fig.~4 in Kaufman {\it et al.} \cite{Kaufman:2016qu}. 

Our goal here is to outline a program which aims at describing essential information theoretical properties of high energy nuclear collisions, including those properties resulting in a Page curve, using concepts from AdS/CFT duality. Obviously this is an ambitious goal, and in the present article we can only outline the first steps. We will provide a list of open issues at the end of this work that may help define a systematic pathway towards a more precise understanding of how the apparent thermal properties of the final state of a relativistic heavy ion collision can be reconciled with the unitarity of the S-matrix in QCD.

There are many quantum field theories for which holographic duals are known to exist. In the realm of gauge theories the original prototype, the ${\cal N}=4$ supersymmetric conformal Yang-Mills theory, is by far the most widely explored example, because for large $N_c$ and large 't Hooft coupling $\lambda$ it has a tractable holographic dual, i.~e.\ classical supergravity in AdS space. The exact holographic dual for QCD is not known, although models that share important features with QCD have been constructed \cite{Witten:1998zw,Sakai:2004cn}. Therefore, any attempt to apply AdS/CFT phenomenology to QCD is problematic from the start and can only be successful when combined with other field theory and string theory techniques which allow to correct for some of the differences. For example, the violation of conformal symmetry can be treated perturbatively in QCD, see, e.g. \cite{Braun:2003rp,Braun:2018mxm,Kumericki:2006xx}, the finiteness of the 't Hooft coupling can be corrected by string perturbation theory, see e.g. \cite{Waeber:2015oka}, and the calculation of $1/N_c$  expansions on the field theory side has evolved into a broad research field of its own \cite{tHooft:1973alw}.

However, QCD has also certain properties, like the confinement/deconfinement transition, which cannot be treated perturbatively and thus require a more incisive approach. As we are interested in the hadronization of the quark-gluon plasma we cannot avoid studying holographic models that incorporate a confinement/deconfinement transition into their basic framework through some versions of the Hawking-Page transition \cite{Hawking:1982dh}. Examples of such models include the ${\cal N}=4$ super-Yang-Mills theory on a compactified spatial volume, such as $S^3$, models including a compactified additional dimension of AdS space \cite{Aharony:2005bm}, and models that break conformal invariance through the introduction of a dilaton field \cite{Gursoy:2008za,Mandal:2011ws}. Although there are reasons to believe that such holographic models can describe many relevant properties of QCD qualitatively, it is far from clear whether they can be refined to decribe these relevant features with sufficient precision to allow for a quantitative comparison with experimental data.

In view of these {\it caveats} the success of holographic methods in modeling certain aspects of relativistic heavy ion collisions came as a pleasant surprise. They have been used extensively to understand rapid thermalization \cite{Lin:2006rf,Balasubramanian:2010ce,Balasubramanian:2011ur,Shuryak:2011aa,Chesler:2010bi,Chesler:2013lia,Hubeny:2013hz}, the fast transition to hydrodynamic flow \cite{Janik:2005zt,Heller:2011ju}, the small value of the specific shear viscosity \cite{Kovtun:2004de}, and much more. In fact, the last two decades have witnessed such continuous progress that today a much larger assortment of powerful techniques and insights exists than ever before. For example, semiclassical quantum gravity and applications to black hole physics have recently conceptually resolved the information paradox \cite{Maldacena:2013xja,Maldacena:2018lmt,Akers:2019nfi,Almheiri:2019qdq,Almheiri:2019psf,Almheiri:2020cfm,Penington:2019kki,Penington:2019npb,Anderson:2020vwi}. All these encouraging successes motivate us to try to understand aspects of the breakup of the QGP into hadrons, which have only rarely been explored using holographic methods. 

Our effort to expand the use of holographic duality to a description of the complete heavy ion collision is also motivated by the recognition that a field theoretical description of hadronization on the quantum level is far too complex to be tractable. Without tractable holographic descriptions that capture salient features of this transition, even if they cannot reproduce all aspects quantitatively, we simply lack a perspective to fully understand the many-body quantum mechanics of a heavy ion collision.

Let us start by reminding the reader of a few salient phenomenological properties of heavy ion collisions that any comprehensive description must address:
\begin{itemize}
\item
There exist two distinct ways in which the particles that are ultimately detected, hadrons, are produced from the evaporating QGP. Beginning immediately after formation of the QGP "fireball", hadrons are emitted from its surface. As the expanding QGP cools down to the pseudocritical temperature $T_c$, hydrodynamics ceases to be valid, and the fireball converts into hadrons throughout the remaining volume. The thermal model \cite{Andronic:2017pug} describes the production rates of all hadrons and nuclei remarkably well with a single universal temperature parameter (the chemical freeze-out temperature $T_{\rm CF}= 156.6 \pm 1.7$ MeV \cite{Andronic:2021dkw}) that is determined to percent-level accuracy. Because of the closeness of this value to the central temperature of the QCD crossover transition determined by lattice QCD ($T_c = 158 \pm 0.6$ MeV \cite{Borsanyi:2020fev} or $T_c = 156.5 \pm 1.5$ MeV \cite{HotQCD:2018pds}) $T_{\rm CF}$ is interpreted as QCD deconfinement temperature. The precision of the coincidence between $T_{\rm CF}$ and $T_c$ itself is remarkable because the QCD crossover transition, as measured by the chiral susceptibility, has an intrinsic width of $15 \pm 1$ MeV \cite{Borsanyi:2020fev}.
\item
The fact that the thermal model works so well is quite astonishing in view of the strong interactions among the constituents of the system before and after the hadronization transition. A focus of the discussion on this point is provided by the hypertriton He$^3_{\Lambda}$ whose yield in Pb-Pb collisions at LHC agrees with the prediction of the thermal model for the universal temperature $T_{\rm CF}$ despite the fact that its binding energy is only 0.4 MeV, i.~e.\ much smaller than $T_{\rm CF}$, and its size is comparable to the size of the collision system. Recently, however, it was found \cite{ALICE:2021puh} that this is no longer true for a smaller collision system (p-Pb) with earlier hadronization time, adding to the conundrum of hadronization in heavy ion collisions. 
\item
Heavy ion collisions create thermal conditions that resemble those prevailing in the early universe, but the time scales are vastly different. In the cosmos, different particle species fell out of thermal equilibrium at different times due to cosmic expansion and cooling. For example, photons and neutrinos have different cosmic background radiation temperatures. The same could be expected for heavy ion collisions but is not observed for Pb-Pb collisions. As we argued in \cite{Muller:2017vnp} one framework that can potentially provide an explanation for the lack of thermal differentiation is the Eigenstate Thermalization Hypothesis (ETH) \cite{Deutsch:1991qu,Srednicki:1994mfb,DAlessio:2015qtq}. The ETH may explain not only the success of the thermal model but also the difference between Pb-Pb and p-Pb collisions as the lifetime of the QGP in p-Pb collisions may be too short to apply the full ETH formalism, which is based on energy eigenstates (see Section \ref{sec:ETH}).
\end{itemize} 

Our manuscript is organized as follows. Section \ref{sec:EE} contains some general remarks about entanglement entropy in the context of relativistic heavy ion collisions, and we introduce the concept of the Page curve. Because the ETH plays an important role in our arguments we present a brief review in Section \ref{sec:ETH} of those aspects that are relevant for us. In Sections \ref{sec:HIC} and \ref{sec:HadronHIC} we will further discuss some relevant aspects of heavy ion collisions and provide more details on the properties mentioned above. In particular, we will argue in Section~\ref{sec:HIC} that the usual numerical AdS calculations in the Poincar\'e patch are insufficient to obtain the Page curve. In Section \ref{sec:Thermalization} we will review these holographic model calculations of the early stages of heavy ion collisions up until apparent thermalization. In this time interval they contain the relevant physics and give phenomenologically satisfactory results. 

To extend the holographic description to later times we propose two distinct hadronization mechanisms: surface radiation and bulk hadronization. In Section \ref{sec:Model_1} we adapt recent ideas \cite{Aharony:2005bm,Mateos:2006nu,Bigazzi:2020phm} for the Hawking-Page phase transition to bulk hadronization. To describe the quantum entanglement of hadrons emitted from the fireball surface requires a very different approach. As hadrons cannot exist within the QGP fireball, its surface acts as a boundary for hadron states. In Section \ref{sec:Model_2} we discuss how hadron emission can be treated as the production of particle-hole states at the QGP surface in analogy to Hawking radiation from the black hole horizon. This allows us to adapt the arguments for the formation of "islands" within black holes which are entangled with the outgoing Hawking radiation to the QGP fireball. The increasing entanglement of the remaining QGP  with the emitted hadrons is holographically mapped onto the black hole evaporation process. This will make it possible to transcribe the arguments for the occurrence of a Page curve to the QGP hadronization process. We synthesize these considerations into a schematic holographic description of the hadronization process in Section \ref{sec:holmod}.

Our arguments are admittedly speculative, but they have the advantage that their underlying assumptions can be confirmed or refuted by detailed calculations. In Section \ref{sec:ToDo} we present a limited list of possible future studies.

\section{Entanglement Entropy}
\label{sec:EE}

In high-energy heavy ion collisions one never measures the complete final state but detects only a subset of all emitted particles. Detector constraints typically limit the detection to hadrons in a certain rapidity interval $\Delta y \sim 1$ that is much smaller than the full beam rapidity range $-y_{\rm beam} \leq y \leq y_{\rm beam}$.\footnote{The acceptance of detectors usually covers a certain pseudorapidity window rather than a rapidity window. At high energy, however, by far the largest number of emitted particles are pions whose mass is smaller than their average transverse momentum, the effect of the difference between pseudorapidity and rapidity is small for most global observables.}  Because the subsystem $\Phi_{\Delta y}$ within the kinematic range $\Delta y$ is entangled with the rest of the final state $\Phi'_{\Delta y} = \Phi\setminus\Phi_{\Delta y}$ outside this rapidity window, it carries a certain entanglement entropy:
\be
S(\Phi_{\Delta y}) = {\rm Tr}[\rho(\Phi_{\Delta y}) \ln\rho(\Phi_{\Delta y})] .
\ee
Here
\be
\rho(\Phi_{\Delta y}) = {\rm Tr}_{\Phi'}[\rho(\Phi)] ,
\ee
denotes the reduced density matrix of the final state constrained to the rapidity window $\Delta y$, $\Phi$ denotes the complete final state, and $\rho(\Phi)$ is the complete final-state density matrix. Unitarity of the S-matrix in QCD ensures that $S(\Phi) = 0$, if the initial state is a pure quantum state, which is satisfied to a high degree of precision in the collision of two heavy ions.

\begin{figure}[htb]
\centering
\hspace{16pt}
\includegraphics[width=0.85\linewidth]{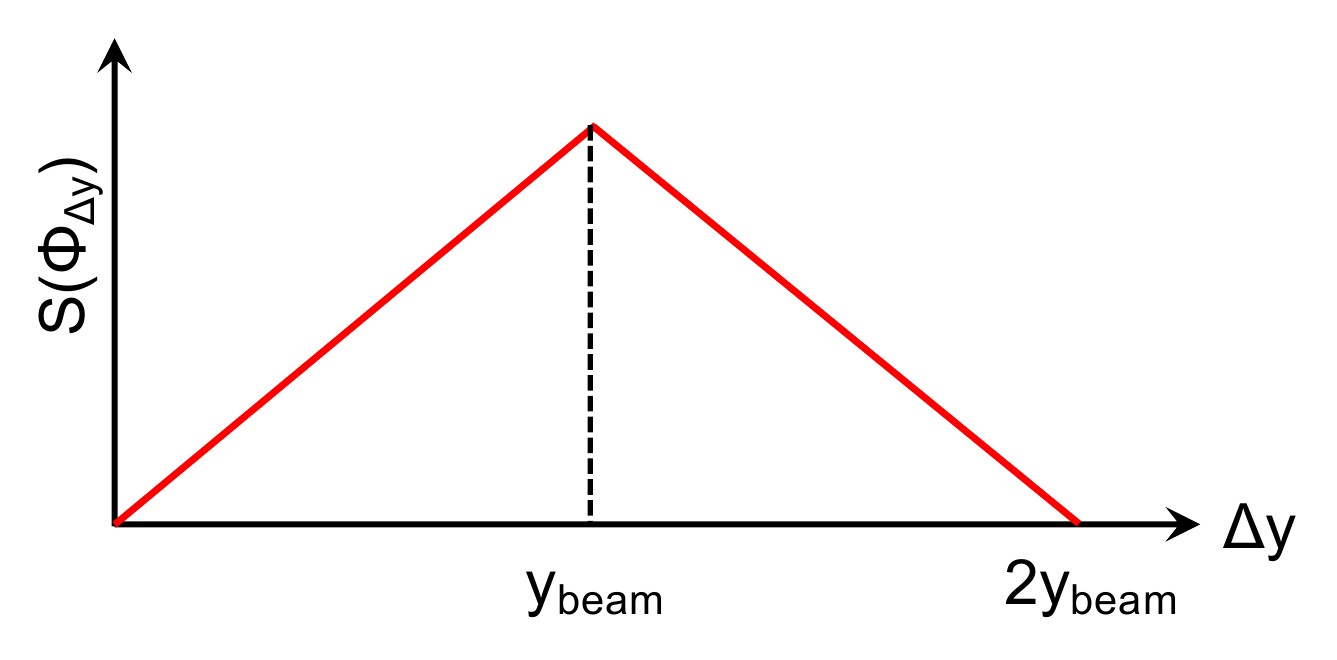}
\caption
{%
Page curve (red solid line) for the entanglement entropy associated with a rapidity window $\Delta y$.} 
\label{fig:Page_Deltay}
\end{figure}

For $\Delta y \ll 2y_{\rm beam}$ the entanglement entropy $S(\Phi_{\Delta y})$ grows with the size of the rapidity window. Boost invariance at midrapidity, realized at high collision energy, dictates that $S(\Phi_{\Delta y})/\Delta y$ is approximately independent of $\Delta y$ for small rapidity windows. The complementarity law of quantum information also dictates that $S(\Phi_{\Delta y}) = S(\Phi'_{\Delta y})$. Accordingly, when $\Delta y$ exceeds half the total rapidity range of the final state, $\Delta y > y_{\rm beam}$, the entanglement entropy has to decrease again and approach zero when $\Delta y \to 2y_{\rm beam}$. The entanglement entropy associated with a rapidity window thus follows a Page curve \cite{Page:1993df,Page:1993wv}, as sketched by the red solid line in Fig.~\ref{fig:Page_Deltay}.

\begin{figure}[htb]
\centering
\hspace{16pt}
\includegraphics[width=0.85\linewidth]{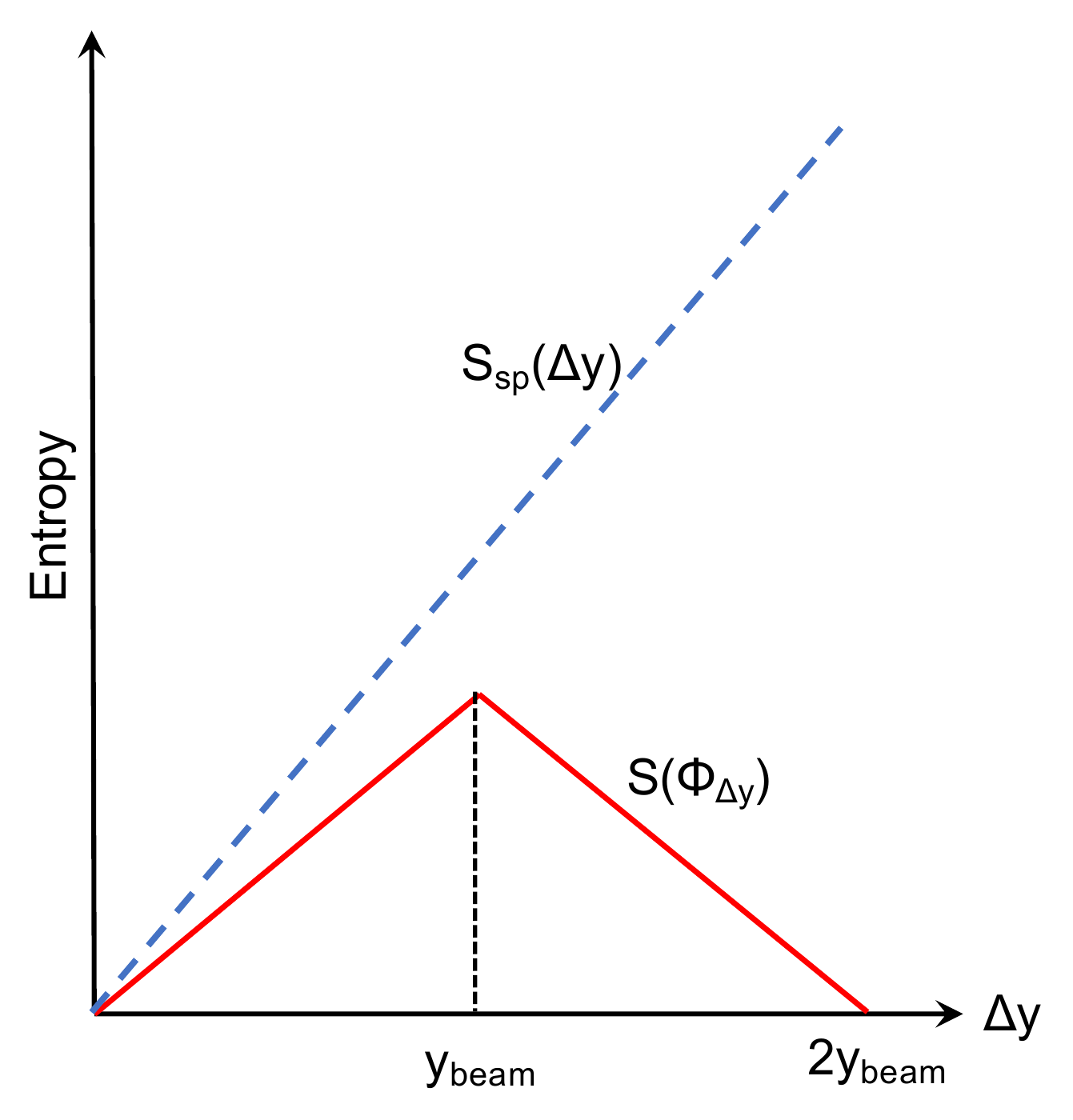}
\caption
{%
Schematic diagram of the final-state entropy $S_{\rm sp}(\Delta y)$ in a rapidity window $\Delta y$ derived from the single-particle momentum distribution, shown as blue dashed line, in comparison with the entanglement entropy in the same window, shown as red solid line.} 
\label{fig:Entropy_Deltay}
\end{figure}

Particle detectors are not capable of measuring all information contained within $\Delta y$, e.g., they generally do not record the relative phases among all emitted particles. Although many modern detectors are capable of recording the full set of kinematic correlations among hundreds or thousands of detected particles, a large part of this information is usually discarded in the data analysis and only selected few-body correlations are retained \footnote{There are exceptions to this statement, e.g., in the measurement of the collective flow vector or in fully resolved jet measurements, but even these analyses ignore the full correlation information that is, in principle, contained in the recorded data.}. The reduction to a subset of data within the rapidity window $\Delta y$ further changes the concept of entropy into a notion that has been aptly called ``entropy of ignorance'' about the full details of the final state \cite{Duan:2020jkz}. What is commonly called ``entropy'' of the final state of a relativistic heavy ion collision is derived from the single-particle distribution in momentum space; here we denote this single-particle entropy as $S_{\rm sp}(\Delta y)$. The single-particle entropy grows at a different rate than the entanglement entropy (see Duan {\it et al.} \cite{Duan:2020jkz} for a detailed analysis in a slightly different context) and continues to grow as $\Delta y$ exceeds $y_{\rm beam}$, as illustrated by the blue dashed line in Fig.~\ref{fig:Entropy_Deltay}.

A number of estimates of the single-particle entropy have been presented in the literature. The estimates are based on the measured spectra of produced particles and calculate the entropy of an equilibrated Boltzmann (or Bose/Fermi) gas with the same momentum and mass distribution. This is usually done per unit of rapidity (at midrapidity), resulting in a measured value of $dS_{\rm sp}/dy = S_{\rm sp}(\Delta y)/\Delta y$, which corresponds to the slope of the dashed line in Fig.~\ref{fig:Entropy_Deltay}. At top RHIC energy, $dS_{\rm sp}/dy \approx 5,000$ \cite{Pal:2003rz,Muller:2005en}; at top LHC energy $dS_{\rm sp}/dy \approx 11,500$ \cite{Hanus:2019fnc}.

\section{The Eigenstate Thermalization Hypothesis}
\label{sec:ETH}

ETH \cite{Deutsch:1991qu,Srednicki:1994mfb,DAlessio:2015qtq} posits that even a single energy eigenstate can appear like a thermal system for most, if not all, observables of practical interest. It is thought to apply generally to systems that exhibit chaos at the quantum level. The ETH represents a generalization of Random Matrix Theory (RMT) \cite{Wigner:1967ran} and, like RMT, is formulated in terms of the matrix elements of an observable ${\cal A}$ in the energy basis:
\be
A_{\alpha\beta} = \langle E_\alpha | {\cal A} | E_\beta \rangle \, .
\ee
The ETH goes beyond RMT by assuming that the off-diagonal elements of the matrix $A_{\alpha\beta}$ can be written as normalized elements $R_{\alpha\beta}$ a random matrix  modified by factors involving the microcanonical entropy $S(E)$ and the spectral function $f(E,\omega)$ of the system:
\be
A_{\alpha\beta} = A(E)\delta_{\alpha\beta} + e^{-S(E)/2} f(E,\omega) R_{\alpha\beta} \, ,
\label{eq:ETH}
\ee
where $E = (E_\alpha + E_\beta)/2$ and $\omega = E_\alpha - E_\beta$. The exponential factor implies that the individual off-diagonal elements are suppressed relative to the diagonal elements for a system with a high energy level density, i.\ e., with $S(E) \gg 1$. Then the diagonal matrix elements alone determine the thermal average of the observable:
\ba
\langle{\cal A}\rangle_T &=& Z(T)^{-1} \int \frac{dE}{E}\, e^{S(E)-E/T}\, A(E) ,
\nn \\
Z(T) &=& \int \frac{dE}{E}\, e^{S(E)-E/T} ,
\label{eq:AT}
\ea
with corrections that are exponentially suppressed.

On the basis of the assumption (\ref{eq:ETH}) it is possible to show that the system prepared in an energy eigenstate behaves like a thermal system. A few particularly noteworthy properties are \cite{DAlessio:2015qtq}:
\begin{itemize}
\item The long-time average of an observable equals the thermal average: $\overline{A}=\langle{\cal A}\rangle_T$.
\item The quantum fluctuations of ${\cal A}$ are equal to the thermal fluctuations with corrections of $O(1/N)$, where $N$ is the number of degrees of freedom of the system.
\item The time correlation function of finite-time expectation values $\langle{\cal A}\rangle_t$ obeys a Kubo relation with the function $f(E,\omega)$ as the spectral density.
\end{itemize}
These properties assure that the system, when monitored through the observable ${\cal A}$, is indistinguishable from a thermal system.

The {\em Thouless energy} $E_{\rm Th}$ is defined as the energy difference $\omega$ below which the factor $f(E,\omega)$ in (\ref{eq:ETH}) can be replaced with a constant for most observables, and ETH reaches the RMT limit. In a dynamically evolving system, the relevant range of energy differences $\omega$ is inversely related to the evolution time by the uncertainty relation.  It is often assumed that the behavior of a chaotic quantum system for times longer than the Thouless time $t_{\rm Th}=\hbar/E_{\rm Th}$ is described by RMT. However, it was found in numerical studies of discrete quantum systems that the time until full ETH behavior is established, $t_{\rm ETH}$, can be much longer, depending on the operator under investigation \cite{Dymarsky:2018sef,Richter:2020bkf}. In fact, can be parametrically longer than $t_{\rm Th}$ by a factor proportional to the system size.

The intuitive interpretation of this latter time scale is that the establishment of quantum entanglement requires causal connection across the whole system, and full entanglement is only reached asymptotically. The most important prediction of ETH for a fully entangled system is that if only a small part of the density operator enters an observable, i.e. if the trace is taken over more than half of the states, the observable behaves as for a system in contact with a heat bath. It is tempting to assume that this property explains the experimental success of the thermal model in relativistic heavy ion collisions mentioned in the Introduction. 

The operator dependence of the onset of ETH behavior might actually allow us to understand why the measured yield of hypertritons in p+Pb collisions differs from the prediction of the thermal model. The question is whether the time of hadronization, $t_{\rm H}$, is larger or smaller than $t_{\rm ETH}$. As the hypertriton wave function is much more extended than that of the proton one should expect a strong form-factor suppression of the corresponding matrix element, resulting in an especially large $t_{\rm ETH}$. At the same time $t_{\rm H}$ is especially small for the proton size fireball produced in p+Pb. Both effects suggest that ETH-behavior has not yet established itself when hypertritons are formed in p+Pb collisions, which would explain the observed suppression compared to the thermal model.     

The current consensus is that ETH applies to most, if not all, systems that exhibit quantum chaos. This conjecture has been confirmed in tractable model systems where precise numerical calculations are possible. It leads to the question whether QCD exhibits chaos at the level of the full quantum field theory. We know that nonabelian gauge theories are chaotic at the classical level and exhibit ergodic properties \cite{Bolte:1999th}. Unfortunately, however, we are still unable to construct highly excited energy eigenstates of a system governed by the laws of QCD and confirm the validity of ETH for such systems. The validity of ETH for QCD thus remains a conjecture.

\section{Time scales in the collision}
\label{sec:HIC}

The phenomenology of thermalization in heavy ion collisions is characterized by multiple different time scales. The initial state is very far from thermal equilibrium, but the time available to reach it is limited by the rapid longitudinal expansion of the created quark-gluon system. Hadronization, a confinement-deconfinement transition, takes place during this short time span and drastically changes the properties of the system. This transition is usually modeled as an early "chemical freezout", to be distinguished from complete kinematic freeze-out, which is assumed to occur later. 

The very large momenta of the colliding nuclei imply large Lorentz factors $\gamma$ in the center-of-mass (CM) system, up to $\gamma\sim O(1000)$. Accordingly, the transverse extent of the colliding nuclei at the initial collision instant in the CM system is much larger (by a factor $\gamma$) than their longitudinal thickness. In addition, transverse fluctuations are substantial, of order 50\% for the energy density \cite{Muller:2011bb}, on length scales between the inverse saturation scale $1/Q_s \sim 0.2$~fm and the nucleon radius (0.7 fm). This generates, in combination with the velocity of light $c$ for ballistic processes or the velocity of sound $v_s=c/\sqrt{3}$ for diffusive processes, additional time scales. The various time scales pose a challenge to any holographic description spanning the entire collision. 

However, upon closer inspection one realizes that all these time and length scales can be traced back to three primary scales related to the underlying field theory (QCD), and the initial and spatial boundary conditions:
\begin{itemize}
    \item The energy density $\varepsilon_0$ initially deposited in the collision; in the glasma model $\varepsilon_0 \sim Q_s^4$, where $Q_s$ is the gluon saturation scale \cite{Lappi:2006hq}.
    \item The initial transverse extent $R_T$ of the collision region, which depends on the nuclear radii $R$ and the impact parameter $b$.
    \item The QCD confinement scale $\Lambda_{\rm QCD}$ reflected in the nucleon radius $r_N$ and the pseudocritical temperature $T_c$.
\end{itemize}

There are three analogous independent scales that can be present in the AdS/CFT dual description:
\begin{itemize}
    \item The energy density $\varepsilon_0$ initially deposited in the shock wave collision.
    \item The initial transverse extension $R_T$ of the collision region, which depends on the transverse width of the shock waves and the impact parameter $b$.
    \item In holographic models, which contain an extraneous length scale, this can set the scale for a transition of the Hawking-Page type~\footnote{We refer to any transition between AdS space with a black brane and thermal AdS space as ``Hawking-Page type'' transition, independent of the mechanism that sets the temperature scale at which the transition occurs.} that marks the boundary between a confined and a deconfined phase. For example, in the strong coupling limit of the simplest holographic model dual to ${\cal N}=4$ super-Yang-Mills theory \cite{Maldacena:1997re}, such a transition occurs when the dual theory is considered in global AdS space \cite{Hawking:1982dh}. Similar transitions exist in more realistic holographic models of QCD even when considered on the Poincar\'e patch.
\end{itemize}
When the dual description is constrained to the Poincar\'e patch of AdS space \cite{Bayona:2005nq}, as it is necessary to describe a heavy ion collision, the third scale must be introduced by some modification of the holographic dual that breaks the conformal symmetry, if hadronization is to be described. Examples include models with imbedded D-branes \cite{Kruczenski:2003uq,Sakai:2004cn} and the Scherk-Schwarz compactification of a dual theory with an additional dimension considered by Witten \cite{Witten:1998zw} and used in the study of Aharony {\it et al.} \cite{Aharony:2005bm}.

In recognition of these primary scales we can divide the progression of a relativistic heavy ion collision into four stages, which are summarized in Table~\ref{tab:times}.

\begin{widetext}
\begin{center}
\begin{table}[htb]
\begin{tabular}{|c|c|c|}
\hline
&&\\
time & HIC phenomenology & Holographic dual model\\
&&\\
\hline
&&\\
Stage I & {\bf Hydrodynamization} & {\bf Numerical AdS simulations}\\
$t\leq 1$~fm/c & hydrodynamic attractors lead from  & entropy production mapped by apparent horizon  \\
& transient large amplitude fluctuations   &  $t\leq 0.2$~fm/c equilibration without fluctuations\\ 
& to viscous hydrodynamic expansion  &  $t \leq 1-2$~fm/c equilibration with fluctuations\\ 
&&\\
\hline 
&&\\
Stage II & {\bf Expansion of the QGP fireball} & {\bf Collision of localized shocks \cite{Waeber:2022tts}}\\
$t\leq O(15)$~fm/c & hadronization at QGP surface & (Sections~\ref{sec:Thermalization} and \ref{sec:Model_1})\\ 
&&\\
\hline 
&&\\
Stage III & {\bf Bulk hadronization} & Smoothed {\bf HP transition} \\ 
$t\approx O(15)$~fm/c & chemical freeze-out & (Sections~\ref{sec:Model_1} and \ref{sec:holmod}) \\
&&\\
\hline 
&&\\
Stage IV & {\bf Expansion of HRG}& {\bf Entangled hadrons}  \\
$t\geq O(15)$~fm/c & kinetic freeze-out & network of ER bridges (Section~\ref{sec:Model_2})\\
&&\\
\hline
\end{tabular}
\label{tab:times}
\caption{The different stages of a heavy ion collision (HIC) and their proposed holographic modelling.}
\end{table}
\end{center}
\end{widetext}

We now explain the reasoning behind Table~\ref{tab:times} and describe the roles the primary scales play in the different stages.

{\bf Stage I:} At very short times neither the confinement scale nor the transverse extent of the reaction region are important. During this stage the local energy density far exceeds the critical density $\varepsilon_c$ at the deconfinement transition, and information about the finite transverse size has not spread widely because of causality. The initial energy density $\varepsilon_0$ is the only relevant scale both on the QFT and holographic side of the duality. Both statements may not apply to very small collision systems, such as those produced in proton-proton or proton-nucleus collisions.

{\bf Stage II:} The resulting hydrodynamic QGP expands during the second stage. During this expansion a small fraction of the QGP hadronizes on the surface of the expanding fireball. By construction, this surface is at the confinement-deconfinement transition temperature $T_c$, which is of the order of $\Lambda_{\rm QCD}$. Thus, the expansion dynamics of the QGP fireball depends on the two scales $\varepsilon_0$ and $R_T$, while the hadronization processes at its surface depend, in addition, on $\Lambda_{QCD}$. Therefore, the latter aspect cannot be described in the standard framework of AdS modeling on the Poincar\'e patch, while the hydrodynamical expansion can. 

{\bf Stage III:} At the end of the second stage the remaining fireball becomes thermodynamically and hydrodynamically unstable and completely hadronizes into a quantum entangled hadron gas. The duration of this process depends on the nature of the phase transition. In models with a very large number of colors and a first-order transition the conversion from a gauge plasma to a hadron (glueball) gas takes an extended period of time because the formation rate of color singlet hadrons is small \cite{Aharony:2005bm}. In QCD, where the transition is a rapid, but smooth crossover, the process is deemed to be so fast that it is usually described as instantaneous. Because time evolution in QCD is unitary no thermal entropy is produced, but ETH applies to this stage, explaining the validity of the ``thermal'' model. (For this to apply the time from the start of the collision until this third stage must be long enough for global ETH behavior to be established \cite{Dymarsky:2018sef}, which may not apply to very small collision systems. Numerical simulations in the AdS Poincar\'e patch are also unable to describe this stage adequately.) 

\begin{figure}[htb]
\centering
\includegraphics[width=0.95\linewidth]{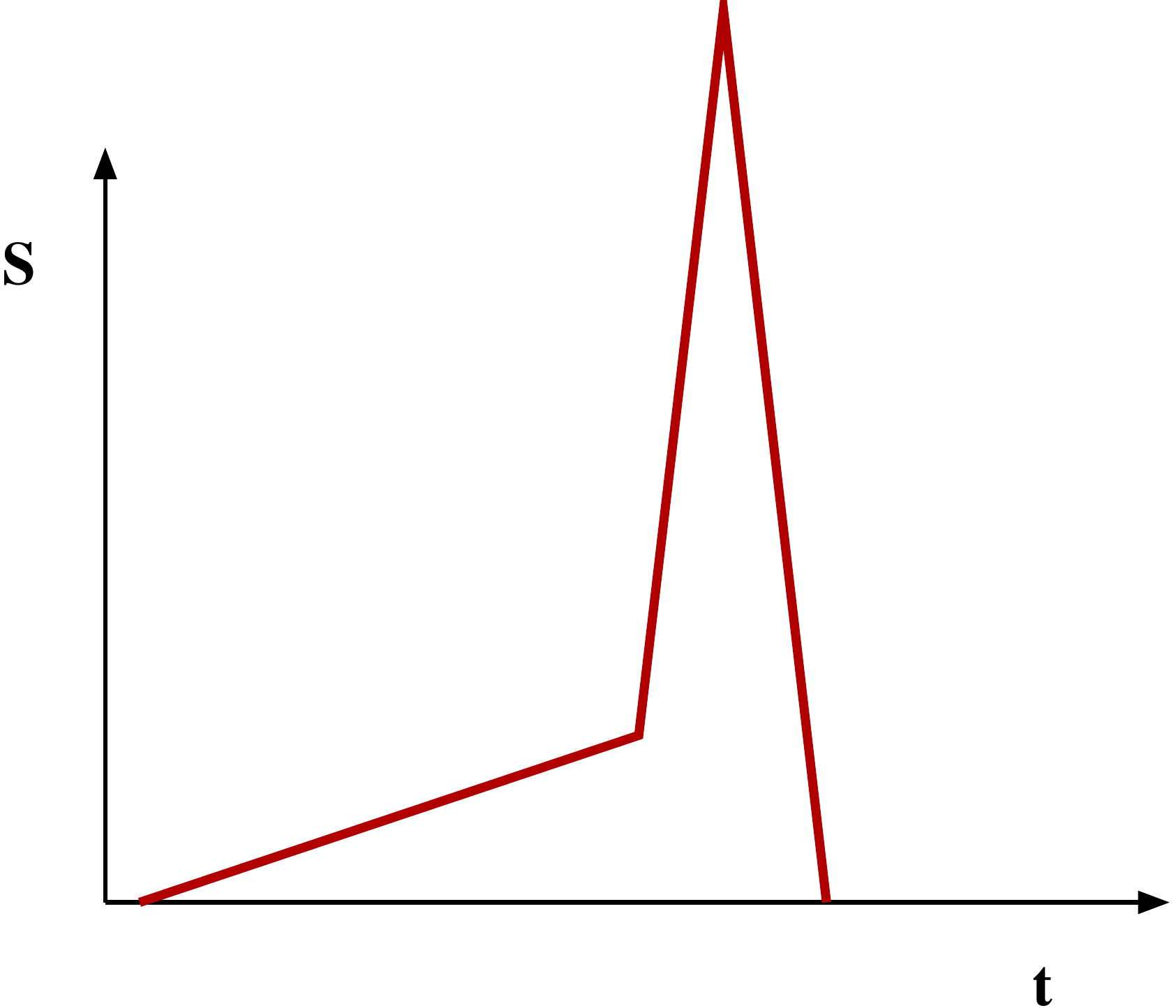}
\caption
{%
Schematic representation of the Page curve for the hadronizing quark-gluon plasma in a relativistic heavy ion collision. The growth rate of the entanglement entropy of the emitted hadrons is low during Stage II because hadrons are only emitted from the surface. The growth rate then increases quickly during Stage III due to bulk hadronization before dropping equally rapidly after half of the QGP has hadronized. At the end of the hadronization transition, the entropy again equals the initial entropy because all hadrons are entangled with each other.}
\label{fig:modPage}
\end{figure}

The transition from (slow) surface hadronization to (rapid) bulk hadronization means that the Page curve for a heavy ion collision has an asymmetric shape. Before the final hadronization transition only a small fraction of the QGP (in Section~\ref{sec:HadronHIC} we will argue for roughly 20\%) has decayed, while the rest decays on a much shorter time scale during Stage III. Figure~\ref{fig:modPage} shows a schematic sketch of the Page curve for this scenario.

{\bf Stage IV:} The final stage of a heavy ion collision, which features an expanding cloud of entangled hadrons, resembles the final stage of a decaying black hole which comprises an expanding cloud of entangled photons without a black hole remnant at the center. Again, numerical simulations limited to the Poincar\'e patch of AdS space are insufficient to capture this entanglement. 

While each stage of a heavy ion collision has a well motivated holographic dual, is is not clear how well holography can model a heavy ion collision in quantitative terms. This is not a concern for the very early stage, when the quark-gluon system is far away from the confinement scale, and a holographic description can be justified by its approximate conformal symmetry and the limited sensitivity of hydrodynamics to  details of the initial state. It is unclear whether tractable holographic models can provide for a quantitative description at later times where more time scales are relevant. 

Even if the holographic model cannot mirror a real heavy ion collision quantitatively, in can potentially provide important conceptual insights with parametric validity. As an example, we consider the final state when the whole system consists of hadrons that are freely streaming toward the detectors. There are two possibilities: Either the hadrons are in thermal equilibrium and described by a density matrix with high entropy, which is the standard assumption made in heavy ion physics, or the hadrons form a highly entangled, nearly pure quantum state with very little entropy, as we argue here. In the first case there exist well established techniques that permit detailed theoretical predictions, whereas any such calculation requires additional assumptions in the latter case. We will argue in the next Section, building on work by van Raamsdonk et al. \cite{VanRaamsdonk:2020ydg}, that the holographic dual of an entangled state of many hadrons differs only in subtle ways from the holographic dual of a QGP fireball and that, therefore, a holographic description can extend smoothly beyond the hadronization transition. 

Another important question is whether the fireball reaches ETH-type behavior before hadronization. This cannot be true for {\it all} observables, as the following argument shows: As already noted, transverse variations in locally conserved quantities equilibrate by diffusion. The associated time scale $R_T^2/D$, where $D$ is the relevant diffusion constant in the QGP, is identical to the one introduced in the previous Section as Thouless time $t_{\rm Th} = 1/E_{\rm th}$. It is useful to estimate this time constant for a typical volume of QGP created in a heavy ion collision. The typical value of a diffusion coefficient in QCD is $D \sim (\pi T)^{-1}$ \cite{Ding:2012sp}. With $R_T \approx 5$ fm and $T \approx 300$ MeV one finds $t_{\rm Th} \sim 120$ fm/c, which is much longer than the lifetime of the QGP. 

The crucial question in this context is, how long it takes until few-particle observables that are amenable to experimental measurement are well approximated by their thermal value. The time it takes for entanglement to propagate throughout the fireball provides for a lower bound. Quantum information propagates with velocity $v_E < c$ yielding this information time scale to be of order $R_T/v_E$. At strong coupling holographic models in $d=4$ dimensions give $v_E \approx 0.62c$ \cite{Liu:2013iza}, which means that the speed of information transport is similar to that for energy transport, which is given by the speed of sound $c_s = c/\sqrt{3}$. This means that entanglement can spread through the entire fireball before the onset of hadronization.

In this work we discuss certain ideas that may ultimately make it possible to extend the successful holographic description of heavy ion collisions beyond the Page time. The crucial feature of any such attempt is the preservation of full quantum coherence which, in turn, requires the addition of novel concepts to the holographic description that become relevant at later times in the collision. As we discussed above, this implies that numerical simulations of AdS gravity  must be expanded to cover the global AdS geometry, not just the Poincar\'e patch.

The underlying idea is that a heavy ion collision proceeds from the highly entangled many-body wave functions of the colliding nuclei to a highly entangled QGP which evolves continuously into a highly entangled multi-hadron state but never to a truly thermal state with large von Neumann entropy. In other words, the collision can be viewed as a unitary S-matrix mapping between the initial and final hadron states. In this language, it is completely irrelevant that in some intermediate state certain local observables look thermal. It is ultimately the interaction of the emitted hadrons with the detectors, which act as incoherent environment, that creates the thermal entropy. This view implies that the establishment of full ETH behavior and thus the success of the thermal model is far less of a mystery, because $t_{\rm Th}$ is much shorter than the time that elapses until final-state hadrons hit the first detectors.

Elucidating the microscopic processes which govern thermalization of many-particle quantum systems is relevant for many subfields of physics. For example, in \cite{Popescu:2006en} it was argued that thermodynamics can be based on the assumption of a pure state wave function of the universe with which all systems investigated in the laboratory are entangled. Clearly, such an assumption requires experimental verification. HICs involving two nuclei colliding in the ultra-high vacuum of the beam-pipe, may be the ``cleanest'' system to obtain such verification or refutation as the evolution of the system can be followed without environmental influence over times many orders of magnitude longer than the intrinsic time scales and with a dynamics which is completely defined by QCD. A huge amount of precise data has already been accumulated and is readily available for theoretical analysis.

It is not clear whether measurements that can detect the presence of complex entanglement features among the emitted hadrons are feasible in heavy ion collisions, just as it is unlikely that a practical test of the entanglement pattern of Hawking radiation could be conducted even if we had access to an isolated evaporating black hole. Examples of observables that might be both, sensitive to the entanglement properties of the emitting system and amenable to measurement, are quantum correlations among hadron spins, especially in small collision systems \cite{Gong:2021bcp}, or among isospins, so-called disoriented chiral condensates \cite{Anselm:1996vm,Mohanty:2005mv}, for which possible indications have been recently observed \cite{ALICE:2021fpb}.  

There is hope that such experiments are possible with atomic physics analogues of black holes \cite{Steinhauer:2015saa,Kolobov:2019qfs}. As QED is also time reversal invariant, QED and QCD  processes should have similar information theoretical properties and, therefore, the experimental verification of ETH behavior and Hawking radiation for QED systems suggest that QCD systems behave in a similar manner. However, while these insights are intellectually satisfying, their practical utility depends on whether or not a holographic description of entanglement and ETH behavior is feasible in practice. This will be discussed in Section \ref{sec:Model_1}.

\section{Holographic Simulations of Early Collision Stages}
\label{sec:Thermalization}

Starting with the pioneering papers \cite{Chesler:2008hg,Chesler:2010bi,Chesler:2013lia} numerical AdS calculations have played an ever more important role to inform our understanding of the early phase of high-energy heavy ion collisions. Over time the numerical techniques have been steadily improved, such that today realistic three-dimensional collisions can be studied \cite{Waeber:2022tts} instead of somewhat schematic collisions of infinite plane shock waves. Also, more and more details and special cases were studied \cite{Heller:2012km,Casalderrey-Solana:2013sxa,Chesler:2015fpa,Ecker:2016thn,Casalderrey-Solana:2016xfq,Endrodi:2018ikq,Waeber:2019nqd,Muller:2020ziz}. In parallel, calculations of the entropy density \cite{Gubser:1998nz},  shear viscosity \cite{Buchel:2008sh}, conductivity \cite{Waeber:2018bea} and inverse equilibration times \cite{Waeber:2018bea} at NLO in the string coupling were found to be in good agreement with QGP phenomenology. As these corrections can be encoded in higher derivative terms of classical gravity, there is a possible path to more quantitative holographic simulations of the early collision stages.

Numerical solutions of the classical Einstein equations with AdS boundary conditions share the difficult problem of having to deal with the diffeomorphism invariance of general relativity. Using a suitable metric as an {\it ansatz} one can significantly reduce the diffeomorphism freedom. Chesler and Yaffe used Eddington-Finkelstein coordinates and managed to rewrite the Einstein equation as a system of nested ordinary differential equations, which they solved using functional methods (Chebychev functions). A nice review of early developments can be found in \cite{Chesler:2015lsa}. In recent years activity has somewhat abated although there is still continuous progress, see e.g. \cite{Waeber:2022tts}. Already early on these simulations showed that AdS dynamics leads to rapid hydrodynamization  \cite{vanderSchee:2013pia,Chesler:2015fpa} with sizeable transverse flow \cite{Chesler:2015wra}. 

The prevailing strategy has been to merge these holographic simulations of the early stages of heavy ion collisions with statistical descriptions of the later stages of the collisions \cite{vanderSchee:2013pia}. Early calculations focused on highly symmetric settings in order to reduce the computational demands, but more recently asymmetric settings have been studied, and by now semi-realistic heavy ion collisions can be simulated \cite{Heller:2012km,Chesler:2015fpa,Ecker:2016thn,Casalderrey-Solana:2016xfq,Endrodi:2018ikq,Waeber:2019nqd,Muller:2020ziz}. In particular, these studies have shown that holographic simulations reproduce relativistic viscous hydrodynamics, validating the hybrid holographic-transport approach, see \cite{Schenke:2021mxx} for a recent review. In the present context it is relevant that hydrodynamization occurs at a fixed proper time \cite{Waeber:2019nqd} and that AdS dynamics and viscous hydrodynamics are indistinguishable for hydrodynamic observables close to hydrodynamization such that the whole QGP fireball shows hydrodynamic behavior at nearly the same proper time. 

Holographic calculations are attractive because they allow to treat problems that are too difficult to solve using QFT techniques. The hope is that the successful application of holographic techniques, which has been realized for the early stages of heavy ion collisions, can be extended to the later collision stages. This hope is nurtured by several observations documented in the literature, which we will discuss now.

The insight that quantum entanglement in the field theory is holographically encoded in features of the higher dimensional geometry was most clearly expressed by Maldacena and Susskind \cite{Maldacena:2013xja} who argued, building on ideas formulated in \cite{Israel:1976ur,Maldacena:2001kr}, that the maximally extended (eternal) AdS-BH geometry is dual to a thermofield double state in the quantum field theory, which means that the quantum states of the field theory in the two asymptotic regions are maximally entangled. More generally, in this picture, the entanglement between the quantum states in two different space-time regions is geometrically represented by an Einstein-Rosen (ER) bridge between two regions in AdS space. We refer the reader to \cite{Almheiri:2019hni} for a review of the many publications building on this paradigm and as well as its relation to recent progress in understanding the role of entanglement for BH decay by Hawking radiation. 

In \cite{Maldacena:2013xja} the label ``ER = EPR'' was coined for this connection, where EPR stands for Einstein-Podolsky-Rosen and, more generally, for any form of entanglement in the quantum field theory. For our discussion Fig.~13 in \cite{Maldacena:2013xja} is most stimulating as it suggests that the emission of entangled Hawking radiation can be described as creation of higher dimensional Einstein-Rosen bridges between the decaying BH and the emitted particles. This picture, however, also displays an obvious problem of this idea: In a high-energy heavy ion collison thousands of hadrons are produced allowing for a plethora of entanglement patterns, far more than the number of simple geometries available for the dual picture. 

In this context the recent ideas advanced by van Raamsdonk {\it et al.} \cite{VanRaamsdonk:2016exw,VanRaamsdonk:2020ydg, VanRaamsdonk:2020tlr, Raamsdonk:2020tin,May:2020tch} are relevant, which posit that one large connected region of conformal field theory (CFT) and several smaller entangled domains of the same CFT have a nearly indistinguishable holographic dual. If it were legitimate to interpret the large region as QGP fireball and the many small domains as individual hadrons this would imply that the crossover transition from QGP to hadron gas could be continuous in a holographic dual description. This notion instills hope that a holographic description of the hadronization process is feasible and practical.\footnote{Our concept inverts van Raamdonk's reasoning. His aim was to show that continuous space can emerge from the entanglement of isolated domains; we want to describe entanglement among isolated components of the final state geometrically.} The dual description would be the disintegration of one large AdS black hole into several small ones as discussed, e.g., in \cite{Maldacena:2013xja} in the contact of black hole evaporation. 

The analogy is now clear. Quarks and gluons are confined to the interior of hadrons. Therefore, a diluted gas of thousands of hadrons can be interpreted in analogy to many isolated conformal field theories (CFT) on the edge. As each of these covers only a small faction of the fireball or expanding hadron cloud, in the bulk AdS direction this leads only to noticeable effects very close to the asymptotic region of AdS space. (In order to distinguish this asymptotic region from spatial boundaries of localized states, we will use the term ``AdS edge''. Such localized systems are usually modeled in the framework of boundary conformal field theory (BCFT) and its holographic dual description \cite{Takayanagi:2011zk}.) Deeper inside the bulk, all hadronic regions will remain connected via a network of ER bridges. As a result, the dual geometry will be very similar to that of the QGP fireball which is just one large AdS-BH. This means that from the information theoretical point of view there is no marked difference between entangled quarks and gluons with large relative momentum in the early fireball phase, a mixed QGP plus hadron gas state at intermediary times, and an entangled purely hadronic state at late times and that, therefore, many properties of this system can be calculated from the dual description valid at early times. In Sections \ref{sec:Model_1} and \ref{sec:Model_2} we will expand on this from the AdS perspective.
 
To summarize, we argued that a successful extension of AdS calculations to the description of hadronization requires their extension from the Poincar\'e patch to the global AdS geometry and the inclusion of the scale $L$. We will suggest in Sections~\ref{sec:Model_1} and \ref{sec:Model_2} simplified models as first steps towards this long-term goal.

\section{Hadronization in Relativistic Heavy Ion Collisions}
\label{sec:HadronHIC}

At very high energies, the valence quarks of two colliding nuclei effectively pass through each other and deposit some of their energy on a time scale much shorter than 1 fm/c, a process that can be modeled as a quantum quench. Following this quench. the deposited energy thermalizes and forms a quark-gluon plasma on a time scale of the order  $\tau_{\rm th} \sim O(1/T)$ where $T$ denotes the temperature after thermalization. The thermalization process can be studied in holographic models, either by energy shell collapse in 5-dimensional Anti-de Sitter space (AdS$_5$) \cite{Balasubramanian:2010ce,Balasubramanian:2011ur} or by numerically solving shock front collisions in AdS$_5$ \cite{Chesler:2010bi,Chesler:2013lia}. 

The quark-gluon plasma then expands hydrodynamically while hadronization occurs in regions where the temperature reaches $T_c \approx 150$ MeV. At high collision energies the evolution of the matter near the center of momentum of the two nuclei is approximately boost invariant and can be described in terms of Milne coordinates $\tau, \eta, x, y$, where $\tau = \sqrt{t^2-z^2}$ is the proper time and $\eta = \frac{1}{2}\ln[(t+z)/(t-z)]$ is called space-time rapidity. Massless particles moving along lines of constant $\eta$ also have rapidity $\eta$ in momentum space.  Boost invariance means that to a good approximation the evolution is a function of $\tau$ and the transverse coordinates $x,y$ only for moderate values $|\eta| \ll \eta_{\rm beam}$, where $\eta_{\rm beam}$ is the beam rapidity in the center-of-momentum frame.

\begin{figure}[htb]
\centering
\includegraphics[width=0.95\linewidth]{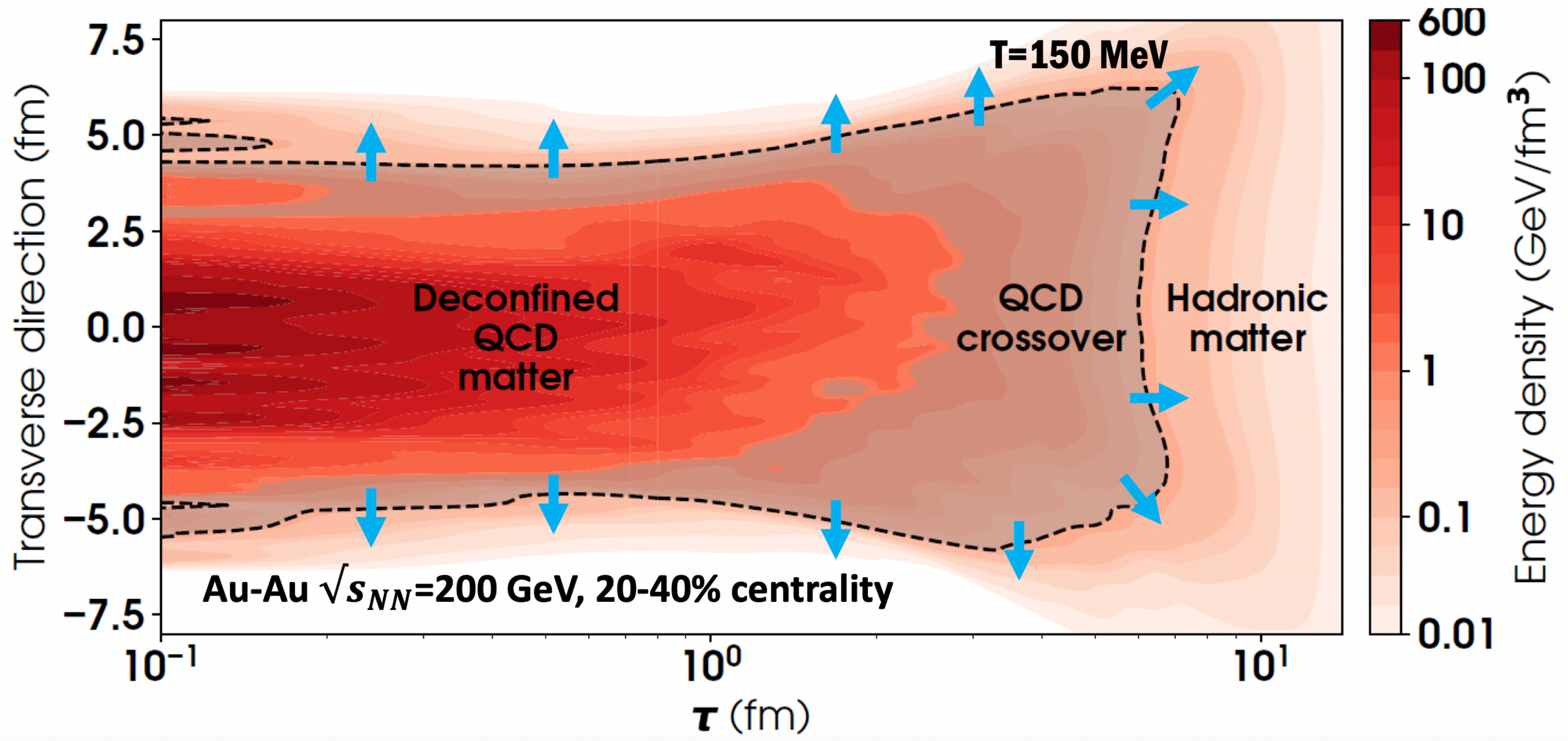}
\caption
{%
Contour plot of the evolution of the energy density in a midcentral Au+Au collision at the highest RHIC energy $\sqrt{s_{\rm NN}} = 200$ GeV. The horizontal axis shows the proper time $\tau$, the vertical axis shows one of the transverse coordinates. The black dashed line delineates the hadronization hypersurface $T = T_c = 150$ MeV \cite{Gale:2020dum}.} 
\label{fig:hydro}
\end{figure}

Hydrodynamics simulations show that the hypersurface where the transition from quark-gluon plasma (deconfined matter) to hadrons (confined matter) occurs is composed of two main domains (see Fig.~\ref{fig:hydro}). One domain is time-like and located approximately at constant $\sqrt{x^2+y^2} \equiv r \approx R$, where $R$ is the nuclear radius, and stretches from $\tau_{\rm th}$ to a time $\tau_c$ when the quark-gluon plasma converts to hadrons in bulk. The bulk hadronization proceeds along a space-like hypersurface given by $r < R$ and $\tau = \tau_c$, which defines the second domain.

The hadronization processes in these two domains are different: On the time-like boundary of the quark-gluon plasma, hadron emission can be visualized as surface radiation. The emission rate is given by
\be
dE_{\rm had}^{\rm surface}/d\tau = \int dx\, dy\, \tau d\eta\, \delta(r-R)\, T^{0i}n_i ,
\ee
where $n_i$ is the outward directed (space-like) normal vector on the hadronization hypersurface. Introducing the notation $S_c = T^{0i}n_i$ for the energy flow density at the hadronization temperature $T_c$, one finds
\be
dE_{\rm had}^{\rm surface}/d\eta = 2\pi R \int_{\tau_{\rm th}}^{\tau_c} \tau\, d\tau\, S_c
\approx \pi R\, \tau_c^2\, S_c ,
\ee
since $\tau_c \gg \tau_{\rm th}$. The energy converted into hadrons on the space-like hadronization surface is similarly given by 
\be
dE_{\rm had}^{\rm bulk}/d\eta = 2\pi \int_0^R r\, dr\, \tau_c\, \varepsilon_c
\approx \pi R^2\, \tau_c\, \varepsilon_c ,
\ee
where $\varepsilon_c$ is the energy density of matter at temperature $T_c$. 

For noninteracting hadrons with mass $m$ at temperature $T = 1/\beta$, neglecting quantum statistics, one has
\ba
S(T,m) &=& \frac{T^4}{4\pi^2} \left( 3(1+\beta m) +(\beta m)^2 \right) e^{-\beta m} \\
\varepsilon(T,m) &=& \frac{T^4}{2\pi^2} \left( 3(\beta m)^2 K_2(\beta m) + (\beta m)^3 K_1(\beta m) \right) . 
\nn
\ea
When summed over all well established hadron species in the Particle Data Book weighted by their statistical degeneracies $d_i$, one finds
\be
\frac{S_c}{\varepsilon_c} \equiv 
\frac{\sum_i d_i S(T_c,m_i)}{\sum_i d_i \varepsilon(T_c,m_i)}
\approx 0.17 .
\ee
Putting everything together and considering that $\tau_c \approx R$, one obtains
\be
\frac{dE_{\rm had}^{\rm surface}/d\eta}{dE_{\rm had}^{\rm bulk}/d\eta} 
\approx \frac{\tau_c\, S_c}{R\, \varepsilon_c} \approx 0.15 - 0.20 .
\ee
This means that approximately $80-85$\% of the hadrons produced by hadronization of the quark-gluon plasma are created during the bulk transition.\footnote{As some regions at the fringe of the nuclear fireball may never become hot enough for deconfinement to occur, a small fraction of the final state hadrons may be produced directly without going through an intervening plasma phase. This phenomenon is described in so-called core-corona models \cite{Werner:2007bf}. The relative magnitude of the corona contribution to hadron production shrinks with increasing size of the collision region and increasing collision energy \cite{Petrovici:2017izo}.} 

\begin{figure}[htb]
\centering
\includegraphics[width=0.9\linewidth]{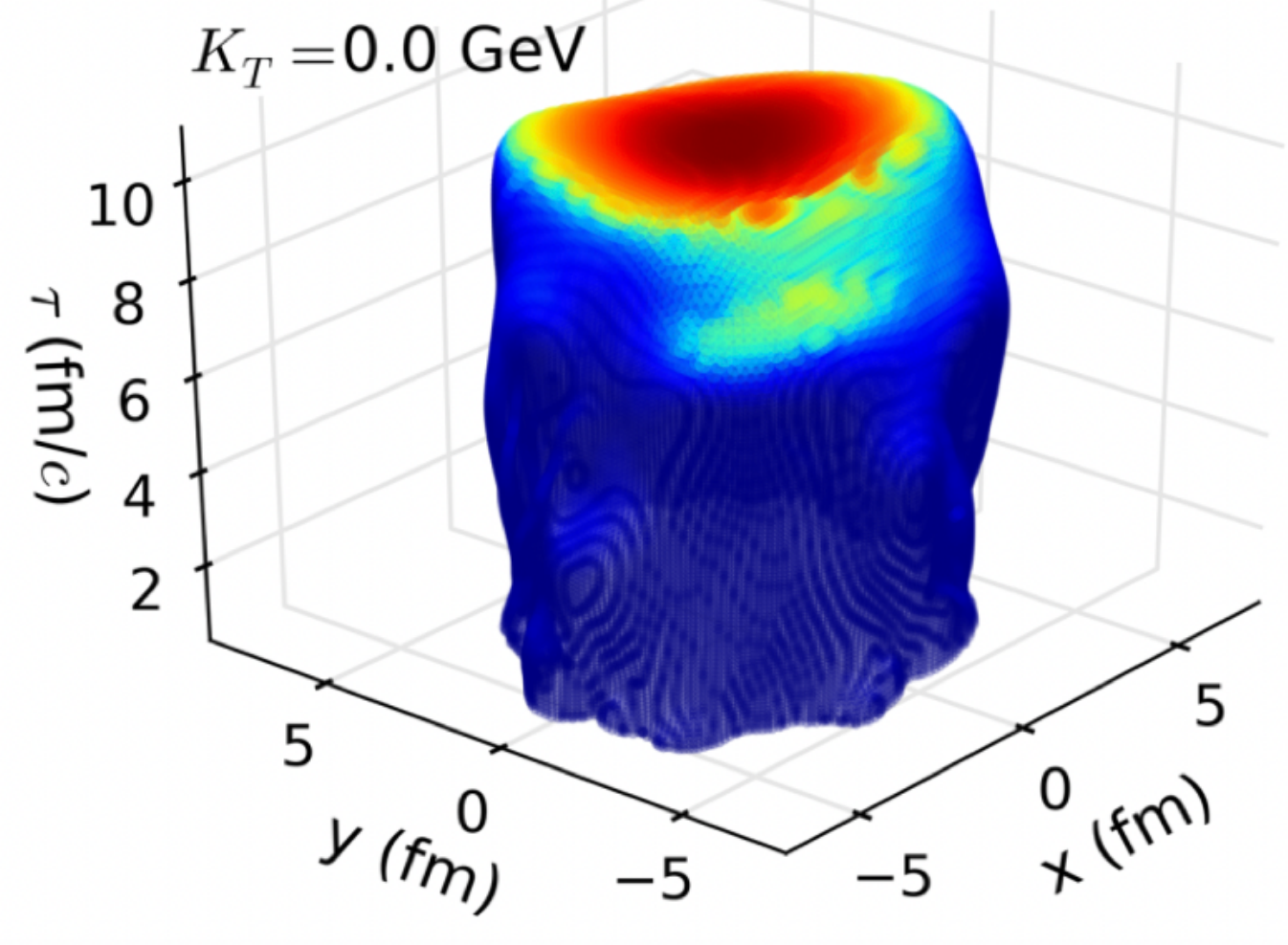}
\\
\vspace{16pt}
\includegraphics[width=0.9\linewidth]{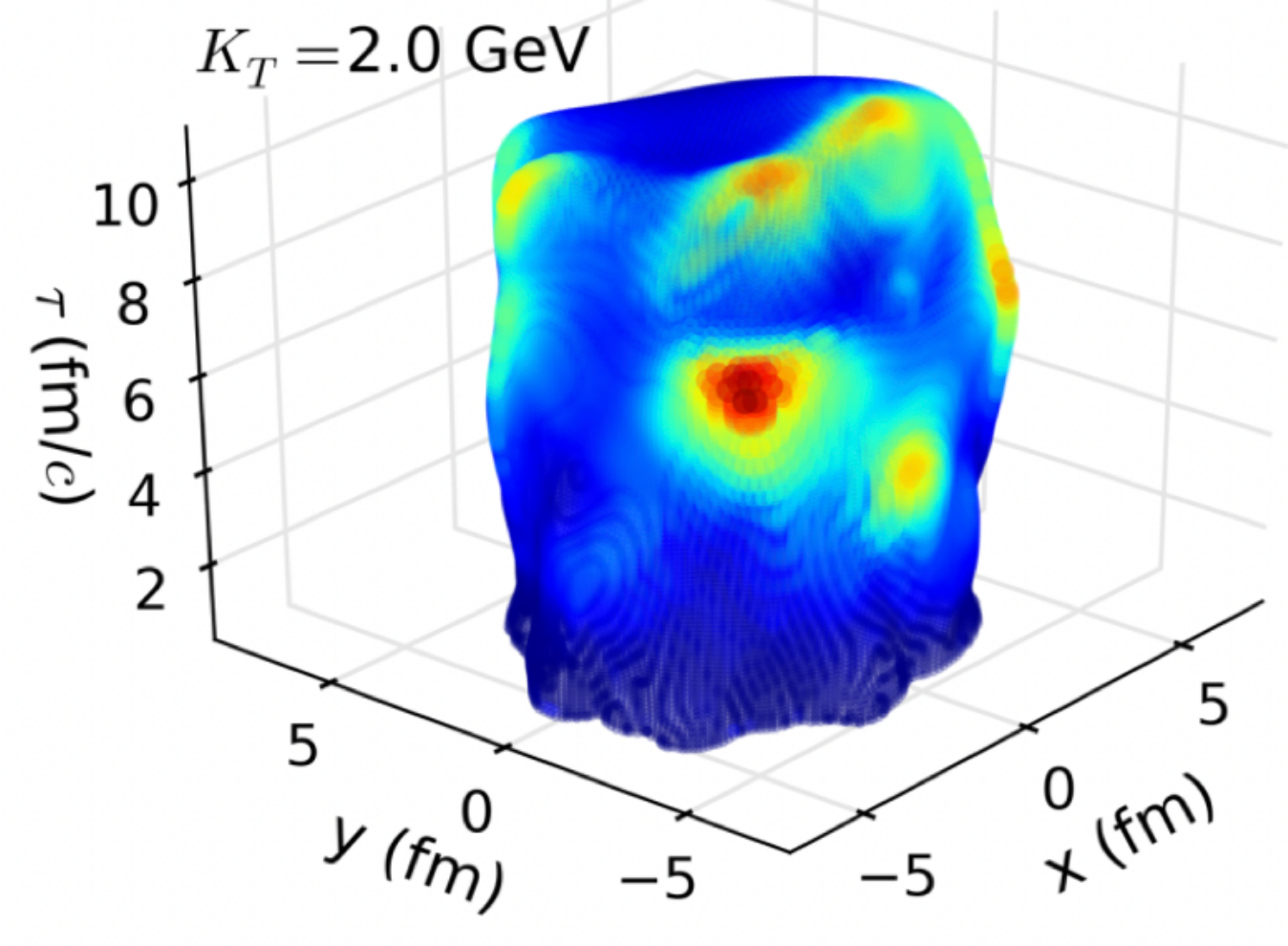}
\caption
{%
Intensity plots of the emission hypersurface for hadron pairs in Au+Au collisions at $\sqrt{s_{\rm NN}} = 200$ GeV (from \cite{Plumberg:2015eia}). Upper panel: Hadron pairs with $K_T = 0$, selectively weighting hadrons from bulk hadronization. Lower panel: Hadron pairs with $K_T = 2$ GeV/c, selectively weighting hadrons from surface radiation. The emission intensity is highest in the dark red regions and lowest in the dark blue regions.} 
\label{fig:hadron_pairs}
\end{figure}
The two hadronization mechanisms can be experimentally separated by measuring the emission of hadron pairs instead of single hadrons. Hadron pairs radiated off the surface of the outward flowing fireball carry a large outward directed total momentum $K_T = p_{T,1}+p_{T,2}$. On the other hand, hadron pairs formed when the QGP hadronizes in bulk have on average $K_T = 0$. This is nicely seen in Fig.~\ref{fig:hadron_pairs}. where the emission hypersurface of hadron pairs at midrapidity for the same Au+Au collisions at RHIC is shown in the $(x,y,t)$ space, where $x,y$ are the transverse coordinates \cite{Plumberg:2015eia}. Just as in Fig.~\ref{fig:hydro} the hypersurface resembles a capped cylinder. The upper panel shows the emission points of hadron pairs with $K_T = 0$; the lower panel shows the emission points with $K_T = 2$ GeV/c. Red color indicates the highest rate of emission; blue color indicates the lowest emission rate.

\section{Hadronization as an AdS phase transition}
\label{sec:Model_1}

The Einstein equations with a negative cosmological constant have two solutions with asymptotic AdS geometry. One is plain AdS space; the other is a black hole (AdS-BH) imbedded in AdS space. The black hole geometry is parametrized by the Schwarzschild horizon (we will use the notation $r_h$). As we shall discuss, $r_h$ is related to the color screening distance in the dual gauge theory. At thermal equilibrium, $r_h$ is uniquely determined by the temperature $T = 1/\beta$ encoded in the period of the Euclidean version of the geometry \cite{Gibbons:1976ue}. 

The Euclidean action plays the role of free energy in the space of geometries; its minimum thus corresponds to the stable equilibrium state at a given temperature. At low values of $T$, plain AdS space ($r_h=0$) has the lowest Euclidean action; above a certain temperature $T_c$, which depends on the details of the dual gravity theory, the AdS-BH geometry has the lowest free energy. As mentioned in a footnote in Section \ref{sec:HIC}, we here refer to such transitions as ``Hawking-Page type'' transitions \cite{Hawking:1982dh}. In the simplest holographic model, the pure $(d+1)$-dimensional Einstein action with negative cosmological constant $\Lambda = -d/L^2$ on global AdS space, the transition is discontinuous with $r_h(T_c) = L$, which means that the Hawking-Page transition is a first-order phase transition. For a detailed derivation and the exact relation between $r_h$ and $T$, see e.~g.\ Ref.~\cite{Witten:1998zw}. 

\begin{figure}[htb]
\centering
\includegraphics[width=\linewidth]{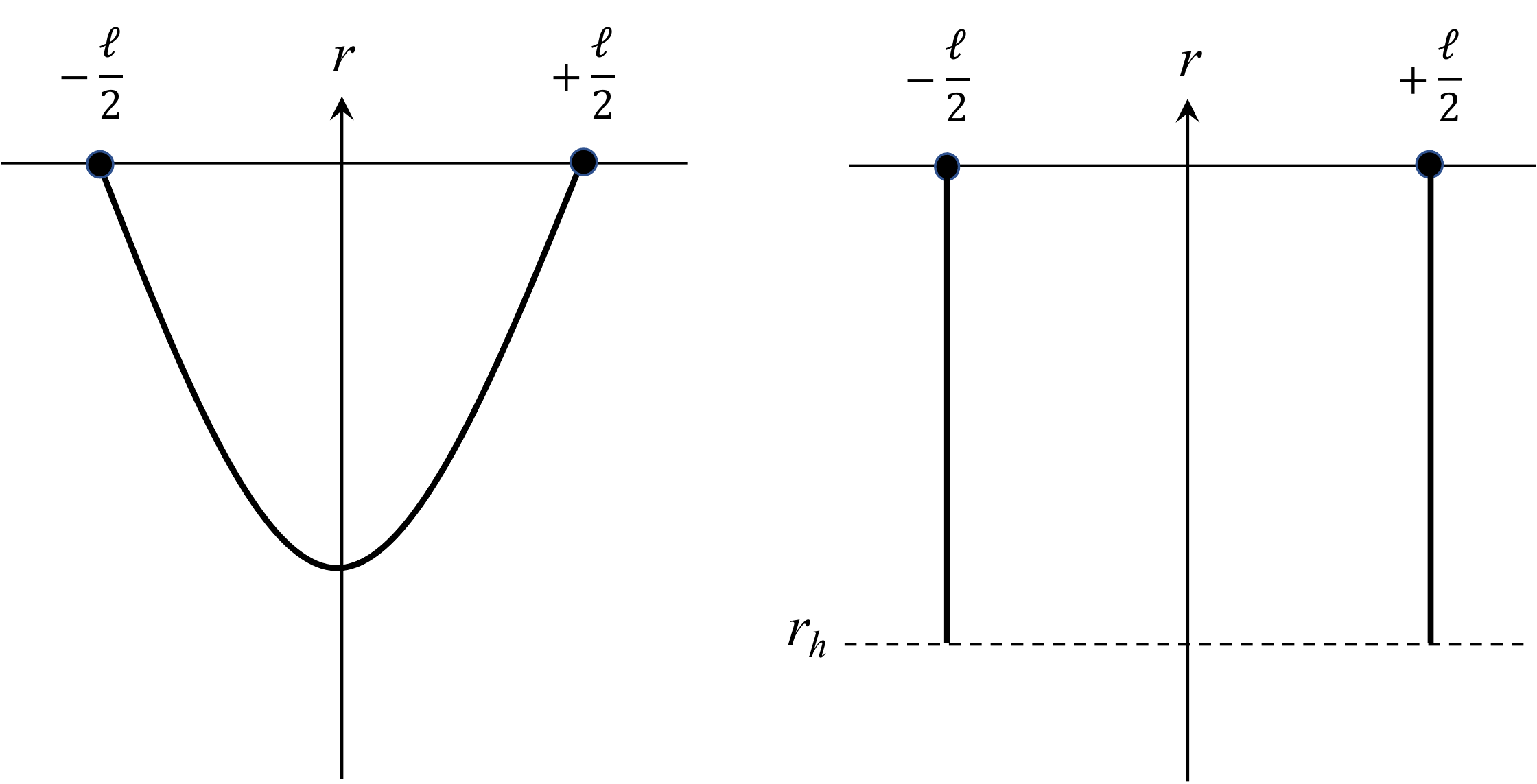}
\caption
{%
String configurations in the AdS bulk connecting a static quark-antiquark pair separated by distance $\ell$. The left panel shows the string in the plain AdS geometry; the right panel shows the two disconnected strings reaching from the quark (antiquark) to the black hole horizon $r_h$ in the AdS-BH geometry.} 
\label{fig:QQstring}
\end{figure}

In the dual gauge theory, the Hawking-Page transition corresponds to the confinement-deconfinement transition \cite{Witten:1998zw,Aharony:2003sx}. This can be seen in multiple ways. The most commonly used argument is that the free energy of the thermal plain AdS geometry is described by a Hagedorn spectrum of gauge-singlet excitations, whereas the free energy of the AdS-BH geometry is proportional to $N_c^2$. Maybe the most intuitive argument is obtained by considering the potential of a heavy quark-antiquark pair, i.~e., static objects carrying color charge in the fundamental representation of the gauge group \cite{Maldacena:1998im}. In the gauge theory, this potential is determined by a Wilson loop connecting the world lines of the quark-antiquark pair. In the holographic dual representation the potential is determined by the Nambu-Goto action of a string connecting the quark-antiquark pair through the bulk (see left panel of Fig.~\ref{fig:QQstring}). 

For the conformally invariant ${\cal N}=4$, large-$N_c$ super-Yang-Mills theory and for the plain AdS geometry in the bulk, the potential is found to be \cite{Maldacena:1998im}
\be
U(\ell) = - \frac{4\pi^2\sqrt{2\lambda}}{\Gamma(1/4)^4\,\ell} ,
\ee
where $\lambda = g^2N_c$ is the 't Hooft coupling and $\ell$ is the quark-antiquark separation. (There is no confining potential because the theory is conformally invariant.) For the AdS-BH geometry one finds that the potential $U(\ell)$ vanishes for separations
\be
\ell > \ell_s \approx 0.869 L^2/r_h ,
\ee
which defines the screening distance of the color force in the gauge theory \cite{Rey:1998bq,Brandhuber:1998bs,Liu:2006nn}. The screening occurs, because for $\ell > \ell_s$ the lowest energy string configuration corresponds to a disconnected pair of strings stretching straight from the quark (antiquark) to the black hole horizon $r_h$ (shown in the right panel of Fig.~\ref{fig:QQstring}). We can thus consider $r_h$ as a geometric parameter related to the color screening length in the gauge theory. Since for $T \gg T_c$ the black hole radius is related to the temperature $T$ at thermal equilibrium by $r_h = (4\pi/n)L^2T$, this translates into a thermal color screening length $\ell_s \approx 0.869/(\pi T)$in $d=4$ space-time dimensions.

\begin{figure}[htb]
\centering
\includegraphics[width=0.95\linewidth]{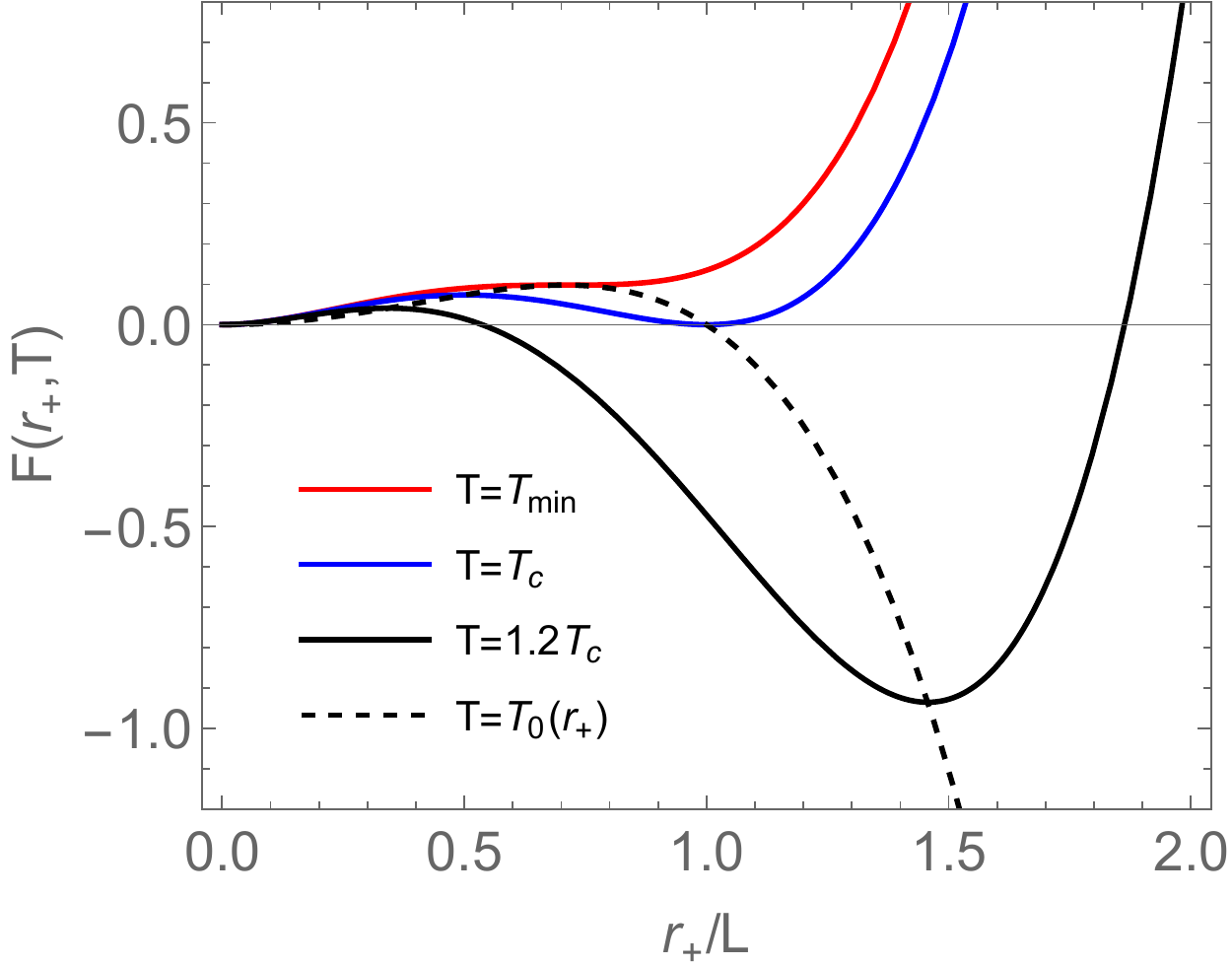}
\caption
{%
The solid curves show the free energy $F(r_h,T)$ for AdS$_5$ as function of the black hole radius $r_h$ for three different temperatures $T$: $T=T_{\rm min}$ (red upper curve), $T=T_c$ (blue middle curve), and $T=1.2T_c$ (black lower curve). See text for the definition of $T_{\rm min}$ and $T_c$. The dashed line shows the equilibrium (on-shell) free energy $F(r_h,T_0(r_h))$ which traces out the extrema in the family of free energy curves. The free energy is shown in units of $L/G$ where $G$ is the gravitational constant.}  
\label{fig:EnergyF}
\end{figure}

It is possible to relax the firm connection between AdS-BH radius $r_h$ and the temperature $T$ by allowing for geometries with a conical singularity \cite{Eune:2013qs}. Such geometries do not correspond to a minimum of the free energy but they allow for a smooth interpolation between the equilibrium AdS-BH geometry and the thermal AdS geometry. This is illustrated in Fig.~\ref{fig:EnergyF} for AdS$_5$. The solid lines show the free energy $F(r_h,T)$ for several different values of $T$, and the dashed line traces the location of the minima. The uppermost solid (red) curve is for the lowest temperature $T_{\rm min}$ for which an AdS-BH solution of Einstein's equations exists, $T_{\rm min} = \sqrt{2}/(\pi L)$, the middle solid (blue) curve represents the free energy at the temperature of the Hawking-Page phase transition, $T_c = 3/(2\pi L)$, and the lowest solid (black) curve shows the free energy for $T = 1.2T_c$. 

As is evident from Fig.~\ref{fig:EnergyF}, the Hawking-Page transition for the conformal, large-$N_c$ super-Yang-Mills theory is a first-order phase transition. This is mirrored in the behavior of the pure non-supersymmetric SU($N_c$) gauge theory, which also exhibits a first-order deconfinement transition for $N_c \geq 3$ \cite{Panero:2009tv}. Gravity dual models that resemble pure SU($N_c$) gauge theory more quantitatively can be constructed by adding a dilaton field to 5-dimensional Einstein gravity, which describes the running of the gauge coupling constant. In general, dilaton gravity duals of confining gauge theories exhibit a first-order deconfinement phase transition similar to that of the conformal super-Yang-Mills theory \cite{Gursoy:2008za}.

On the other hand, the deconfinement transition in QCD is known to be a smooth crossover \cite{HotQCD:2018pds,Borsanyi:2020fev}. The thermal properties of dual dilaton gravity models including fundamental matter (quarks) that mimic the running coupling and chiral properties of QCD have been studied \cite{Mandal:2011ws}, and models that can change from a first-order phase transition to a crossover transition have been constructed \cite{Attems:2018gou}. 

The dynamics of the bulk transition depends on the rate at which the temperature drops. If the cooling rate is slow compared with the microscopic times scales, the transition will proceed at or near $T_c$ through a mixed phase via bubble formation \cite{Aharony:2005bm,Bigazzi:2020phm,Janik:2021jbq}. In the case of rapid cooling, the transition occurs via a Gregory-Laflamme instability \cite{Gregory:1993vy} from a supercooled phase at $T_{\rm min}$ \cite{Hubeny:2002xn,Mandal:2011ws,Buchel:2015gxa,Dias:2016eto,Yaffe:2017axl}. The second scenario prevails in the limit $N_c \to \infty$, but for holographic models of QCD with $N_c=3$ it is likely that the first scenario is realized under the conditions of a heavy ion collision. For a smooth crossover certainly the first scenario is realized.

\section{Hadron emission as analogue of Hawking radiation}
\label{sec:Model_2}

\begin{figure}[htb]
\centering
\includegraphics[width=0.95\linewidth]{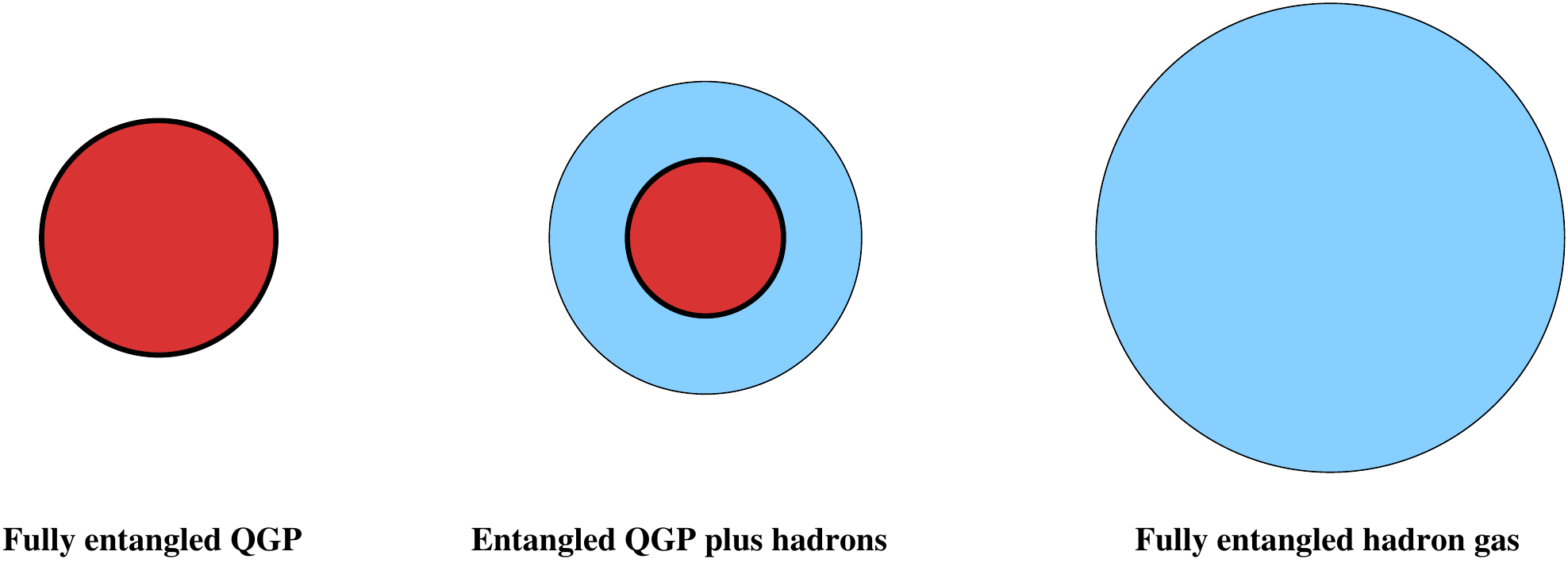}
\caption
{%
Sketch of the transition from a highly entangled QGP fireball to a highly entangled hadron gas based on the fact that the time-reversal invariance of QCD forbids creation of von Neumann entropy.} 
\label{fig:model1}
\end{figure}

Decades of work by many theorists were needed to reach the present level of understanding of the black hole information puzzle. Understanding the mechanisms ruling the generation of entanglement in hadronization is most probably a problem of comparable difficulty. Therefore, it is tempting to profit from the insights of the black hole community by constructing a model which treats hadronization in analogy to Hawking radiation. In this analogy the hadrons correspond to the Hawking radiation outside of the black hole horizon as hadrons cannot exist within the QGP, so its surface presents a horizon for them. Photon pair creation at the black hole horizon corresponds to hadronic particle-hole production at the QGP surface where the ingoing hole state gets absorbed but transfers its entanglement with the outgoing hadron to the interior of the (shrinking) QGP fireball in analogy to island formation in black hole decay \cite{Almheiri:2020cfm}, see the sketch in Fig.\ref{fig:model1}.  

The main idea motivating this analogy is that black hole decay leads to a Page curve for the entropy of Hawking radiation outside of the horizon which is what we also expect for the hadronic state outside of the QGP. Complete entanglement among the hadrons is only reached at the end of hadronization. 

There exist various phenomenological observations on the QCD side which could fit into our model. As explained in the Introduction the uniform temperature of all observed hadron yields, which is well described by the thermal hadron resonance gas model, is difficult to understand. A possible line of argument was suggested by us in Ref.~\cite{Muller:2017vnp} based on the phenomenologically very successful quark recombination model \cite{Fries:2008hs}. In this model quarks and gluons from the fireball coalesce at its surface to form hadrons. In doing so their energies $E_i$ add up and their probability densities multiply, generating for any Fock-state of the hadron $h$ with energy $E_h$ the common factor \cite{Muller:2005pv} 
\begin{equation}
\prod_i e^{-E_i/T_{\rm ch}} = e^{-(\sum_i E_i)/T_{\rm ch}} = e^{-E_h/T_{\rm ch}} ,
\end{equation}
where the subscript ``ch'' indicates that the temperature parameter is derived from the chemical composition of the hadron gas.

An {\em ad hoc} feature of this model, dictated by phenomenology, is that the temperature of the QGP fireball and that of all produced hadrons has the same value $T_{\rm ch}$. This feature is not easily understood without entanglement, because different hadrons scatter differently and thus are not expected to decouple at the same time from the expanding fireball. This could be naturally explained if a highly entangled QGP state and a highly entangled hadronic state are basically indistinguishable as we argued above based on the ideas of van Raamsdonk \cite{VanRaamsdonk:2020ydg}.

\section{A holographic picture of hadronization}
\label{sec:holmod}

As long as information about the state of a time reversal invariant quantum system is not lost by any kind of measurement and the associated (partial) collapse of the complete many-body wave function, it must be possible ``to run the movie backwards''. The highly complex initial state $\Phi_\mathrm{i}$ of a relativistic heavy ion collision has nearly zero entropy.\footnote{The ground state of a colliding nucleus is unique, and any interaction of the nucleus with the accelerator structure is completely negligible on the scale of nuclear excitations. Finally, although the two nuclei can be Coulomb excited on their approach to each other, the excitation is coherent and does not change the fact that the nuclear quantum state is pure.} In the parton (quark-gluon) basis, this state is characterized by a density matrix with two blocks describing the two nuclei approaching each other. This entangled initial state evolves by a unitary transformation into another highly entangled final state $\Phi_\mathrm{f}$ characterized by a full density matrix in the parton basis. Eventually, this final many-parton quantum state is projected onto hadron states by experimental measurements, which identify the asymptotic eigenstates of the many-parton system. This happens at a time of order $O(10^{-9}~{\rm s})$, much longer than the duration of the nuclear reaction which is of order $O(10^{-23}~{\rm s})$. Since the detector acts as a heat bath, this leads to decoherence and thus to entropy production. 

For a realistic holographic description of a relativistic heavy ion collision that includes hadronization it is crucial to describe entanglement at each time. This is not possible in numerical solutions of the classical Einstein equations, which depend only on the classical, local energy momentum tensor on the AdS edge.\footnote{We remind the reader that use the term ``AdS edge'' to avoid confusion of the asymptotic region of AdS space with the boundary of the QCD fireball.} It is thus unclear whether an approximately valid AdS model that keeps track of entanglement during a transition of the Hawking-Page type exists. We argue in this Section that it does. 

Our argument rests crucially on the fact that for a many-particle quantum state, such as the final state of a heavy ion collision, the effects of entanglement are only relevant if the complete wave function is considered, but are negligible for observables involving only a few hadrons. This is a consequence of the "monogamy of entanglement" which is well established in quantum information theory. 

\begin{widetext}
\phantom{Lots of text}
\begin{figure}[htb]
\centering
\includegraphics[width=0.4\linewidth]{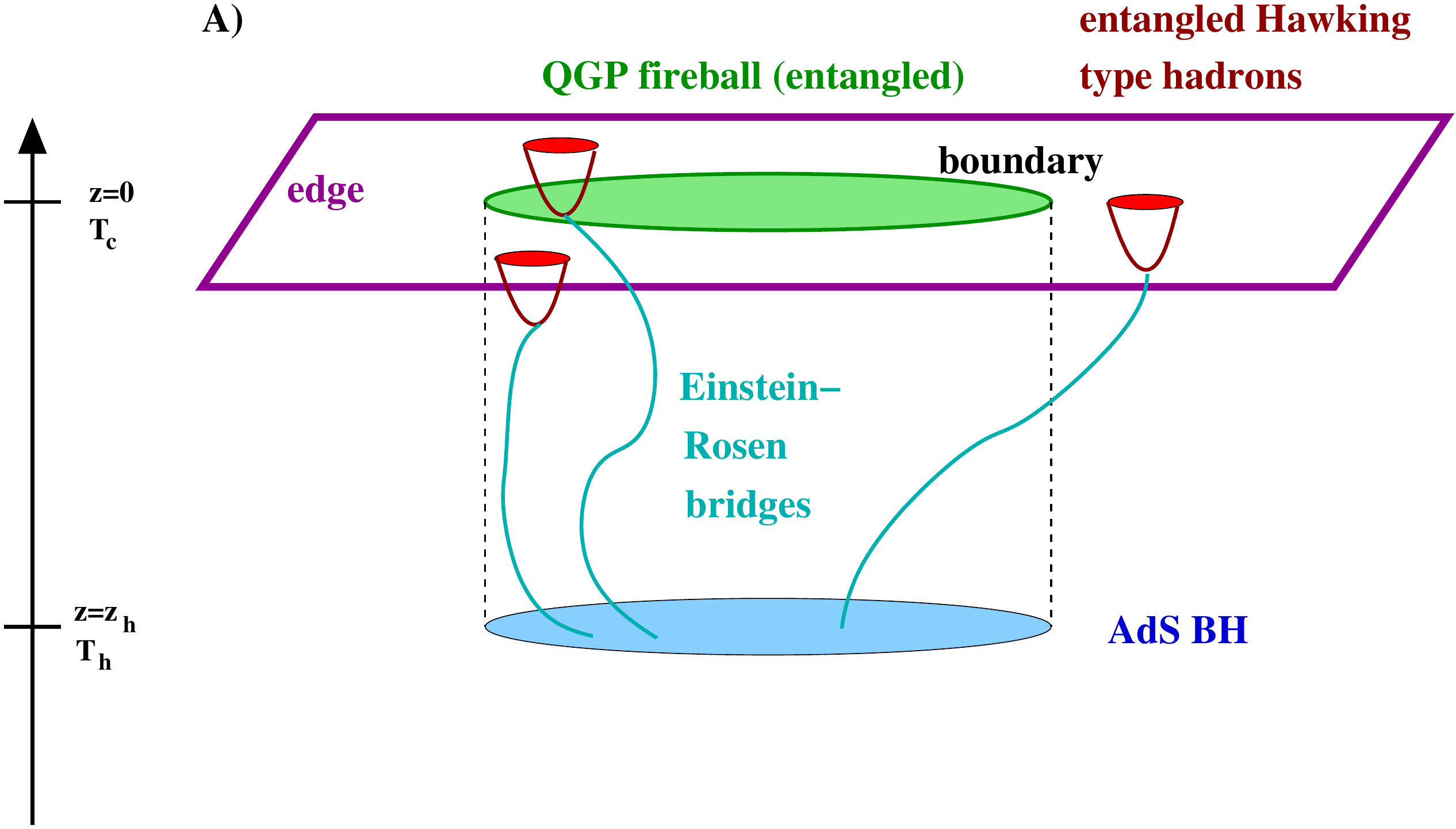}
\hspace{0.10\linewidth}
\includegraphics[width=0.4\linewidth]{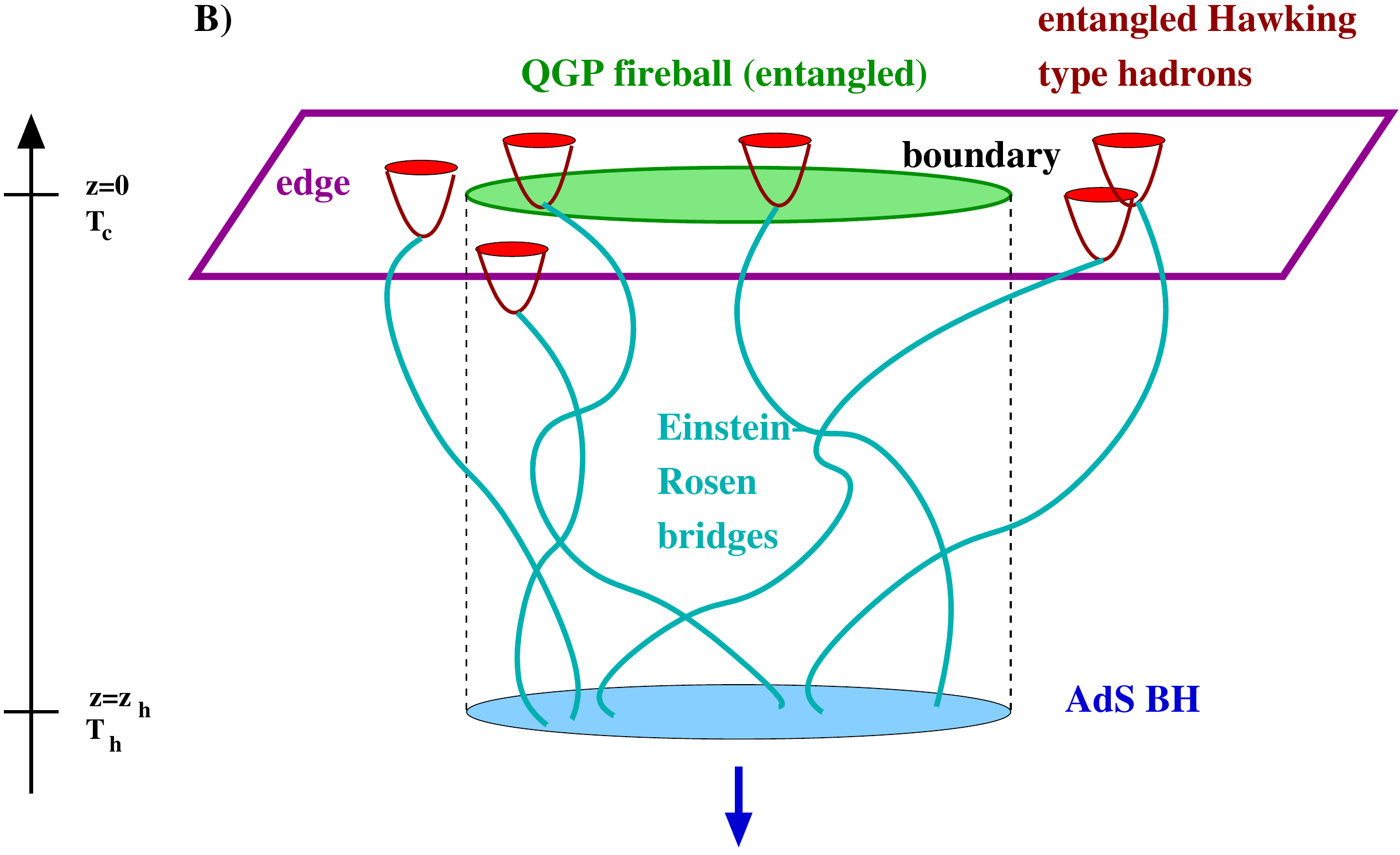}
\includegraphics[width=0.4\linewidth]{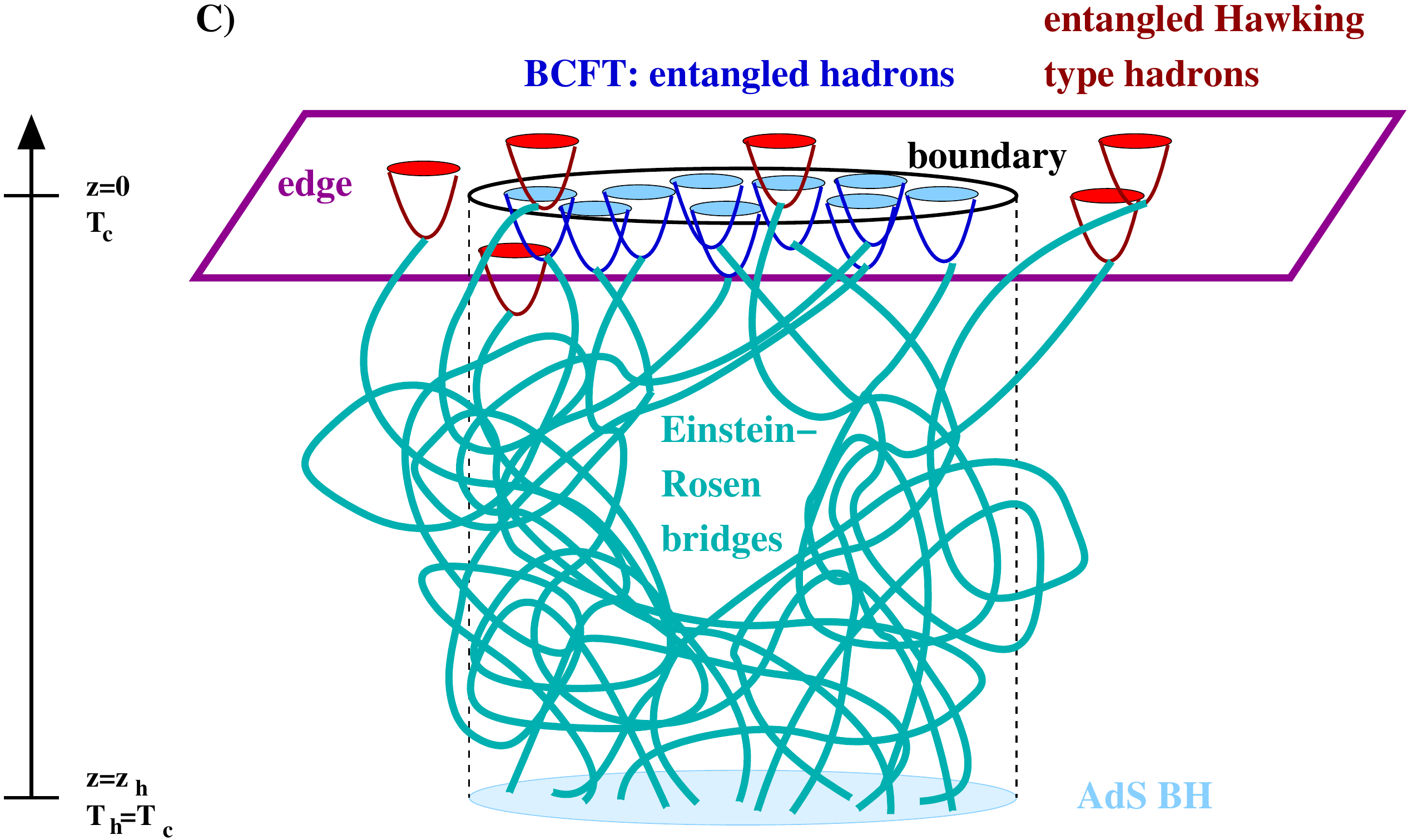}
\hspace{0.10\linewidth}
\includegraphics[width=0.4\linewidth]{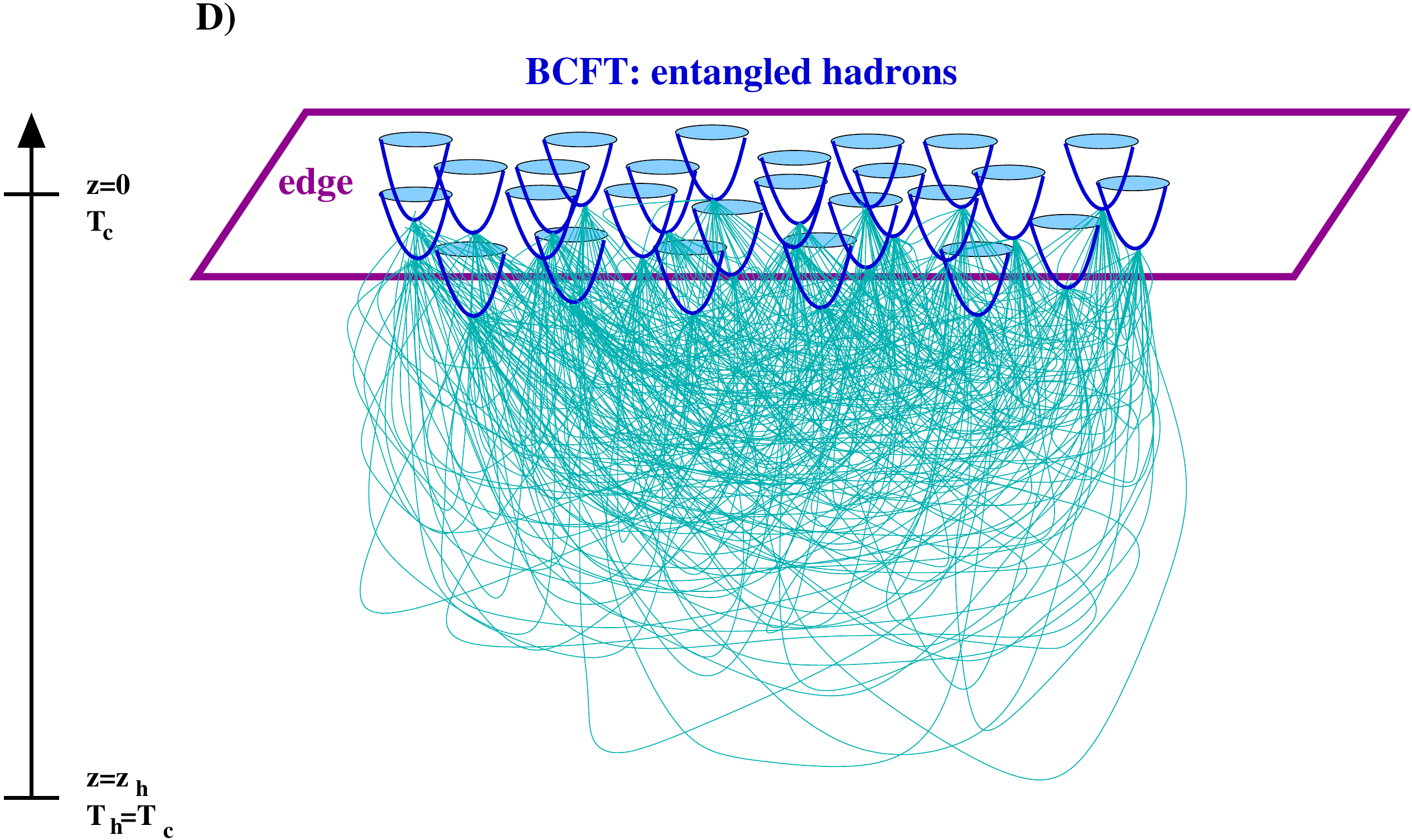}
\caption
{Illustration of the different stages of a high energy heavy ion collison. {\bf A)} After the QCD fireball is formed hadrons are emitted from its surface in analogy to Hawking radiation. {\bf B)}  This leads to volume growth of the whole QCD system on the edge. In parallel the fireball cools, e.g. the dual black hole sinks deeper into the AdS throat. However, the 3-dimensional volume of the QGP fireball and thus the 3-dimensional volume of the AdS BH remains roughly constant, see Fig.~\ref{fig:hydro}. {\bf C)} When the temperature of the fireball, which is identical to that of the AdS black hole, reaches $T_c$ the Hawking-Page-like transition occurs corresponding to complete hadronization of the remaining fireball on the AdS edge. The picture illustrates the moment of the transition. 
{\bf D)} Due to the monogamy of entanglement any pair of hadrons in the state after the hadronization transition can only share a very small fraction of entanglement, on average proportional to $O(1/N_h)$ with $N_h$ being the number of hadrons.} 
\label{fig:hadron_bild_1}
\end{figure}
\end{widetext}

\begin{figure}[htb]
\centering
\includegraphics[width=0.95\linewidth]{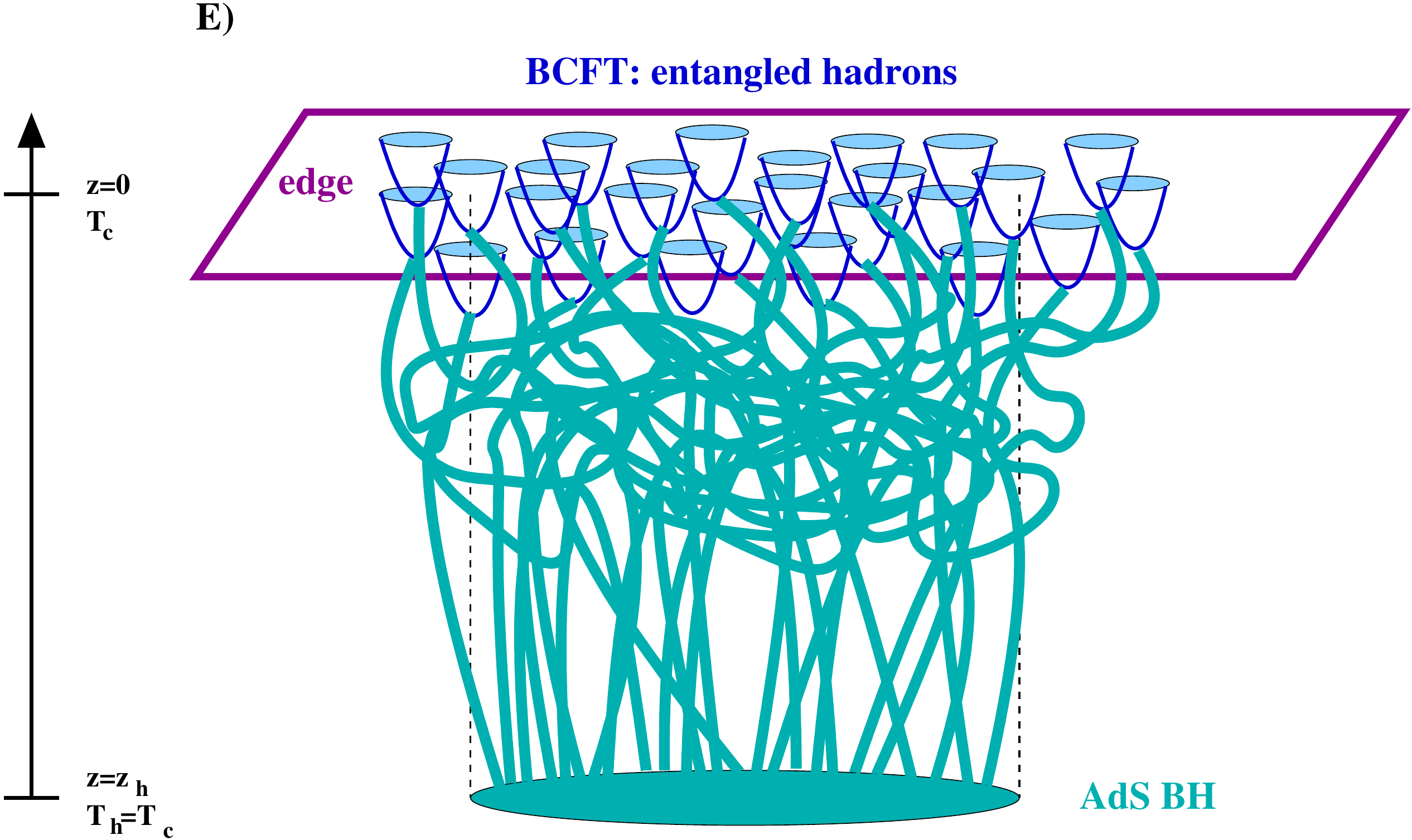}
\caption
{{\bf E)} As the Hawking-Page-like transition for QCD at small baryon number density is a cross-over the difference between the situation just before (at $T=T_c+\epsilon$, see Fig.~\ref{fig:hadron_bild_1}{\bf C}) and after (at $T=T_c-\epsilon$, see Fig.~\ref{fig:hadron_bild_1}{\bf D}) is small. Therefore, {\bf E)}, which is strictly valid only above $T_c$, is expected to provide a good approximation. Note that because hadrons on the AdS edge are maximally entangled with states on the horizon, they are uncorrelated with all other edge hadrons such that all of them form a thermal ensemble with temperature $T_c$. This is just one of the crucial elements of ETH. 
Note also, that the difference in this respect to Fig.~\ref{fig:hadron_bild_1}{\bf D} is only of order  $O(1/N_h)$ and thus negligible. For more discussions, please see the main text.}
\label{fig:hadron_bild_2}
\end{figure}

There exists a fundamental constraint, closely related to the no-cloning theorem \cite{Wootters:1982zz}, which states that quantum entanglement cannot be freely shared among many objects \cite{Wootters:1997id,Coffman:1999jd,Osborne2006xx}. This has been analyzed in detail for systems of qubits for which quantitatively precise statements can be made and formal proofs are possible. One can define a quantity $\tau(\rho_{AB})$, called "tangle", which quantifies entanglement between the elements of bi-partitions of multi-particle quantum states  $A$, $B$, described by a density matrix $\rho_{AB}$. $\tau$ can have values between 0 (no entanglement) and 1 (complete entanglement). For a quantum state of $n$ subsystems $A_1$, $A_2$, ...$A_n$ the following constraint holds 
\cite{Osborne2006xx}:
\begin{equation}
\sum_{k=2}^n \tau(\rho_{A_1,A_k}) \leq  \tau(\rho_{A_1,(A_2A_3...A_n)}) \leq 1 .
\label{eq:tangle}
\end{equation}
We identify the $A_i$ in our case with the $N_h$ individual hadron states and assume that a bound of the type (\ref{eq:tangle}) exists that limits the average entanglement between any two hadrons to a value of order $O(1/N_h)$. We will use this argument only in a very generic manner, arguing that for, e.g. a heavy ion collision at LHC, in which thousands of hadrons are produced, the effects of entanglement are negligible as long as only few-hadron observables are measured.   

As we have argued in Section~\ref{sec:HadronHIC} heavy ion phenomenology provides compelling evidence for the existence of two hadronization mechanisms: the emission of individual hadrons from the QGP fireball surface ($\sim 20$\%) and the instantaneous hadronization of the remaining fireball at $T_c$ ($\sim 80$\%). Both mechanisms are characterized by the same temperature $T_c$. We described specific models in the AdS/CFT context in Section~\ref{sec:Model_1} and \ref{sec:Model_2}. In this Section we combine these models to an overall scenario. 

On one hand, the presence of two mechanisms with quite different time dependence complicates our endeavour compared with the usual black hole information puzzle. On the other hand, we can make use of detailed  phenomenological knowledge based on a huge amount of high-precision data amassed by heavy ion experiments. We hope that this advantage outweighs the disadvantages. 

There exist many concepts in the literature which could be adapted to our situation. For example, Maldacena and Susskind suggested in \cite{Maldacena:2013xja} (see Fig.~13) a geometric, holographic interpretation of the entanglement of Hawking radiation in terms of ER bridges for a Minkowski space black hole, which is somewhat similar to ours. The difference between their picture and ours is that we do not study a horizon in the edge but in the bulk of AdS space and that we treat hadrons as entangled with the AdS horizon dual to the QGP fireball rather than photons as entangled with a black hole horizon. 

Figures~\ref{fig:hadron_bild_1} and \ref{fig:hadron_bild_2} present conceptual illustrations of our ideas. We will outline in Section~\ref{sec:ToDo} some steps that could be taken to make these more quantitative. 

As shown in Fig.~\ref{fig:hydro}, the transverse size of the QGP fireball stays nearly the same as function of time, as the internal cooling caused by the hydrodynamic expansion is nearly compensated by evaporation from the surface. The surface temperature of the QGP fireball stays fixed at $T_c$ during this period while the volume of the hadron resonance gas (HRG) outside of the QGP fireball grows continuously. When the QGP fireball temperature reaches $T_c$ throughout, the remaining volume hadronizes quickly. In the AdS dual description, the spatially bounded region on the AdS edge corresponding to the QGP fireball extends into the AdS bulk until it reaches an also spatially bounded BH horizon. It appears plausible that this region in AdS space, which is schematically depicted as a cylinder in Fig.~\ref{fig:hadron_bild_1} {\bf A)}, should be modeled as the holographic dual of a BCFT \cite{Takayanagi:2011zk}. The emitted hadrons are entangled with states on the BH horizon via ER bridges, as illustrated in Fig.~13 of Ref.~\cite{Maldacena:2013xja}, such that no entropy is produced. 

As discussed in Section~\ref{sec:Intro}, hadronization in heavy ion collisions is usually modeled as ``chemical freeze out'', which is hypothesized to explain the observation that all hadron yields agree with the predictions of the thermal model although hadronic interactions continue in the expanding hadron gas. As we argue that the success of the thermal model is a consequence of quantum coherence, e.~g.\ in the form of ETH, we do not need any such additional mechanism. However, even our mechanism requires a rapid transition from the hydrodynamic QGP phase to the hadronic phase. In fact, microscopic simulations indicate a rapid growth of the specific viscosity at temperatures below 150 MeV signalling a rapid breakdown of hydrodynamics when the expanding fireball cools below $T_c$ \cite{Yang:2022yxa}.  

Holographically, this transition corresponds to the transition between Figs.~\ref{fig:hadron_bild_1} B and D, where Fig.\ref{fig:hadron_bild_1} C is a schematic representation of an intermediate stage. We do not yet have a reliable quantitative description of these intermediate stages, neither in QCD nor in holography, but one future goal stated in Section \ref{sec:ToDo} is to develop such a description in the holographic model of the transition.

With respect to phenomenology, however, we can argue that the mapping to a purely hadronic late stage should be a smooth one, i.~e.\ a cross-over rather than a first order phase transition, in line with the fact that the deconfinement/confinement phase transition of QCD is a cross-over.


We conclude this Section with the following remark: If a holographic dual of QCD exists, it must be possible to develop a quantitative description of the hadronization of a quark-gluon plasma. At present we do not know whether this is the case.  However, even if if it does not or turns out to be unachievable in practice, much could be learned about the quantum physics of the confinement transition by investigation of a holographic model that incorporates salient features of the quark-hadron transition in QCD. In the next Section we will discuss some steps that can be taken in this direction.

\section{(Some) Open Questions}
\label{sec:ToDo}

As already emphasized our reasoning is speculative and must be consolidated or refuted by detailed investigations. Because the problem of decoherence and thermalization of many-particle quantum states is a very generic one, there exists a large and rapidly expanding literature on it, where the ETH plays a prominent role. This leads to our first set of open questions:
\begin{itemize}
    \item Does QCD exhibit ETH behavior?
\end{itemize}
We interpret the success of the thermal model as heuristic indication that it does show ETH behavior, but it would be desirable to affirm this with rigorous methods, such that the success of the thermal model becomes a prediction rather than merely an observation. To make contact with the experimental data, it is not only necessary to confirm ETH properties in principle, but to also establish the time scale at which ETH behavior becomes manifest (see \cite{Dymarsky:2018sef}).

In fact, this question is much discussed in quantum information theory and answering it could become an early success of quantum simulators (see, e.~g., Ref.~\cite{Schuckert:2020qeo} and references therein). There are two crucial advantages of quantum computing in this context. First, for the arguments given in Section~\ref{sec:ETH} the extensive nature of the microcanonical entropy leads to  a strong exponential suppression of all off-diagonal matrix elements except for very small systems, which are easiest to simulate. Second, the simulation of realistic hadrons is not required to test the validity of ETH as long as the Hamiltonian is well enough modeled. Also, showing ETH behavior for the gauge group SU(2), which is easier to simulate, should be sufficient. 

Another promising approach is lattice gauge theory. Just as RMT behavior was established for the QCD Dirac operator by simulations on small lattices with classical computers \cite{Berbenni-Bitsch:1997zmi,Verbaarschot:2000dy} the same may be possible for ETH. 
\begin{itemize}
    \item Is ETH necessary or only sufficient to explain success of the thermal model? If it is necessary the relevant question becomes: How long does it take until few particle observables look in good approximation thermal?
\end{itemize}
Recently a number of observations suggest that ETH is less universal than often assumed and may not even be necessary for a system to thermalize \cite{Harrow:2022znr}. In Ref.~\cite{Majidy:2022kzx} it was shown that non-Abelian charges reduce entropy-production rates and may enhance finite-size deviations from eigenstate thermalization. The occurrence of many-body scars may also negate ergodic behavior in certain systems \cite{Turner2018:weak} and slow down thermalization \cite{Michailidis:2019tgg}.

In general, one should be aware that ETH postulates very restrictive constraints for matrix elements of generic operators between energy eigenstates. It could well be that less restrictive requirements already imply the emergence of thermal behavior to such an extent that the differences are unobservable in realistic experiments. For example, it would be interesting to understand whether monogamy of entanglement in many-particle systems \cite{Wootters:1982zz,Wootters:1997id,Coffman:1999jd,Osborne2006xx} is already sufficient to mimic thermalization as long as only small subsystems of the complete, entangled quantum states are observed. Obviously, the search for alternatives to, and generalizations of, the ETH paradigm is a vast and rapidly developing area of research.

A second set of questions revolves around the search for useful holographic models of the quark-hadron transition.
\begin{itemize}
    \item Do recent advances in numerical solutions of gravity in AdS space make it possible to better understand the dynamics of a Hawking-Page type transition?
\end{itemize}
Over the years much effort has been invested to better understand the phenomenological implications of the original Hawking-Page transition without leading to definitive conclusions \cite{Hubeny:2002xn,Mandal:2011ws,Buchel:2015gxa,Dias:2016eto,Yaffe:2017axl,Janik:2021jbq}. 
In the large-$N_c$ limit, the HP transition will proceed by spinodal decomposition of a deeply supercooled phase unless the cooling process is very slow. One interesting question is whether the system re-equilibrates after the transition is complete \cite{Yaffe:2017axl}; another is whether quantum gravity corrections at finite 't Hooft coupling \cite{Gubser:1998nz} help to smoothen out the HP transition. In a very slowly cooling scenario, the transition will proceed via bubble nucleation. It is an interesting question which of these scenarios is applicable to pure SUN($N_c$) gauge theories at moderate $N_c$ where the equilibrium deconfinement transition is known to be of first order. 

Progress in the numerical treatment of matrix models \cite{Hanada:2022wcq,Pateloudis:2022ijr,Buividovich:2022udc,Buividovich:2018scl} as well as AdS/CFT initial value problems \cite{Waeber:2022tts} may make it possible to explore the real-time dynamics of Hawking-Page transitions.  However, none of these efforts addresses the question how the entanglement structure of the quasi-thermal QGP is reflected in the entanglement structure of the produced hadrons. In the holographic dual this raises the challenge of computing the geometric form of an ER bridge between the asymptotic (``edge'') regions of a holographic thermofield double state for general initial conditions. 
\begin{itemize}
    \item Would holographic models in which the transition is a smooth crossover as in QCD allow for an approximate identification of the HP transition with the hadronization transition of a quark-gluon plasma?
\end{itemize}
The possible application of the HP transition in nonconformal dilaton-gravity models \cite{Gursoy:2008za,Gursoy:2009jd,Megias:2010ku} as a holographic model of the dynamical deconfinement transition in QCD was explored in Ref.~\cite{Gursoy:2013zxa}. More recently, the dynamical nature of the transition including domain formation \cite{Attems:2019yqn} and domain wall propagation \cite{Ecker:2021ukv} have been investigated. Even shock wave collisions were studied in a holographic model that parametrizes the evolution of the quark-hadron transition from a smooth crossover to a first-order phase transition with increasing baryon chemical potential \cite{Attems:2020qkg}. 
\begin{itemize}
    \item Can hadron emission from the surface of a quark-gluon plasma be modeled as splitting quenches?
\end{itemize}
There exist low-dimensional toy models for the splitting of one domain of conformal field theory into two \cite{Shimaji:2018czt,Caputa:2019avh} or the formation of an Einstein-Rosen (ER) bridge between two horizons \cite{Anderson:2020vwi}. Can these models be generalized to multiple black hole splittings, correponding to the production of multiple hadrons and to higher dimensions? Is AdS/BCFT the adequate framework to describe hadronization? If it is, does it imply, in the spirit of Ref.~\cite{VanRaamsdonk:2020ydg}, that the holographic description of the quark-gluon plasma and an interacting hadron gas are quite similar? In fact this is exactly what is observed on the field theoretical side of the duality. The properties of a hadron resonance gas and the QGP are indistinguishable close to hadronization (see, e.~g., \cite{Bellwied:2021nrt}.

Finally, one would like to know whether there are experimental signatures of the entanglement structure among hadrons emitted in the decay of  quark-gluon plasma.
\begin{itemize}
\item How can the validity of the analogy between hadronization and black hole decay proposed in Section~\ref{sec:Model_1} be tested?  
\end{itemize}
As discussed in Section~\ref{sec:holmod} the monogamy property of entanglement \cite{Osborne2006xx} implies that the states of any pair of hadrons among the $N$ emitted hadrons is only entangled at the $O(1/N)$ level. If one would know what the most accessible entanglement signature is, one could test this conclusion as a function of the QGP fireball size. One possibility is to look for quantum correlations among the spins of hadrons \cite{Gong:2021bcp}.

\section{Conclusions}
\label{sec:conclusions}
  
The application of concepts from the AdS/CFT duality to describe the early stage, spanning about $1-2$ fm/$c$, of high energy heavy ion collisions as they are studied experimentally at LHC and RHIC has been remarkably successful and brought many important insights. This suggests that one should try to extend the dual holographic description to later times including the hadronization stage. As we explained, any such effort has to face the full complexity of the black hole information problem in the dual desription. 

The time evolution of both the AdS black hole and QCD is unitary. No information gets lost and no entropy is produced. Consequently, thermal ensemble in the strict sense cannot be produced. However, what can be produced is a system for which all realistic measurements produce results that are indistinguishable from strictly thermal ones. Recent progress on black hole evaporation has substantiated this conclusion. For QCD the corresponding field theoretical calculation, which would have to treat quantum entanglement during each step of the collision appears impractical. If, however, the AdS- and QCD-based descriptions of a heavy ion collision were holographically dual up to calculable corrections, a QCD-based calculation would be unnecessary, and one could simply adapt the insights from black holes to heavy ion collisions. 

Here we focused on the question whether and how the AdS dictionary could be extended to times beyond the hadronization of the quark-gluon plasma and how the phenomenological properties of heavy ion collisions up to and beyond the Page time could be represented holographically. In doing so, we have not reached a conclusive answer, but we also did not encounter obvious roadblocks, and we formulated topics for future work that could help answer some of the many remaining questions. 

{\it Acknowledgments:}
We thank M.~Kaminski, L.~Yaffe, and S.~Waeber for many valuable discussions and comments on a draft of this manuscript. We also thank T.~Hartman for insightful discussions on Euclidean wormholes. B.~M.\ acknowledges support from the U.S. Department of Energy Office of Science (Grant DE-FG02-05ER41367) and from Yale University during a sabbatical stay in Spring 2022.

\bibliographystyle{apsrev4-1}
\bibliography{QGP_Hadron}  

\begin{thebibliography}{146}%
\makeatletter
\providecommand \@ifxundefined [1]{%
 \@ifx{#1\undefined}
}%
\providecommand \@ifnum [1]{%
 \ifnum #1\expandafter \@firstoftwo
 \else \expandafter \@secondoftwo
 \fi
}%
\providecommand \@ifx [1]{%
 \ifx #1\expandafter \@firstoftwo
 \else \expandafter \@secondoftwo
 \fi
}%
\providecommand \natexlab [1]{#1}%
\providecommand \enquote  [1]{``#1''}%
\providecommand \bibnamefont  [1]{#1}%
\providecommand \bibfnamefont [1]{#1}%
\providecommand \citenamefont [1]{#1}%
\providecommand \href@noop [0]{\@secondoftwo}%
\providecommand \href [0]{\begingroup \@sanitize@url \@href}%
\providecommand \@href[1]{\@@startlink{#1}\@@href}%
\providecommand \@@href[1]{\endgroup#1\@@endlink}%
\providecommand \@sanitize@url [0]{\catcode `\\12\catcode `\$12\catcode
  `\&12\catcode `\#12\catcode `\^12\catcode `\_12\catcode `\%12\relax}%
\providecommand \@@startlink[1]{}%
\providecommand \@@endlink[0]{}%
\providecommand \url  [0]{\begingroup\@sanitize@url \@url }%
\providecommand \@url [1]{\endgroup\@href {#1}{\urlprefix }}%
\providecommand \urlprefix  [0]{URL }%
\providecommand \Eprint [0]{\href }%
\providecommand \doibase [0]{http://dx.doi.org/}%
\providecommand \selectlanguage [0]{\@gobble}%
\providecommand \bibinfo  [0]{\@secondoftwo}%
\providecommand \bibfield  [0]{\@secondoftwo}%
\providecommand \translation [1]{[#1]}%
\providecommand \BibitemOpen [0]{}%
\providecommand \bibitemStop [0]{}%
\providecommand \bibitemNoStop [0]{.\EOS\space}%
\providecommand \EOS [0]{\spacefactor3000\relax}%
\providecommand \BibitemShut  [1]{\csname bibitem#1\endcsname}%
\let\auto@bib@innerbib\@empty
\bibitem [{\citenamefont {Page}(1993{\natexlab{a}})}]{Page:1993wv}%
  \BibitemOpen
  \bibfield  {author} {\bibinfo {author} {\bibfnamefont {D.~N.}\ \bibnamefont
  {Page}},\ }\href {\doibase 10.1103/PhysRevLett.71.3743} {\bibfield  {journal}
  {\bibinfo  {journal} {Phys. Rev. Lett.}\ }\textbf {\bibinfo {volume} {71}},\
  \bibinfo {pages} {3743} (\bibinfo {year} {1993}{\natexlab{a}})},\ \Eprint
  {http://arxiv.org/abs/hep-th/9306083} {arXiv:hep-th/9306083} \BibitemShut
  {NoStop}%
\bibitem [{\citenamefont {Page}(2013)}]{Page:2013dx}%
  \BibitemOpen
  \bibfield  {author} {\bibinfo {author} {\bibfnamefont {D.~N.}\ \bibnamefont
  {Page}},\ }\href {\doibase 10.1088/1475-7516/2013/09/028} {\bibfield
  {journal} {\bibinfo  {journal} {JCAP}\ }\textbf {\bibinfo {volume} {09}},\
  \bibinfo {pages} {028} (\bibinfo {year} {2013})},\ \Eprint
  {http://arxiv.org/abs/1301.4995} {arXiv:1301.4995 [hep-th]} \BibitemShut
  {NoStop}%
\bibitem [{\citenamefont {Page}(1993{\natexlab{b}})}]{Page:1993df}%
  \BibitemOpen
  \bibfield  {author} {\bibinfo {author} {\bibfnamefont {D.~N.}\ \bibnamefont
  {Page}},\ }\href {\doibase 10.1103/PhysRevLett.71.1291} {\bibfield  {journal}
  {\bibinfo  {journal} {Phys. Rev. Lett.}\ }\textbf {\bibinfo {volume} {71}},\
  \bibinfo {pages} {1291} (\bibinfo {year} {1993}{\natexlab{b}})},\ \Eprint
  {http://arxiv.org/abs/gr-qc/9305007} {arXiv:gr-qc/9305007} \BibitemShut
  {NoStop}%
\bibitem [{\citenamefont {Kaufman}\ \emph {et~al.}(2016)\citenamefont
  {Kaufman}, \citenamefont {Tai}, \citenamefont {Lukin}, \citenamefont
  {Rispoli}, \citenamefont {Schittko}, \citenamefont {Preiss},\ and\
  \citenamefont {Greiner}}]{Kaufman:2016qu}%
  \BibitemOpen
  \bibfield  {author} {\bibinfo {author} {\bibfnamefont {A.~M.}\ \bibnamefont
  {Kaufman}}, \bibinfo {author} {\bibfnamefont {M.~E.}\ \bibnamefont {Tai}},
  \bibinfo {author} {\bibfnamefont {A.}~\bibnamefont {Lukin}}, \bibinfo
  {author} {\bibfnamefont {M.}~\bibnamefont {Rispoli}}, \bibinfo {author}
  {\bibfnamefont {R.}~\bibnamefont {Schittko}}, \bibinfo {author}
  {\bibfnamefont {P.~M.}\ \bibnamefont {Preiss}}, \ and\ \bibinfo {author}
  {\bibfnamefont {M.}~\bibnamefont {Greiner}},\ }\href@noop {} {\bibfield
  {journal} {\bibinfo  {journal} {Science}\ }\textbf {\bibinfo {volume}
  {353}},\ \bibinfo {pages} {794} (\bibinfo {year} {2016})}\BibitemShut
  {NoStop}%
\bibitem [{\citenamefont {Witten}(1998)}]{Witten:1998zw}%
  \BibitemOpen
  \bibfield  {author} {\bibinfo {author} {\bibfnamefont {E.}~\bibnamefont
  {Witten}},\ }\href {\doibase 10.4310/ATMP.1998.v2.n3.a3} {\bibfield
  {journal} {\bibinfo  {journal} {Adv. Theor. Math. Phys.}\ }\textbf {\bibinfo
  {volume} {2}},\ \bibinfo {pages} {505} (\bibinfo {year} {1998})},\ \Eprint
  {http://arxiv.org/abs/hep-th/9803131} {arXiv:hep-th/9803131} \BibitemShut
  {NoStop}%
\bibitem [{\citenamefont {Sakai}\ and\ \citenamefont
  {Sugimoto}(2005)}]{Sakai:2004cn}%
  \BibitemOpen
  \bibfield  {author} {\bibinfo {author} {\bibfnamefont {T.}~\bibnamefont
  {Sakai}}\ and\ \bibinfo {author} {\bibfnamefont {S.}~\bibnamefont
  {Sugimoto}},\ }\href {\doibase 10.1143/PTP.113.843} {\bibfield  {journal}
  {\bibinfo  {journal} {Prog. Theor. Phys.}\ }\textbf {\bibinfo {volume}
  {113}},\ \bibinfo {pages} {843} (\bibinfo {year} {2005})},\ \Eprint
  {http://arxiv.org/abs/hep-th/0412141} {arXiv:hep-th/0412141} \BibitemShut
  {NoStop}%
\bibitem [{\citenamefont {Braun}\ \emph {et~al.}(2003)\citenamefont {Braun},
  \citenamefont {Korchemsky},\ and\ \citenamefont {M\"uller}}]{Braun:2003rp}%
  \BibitemOpen
  \bibfield  {author} {\bibinfo {author} {\bibfnamefont {V.~M.}\ \bibnamefont
  {Braun}}, \bibinfo {author} {\bibfnamefont {G.~P.}\ \bibnamefont
  {Korchemsky}}, \ and\ \bibinfo {author} {\bibfnamefont {D.}~\bibnamefont
  {M\"uller}},\ }\href {\doibase 10.1016/S0146-6410(03)90004-4} {\bibfield
  {journal} {\bibinfo  {journal} {Prog. Part. Nucl. Phys.}\ }\textbf {\bibinfo
  {volume} {51}},\ \bibinfo {pages} {311} (\bibinfo {year} {2003})},\ \Eprint
  {http://arxiv.org/abs/hep-ph/0306057} {arXiv:hep-ph/0306057} \BibitemShut
  {NoStop}%
\bibitem [{\citenamefont {Braun}\ \emph {et~al.}(2019)\citenamefont {Braun},
  \citenamefont {Manashov}, \citenamefont {Moch},\ and\ \citenamefont
  {Strohmaier}}]{Braun:2018mxm}%
  \BibitemOpen
  \bibfield  {author} {\bibinfo {author} {\bibfnamefont {V.~M.}\ \bibnamefont
  {Braun}}, \bibinfo {author} {\bibfnamefont {A.~N.}\ \bibnamefont {Manashov}},
  \bibinfo {author} {\bibfnamefont {S.~O.}\ \bibnamefont {Moch}}, \ and\
  \bibinfo {author} {\bibfnamefont {M.}~\bibnamefont {Strohmaier}},\ }\href
  {\doibase 10.1016/j.physletb.2019.04.027} {\bibfield  {journal} {\bibinfo
  {journal} {Phys. Lett. B}\ }\textbf {\bibinfo {volume} {793}},\ \bibinfo
  {pages} {78} (\bibinfo {year} {2019})},\ \Eprint
  {http://arxiv.org/abs/1810.04993} {arXiv:1810.04993 [hep-th]} \BibitemShut
  {NoStop}%
\bibitem [{\citenamefont {Kumericki}\ \emph {et~al.}(2007)\citenamefont
  {Kumericki}, \citenamefont {Mueller}, \citenamefont {Passek-Kumericki},\ and\
  \citenamefont {Schafer}}]{Kumericki:2006xx}%
  \BibitemOpen
  \bibfield  {author} {\bibinfo {author} {\bibfnamefont {K.}~\bibnamefont
  {Kumericki}}, \bibinfo {author} {\bibfnamefont {D.}~\bibnamefont {Mueller}},
  \bibinfo {author} {\bibfnamefont {K.}~\bibnamefont {Passek-Kumericki}}, \
  and\ \bibinfo {author} {\bibfnamefont {A.}~\bibnamefont {Schafer}},\ }\href
  {\doibase 10.1016/j.physletb.2007.02.071} {\bibfield  {journal} {\bibinfo
  {journal} {Phys. Lett. B}\ }\textbf {\bibinfo {volume} {648}},\ \bibinfo
  {pages} {186} (\bibinfo {year} {2007})},\ \Eprint
  {http://arxiv.org/abs/hep-ph/0605237} {arXiv:hep-ph/0605237} \BibitemShut
  {NoStop}%
\bibitem [{\citenamefont {Waeber}\ \emph {et~al.}(2015)\citenamefont {Waeber},
  \citenamefont {Sch\"afer}, \citenamefont {Vuorinen},\ and\ \citenamefont
  {Yaffe}}]{Waeber:2015oka}%
  \BibitemOpen
  \bibfield  {author} {\bibinfo {author} {\bibfnamefont {S.}~\bibnamefont
  {Waeber}}, \bibinfo {author} {\bibfnamefont {A.}~\bibnamefont {Sch\"afer}},
  \bibinfo {author} {\bibfnamefont {A.}~\bibnamefont {Vuorinen}}, \ and\
  \bibinfo {author} {\bibfnamefont {L.~G.}\ \bibnamefont {Yaffe}},\ }\href
  {\doibase 10.1007/JHEP11(2015)087} {\bibfield  {journal} {\bibinfo  {journal}
  {JHEP}\ }\textbf {\bibinfo {volume} {11}},\ \bibinfo {pages} {087} (\bibinfo
  {year} {2015})},\ \Eprint {http://arxiv.org/abs/1509.02983} {arXiv:1509.02983
  [hep-th]} \BibitemShut {NoStop}%
\bibitem [{\citenamefont {'t~Hooft}(1974)}]{tHooft:1973alw}%
  \BibitemOpen
  \bibfield  {author} {\bibinfo {author} {\bibfnamefont {G.}~\bibnamefont
  {'t~Hooft}},\ }\href {\doibase 10.1016/0550-3213(74)90154-0} {\bibfield
  {journal} {\bibinfo  {journal} {Nucl. Phys. B}\ }\textbf {\bibinfo {volume}
  {72}},\ \bibinfo {pages} {461} (\bibinfo {year} {1974})}\BibitemShut
  {NoStop}%
\bibitem [{\citenamefont {Hawking}\ and\ \citenamefont
  {Page}(1983)}]{Hawking:1982dh}%
  \BibitemOpen
  \bibfield  {author} {\bibinfo {author} {\bibfnamefont {S.~W.}\ \bibnamefont
  {Hawking}}\ and\ \bibinfo {author} {\bibfnamefont {D.~N.}\ \bibnamefont
  {Page}},\ }\href {\doibase 10.1007/BF01208266} {\bibfield  {journal}
  {\bibinfo  {journal} {Commun. Math. Phys.}\ }\textbf {\bibinfo {volume}
  {87}},\ \bibinfo {pages} {577} (\bibinfo {year} {1983})}\BibitemShut
  {NoStop}%
\bibitem [{\citenamefont {Aharony}\ \emph {et~al.}(2006)\citenamefont
  {Aharony}, \citenamefont {Minwalla},\ and\ \citenamefont
  {Wiseman}}]{Aharony:2005bm}%
  \BibitemOpen
  \bibfield  {author} {\bibinfo {author} {\bibfnamefont {O.}~\bibnamefont
  {Aharony}}, \bibinfo {author} {\bibfnamefont {S.}~\bibnamefont {Minwalla}}, \
  and\ \bibinfo {author} {\bibfnamefont {T.}~\bibnamefont {Wiseman}},\ }\href
  {\doibase 10.1088/0264-9381/23/7/001} {\bibfield  {journal} {\bibinfo
  {journal} {Class. Quant. Grav.}\ }\textbf {\bibinfo {volume} {23}},\ \bibinfo
  {pages} {2171} (\bibinfo {year} {2006})},\ \Eprint
  {http://arxiv.org/abs/hep-th/0507219} {arXiv:hep-th/0507219} \BibitemShut
  {NoStop}%
\bibitem [{\citenamefont {G{\"u}rsoy}\ \emph
  {et~al.}(2009{\natexlab{a}})\citenamefont {G{\"u}rsoy}, \citenamefont
  {Kiritsis}, \citenamefont {Mazzanti},\ and\ \citenamefont
  {Nitti}}]{Gursoy:2008za}%
  \BibitemOpen
  \bibfield  {author} {\bibinfo {author} {\bibfnamefont {U.}~\bibnamefont
  {G{\"u}rsoy}}, \bibinfo {author} {\bibfnamefont {E.}~\bibnamefont
  {Kiritsis}}, \bibinfo {author} {\bibfnamefont {L.}~\bibnamefont {Mazzanti}},
  \ and\ \bibinfo {author} {\bibfnamefont {F.}~\bibnamefont {Nitti}},\ }\href
  {\doibase 10.1088/1126-6708/2009/05/033} {\bibfield  {journal} {\bibinfo
  {journal} {JHEP}\ }\textbf {\bibinfo {volume} {05}},\ \bibinfo {pages} {033}
  (\bibinfo {year} {2009}{\natexlab{a}})},\ \Eprint
  {http://arxiv.org/abs/0812.0792} {arXiv:0812.0792 [hep-th]} \BibitemShut
  {NoStop}%
\bibitem [{\citenamefont {Mandal}\ and\ \citenamefont
  {Morita}(2011)}]{Mandal:2011ws}%
  \BibitemOpen
  \bibfield  {author} {\bibinfo {author} {\bibfnamefont {G.}~\bibnamefont
  {Mandal}}\ and\ \bibinfo {author} {\bibfnamefont {T.}~\bibnamefont
  {Morita}},\ }\href {\doibase 10.1007/JHEP09(2011)073} {\bibfield  {journal}
  {\bibinfo  {journal} {JHEP}\ }\textbf {\bibinfo {volume} {09}},\ \bibinfo
  {pages} {073} (\bibinfo {year} {2011})},\ \Eprint
  {http://arxiv.org/abs/1107.4048} {arXiv:1107.4048 [hep-th]} \BibitemShut
  {NoStop}%
\bibitem [{\citenamefont {Lin}\ and\ \citenamefont
  {Shuryak}(2008)}]{Lin:2006rf}%
  \BibitemOpen
  \bibfield  {author} {\bibinfo {author} {\bibfnamefont {S.}~\bibnamefont
  {Lin}}\ and\ \bibinfo {author} {\bibfnamefont {E.}~\bibnamefont {Shuryak}},\
  }\href {\doibase 10.1103/PhysRevD.77.085013} {\bibfield  {journal} {\bibinfo
  {journal} {Phys. Rev. D}\ }\textbf {\bibinfo {volume} {77}},\ \bibinfo
  {pages} {085013} (\bibinfo {year} {2008})},\ \Eprint
  {http://arxiv.org/abs/hep-ph/0610168} {arXiv:hep-ph/0610168} \BibitemShut
  {NoStop}%
\bibitem [{\citenamefont {Balasubramanian}\ \emph
  {et~al.}(2011{\natexlab{a}})\citenamefont {Balasubramanian}, \citenamefont
  {Bernamonti}, \citenamefont {de~Boer}, \citenamefont {Copland}, \citenamefont
  {Craps}, \citenamefont {Keski-Vakkuri}, \citenamefont {M{\"u}ller},
  \citenamefont {Sch{\"a}fer}, \citenamefont {Shigemori},\ and\ \citenamefont
  {Staessens}}]{Balasubramanian:2010ce}%
  \BibitemOpen
  \bibfield  {author} {\bibinfo {author} {\bibfnamefont {V.}~\bibnamefont
  {Balasubramanian}}, \bibinfo {author} {\bibfnamefont {A.}~\bibnamefont
  {Bernamonti}}, \bibinfo {author} {\bibfnamefont {J.}~\bibnamefont {de~Boer}},
  \bibinfo {author} {\bibfnamefont {N.}~\bibnamefont {Copland}}, \bibinfo
  {author} {\bibfnamefont {B.}~\bibnamefont {Craps}}, \bibinfo {author}
  {\bibfnamefont {E.}~\bibnamefont {Keski-Vakkuri}}, \bibinfo {author}
  {\bibfnamefont {B.}~\bibnamefont {M{\"u}ller}}, \bibinfo {author}
  {\bibfnamefont {A.}~\bibnamefont {Sch{\"a}fer}}, \bibinfo {author}
  {\bibfnamefont {M.}~\bibnamefont {Shigemori}}, \ and\ \bibinfo {author}
  {\bibfnamefont {W.}~\bibnamefont {Staessens}},\ }\href {\doibase
  10.1103/PhysRevLett.106.191601} {\bibfield  {journal} {\bibinfo  {journal}
  {Phys. Rev. Lett.}\ }\textbf {\bibinfo {volume} {106}},\ \bibinfo {pages}
  {191601} (\bibinfo {year} {2011}{\natexlab{a}})},\ \Eprint
  {http://arxiv.org/abs/1012.4753} {arXiv:1012.4753 [hep-th]} \BibitemShut
  {NoStop}%
\bibitem [{\citenamefont {Balasubramanian}\ \emph
  {et~al.}(2011{\natexlab{b}})\citenamefont {Balasubramanian}, \citenamefont
  {Bernamonti}, \citenamefont {de~Boer}, \citenamefont {Copland}, \citenamefont
  {Craps}, \citenamefont {Keski-Vakkuri}, \citenamefont {M{\"u}ller},
  \citenamefont {Sch{\"a}fer}, \citenamefont {Shigemori},\ and\ \citenamefont
  {Staessens}}]{Balasubramanian:2011ur}%
  \BibitemOpen
  \bibfield  {author} {\bibinfo {author} {\bibfnamefont {V.}~\bibnamefont
  {Balasubramanian}}, \bibinfo {author} {\bibfnamefont {A.}~\bibnamefont
  {Bernamonti}}, \bibinfo {author} {\bibfnamefont {J.}~\bibnamefont {de~Boer}},
  \bibinfo {author} {\bibfnamefont {N.}~\bibnamefont {Copland}}, \bibinfo
  {author} {\bibfnamefont {B.}~\bibnamefont {Craps}}, \bibinfo {author}
  {\bibfnamefont {E.}~\bibnamefont {Keski-Vakkuri}}, \bibinfo {author}
  {\bibfnamefont {B.}~\bibnamefont {M{\"u}ller}}, \bibinfo {author}
  {\bibfnamefont {A.}~\bibnamefont {Sch{\"a}fer}}, \bibinfo {author}
  {\bibfnamefont {M.}~\bibnamefont {Shigemori}}, \ and\ \bibinfo {author}
  {\bibfnamefont {W.}~\bibnamefont {Staessens}},\ }\href {\doibase
  10.1103/PhysRevD.84.026010} {\bibfield  {journal} {\bibinfo  {journal} {Phys.
  Rev. D}\ }\textbf {\bibinfo {volume} {84}},\ \bibinfo {pages} {026010}
  (\bibinfo {year} {2011}{\natexlab{b}})},\ \Eprint
  {http://arxiv.org/abs/1103.2683} {arXiv:1103.2683 [hep-th]} \BibitemShut
  {NoStop}%
\bibitem [{\citenamefont {Shuryak}(2012)}]{Shuryak:2011aa}%
  \BibitemOpen
  \bibfield  {author} {\bibinfo {author} {\bibfnamefont {E.}~\bibnamefont
  {Shuryak}},\ }\href {\doibase 10.1088/0954-3899/39/5/054001} {\bibfield
  {journal} {\bibinfo  {journal} {J. Phys. G}\ }\textbf {\bibinfo {volume}
  {39}},\ \bibinfo {pages} {054001} (\bibinfo {year} {2012})},\ \Eprint
  {http://arxiv.org/abs/1112.2573} {arXiv:1112.2573 [hep-ph]} \BibitemShut
  {NoStop}%
\bibitem [{\citenamefont {Chesler}\ and\ \citenamefont
  {Yaffe}(2011)}]{Chesler:2010bi}%
  \BibitemOpen
  \bibfield  {author} {\bibinfo {author} {\bibfnamefont {P.~M.}\ \bibnamefont
  {Chesler}}\ and\ \bibinfo {author} {\bibfnamefont {L.~G.}\ \bibnamefont
  {Yaffe}},\ }\href {\doibase 10.1103/PhysRevLett.106.021601} {\bibfield
  {journal} {\bibinfo  {journal} {Phys. Rev. Lett.}\ }\textbf {\bibinfo
  {volume} {106}},\ \bibinfo {pages} {021601} (\bibinfo {year} {2011})},\
  \Eprint {http://arxiv.org/abs/1011.3562} {arXiv:1011.3562 [hep-th]}
  \BibitemShut {NoStop}%
\bibitem [{\citenamefont {Chesler}\ and\ \citenamefont
  {Yaffe}(2014)}]{Chesler:2013lia}%
  \BibitemOpen
  \bibfield  {author} {\bibinfo {author} {\bibfnamefont {P.~M.}\ \bibnamefont
  {Chesler}}\ and\ \bibinfo {author} {\bibfnamefont {L.~G.}\ \bibnamefont
  {Yaffe}},\ }\href {\doibase 10.1007/JHEP07(2014)086} {\bibfield  {journal}
  {\bibinfo  {journal} {JHEP}\ }\textbf {\bibinfo {volume} {07}},\ \bibinfo
  {pages} {086} (\bibinfo {year} {2014})},\ \Eprint
  {http://arxiv.org/abs/1309.1439} {arXiv:1309.1439 [hep-th]} \BibitemShut
  {NoStop}%
\bibitem [{\citenamefont {Hubeny}\ \emph {et~al.}(2013)\citenamefont {Hubeny},
  \citenamefont {Rangamani},\ and\ \citenamefont {Tonni}}]{Hubeny:2013hz}%
  \BibitemOpen
  \bibfield  {author} {\bibinfo {author} {\bibfnamefont {V.~E.}\ \bibnamefont
  {Hubeny}}, \bibinfo {author} {\bibfnamefont {M.}~\bibnamefont {Rangamani}}, \
  and\ \bibinfo {author} {\bibfnamefont {E.}~\bibnamefont {Tonni}},\ }\href
  {\doibase 10.1007/JHEP05(2013)136} {\bibfield  {journal} {\bibinfo  {journal}
  {JHEP}\ }\textbf {\bibinfo {volume} {05}},\ \bibinfo {pages} {136} (\bibinfo
  {year} {2013})},\ \Eprint {http://arxiv.org/abs/1302.0853} {arXiv:1302.0853
  [hep-th]} \BibitemShut {NoStop}%
\bibitem [{\citenamefont {Janik}\ and\ \citenamefont
  {Peschanski}(2006)}]{Janik:2005zt}%
  \BibitemOpen
  \bibfield  {author} {\bibinfo {author} {\bibfnamefont {R.~A.}\ \bibnamefont
  {Janik}}\ and\ \bibinfo {author} {\bibfnamefont {R.~B.}\ \bibnamefont
  {Peschanski}},\ }\href {\doibase 10.1103/PhysRevD.73.045013} {\bibfield
  {journal} {\bibinfo  {journal} {Phys. Rev. D}\ }\textbf {\bibinfo {volume}
  {73}},\ \bibinfo {pages} {045013} (\bibinfo {year} {2006})},\ \Eprint
  {http://arxiv.org/abs/hep-th/0512162} {arXiv:hep-th/0512162} \BibitemShut
  {NoStop}%
\bibitem [{\citenamefont {Heller}\ \emph
  {et~al.}(2012{\natexlab{a}})\citenamefont {Heller}, \citenamefont {Janik},\
  and\ \citenamefont {Witaszczyk}}]{Heller:2011ju}%
  \BibitemOpen
  \bibfield  {author} {\bibinfo {author} {\bibfnamefont {M.~P.}\ \bibnamefont
  {Heller}}, \bibinfo {author} {\bibfnamefont {R.~A.}\ \bibnamefont {Janik}}, \
  and\ \bibinfo {author} {\bibfnamefont {P.}~\bibnamefont {Witaszczyk}},\
  }\href {\doibase 10.1103/PhysRevLett.108.201602} {\bibfield  {journal}
  {\bibinfo  {journal} {Phys. Rev. Lett.}\ }\textbf {\bibinfo {volume} {108}},\
  \bibinfo {pages} {201602} (\bibinfo {year} {2012}{\natexlab{a}})},\ \Eprint
  {http://arxiv.org/abs/1103.3452} {arXiv:1103.3452 [hep-th]} \BibitemShut
  {NoStop}%
\bibitem [{\citenamefont {Kovtun}\ \emph {et~al.}(2005)\citenamefont {Kovtun},
  \citenamefont {Son},\ and\ \citenamefont {Starinets}}]{Kovtun:2004de}%
  \BibitemOpen
  \bibfield  {author} {\bibinfo {author} {\bibfnamefont {P.}~\bibnamefont
  {Kovtun}}, \bibinfo {author} {\bibfnamefont {D.~T.}\ \bibnamefont {Son}}, \
  and\ \bibinfo {author} {\bibfnamefont {A.~O.}\ \bibnamefont {Starinets}},\
  }\href {\doibase 10.1103/PhysRevLett.94.111601} {\bibfield  {journal}
  {\bibinfo  {journal} {Phys. Rev. Lett.}\ }\textbf {\bibinfo {volume} {94}},\
  \bibinfo {pages} {111601} (\bibinfo {year} {2005})},\ \Eprint
  {http://arxiv.org/abs/hep-th/0405231} {arXiv:hep-th/0405231} \BibitemShut
  {NoStop}%
\bibitem [{\citenamefont {Maldacena}\ and\ \citenamefont
  {Susskind}(2013)}]{Maldacena:2013xja}%
  \BibitemOpen
  \bibfield  {author} {\bibinfo {author} {\bibfnamefont {J.}~\bibnamefont
  {Maldacena}}\ and\ \bibinfo {author} {\bibfnamefont {L.}~\bibnamefont
  {Susskind}},\ }\href {\doibase 10.1002/prop.201300020} {\bibfield  {journal}
  {\bibinfo  {journal} {Fortsch. Phys.}\ }\textbf {\bibinfo {volume} {61}},\
  \bibinfo {pages} {781} (\bibinfo {year} {2013})},\ \Eprint
  {http://arxiv.org/abs/1306.0533} {arXiv:1306.0533 [hep-th]} \BibitemShut
  {NoStop}%
\bibitem [{\citenamefont {Maldacena}\ and\ \citenamefont
  {Qi}(2018)}]{Maldacena:2018lmt}%
  \BibitemOpen
  \bibfield  {author} {\bibinfo {author} {\bibfnamefont {J.}~\bibnamefont
  {Maldacena}}\ and\ \bibinfo {author} {\bibfnamefont {X.-L.}\ \bibnamefont
  {Qi}},\ }\href@noop {} {\  (\bibinfo {year} {2018})},\ \Eprint
  {http://arxiv.org/abs/1804.00491} {arXiv:1804.00491 [hep-th]} \BibitemShut
  {NoStop}%
\bibitem [{\citenamefont {Akers}\ \emph {et~al.}(2020)\citenamefont {Akers},
  \citenamefont {Engelhardt},\ and\ \citenamefont {Harlow}}]{Akers:2019nfi}%
  \BibitemOpen
  \bibfield  {author} {\bibinfo {author} {\bibfnamefont {C.}~\bibnamefont
  {Akers}}, \bibinfo {author} {\bibfnamefont {N.}~\bibnamefont {Engelhardt}}, \
  and\ \bibinfo {author} {\bibfnamefont {D.}~\bibnamefont {Harlow}},\ }\href
  {\doibase 10.1007/JHEP08(2020)032} {\bibfield  {journal} {\bibinfo  {journal}
  {JHEP}\ }\textbf {\bibinfo {volume} {08}},\ \bibinfo {pages} {032} (\bibinfo
  {year} {2020})},\ \Eprint {http://arxiv.org/abs/1910.00972} {arXiv:1910.00972
  [hep-th]} \BibitemShut {NoStop}%
\bibitem [{\citenamefont {Almheiri}\ \emph
  {et~al.}(2020{\natexlab{a}})\citenamefont {Almheiri}, \citenamefont
  {Hartman}, \citenamefont {Maldacena}, \citenamefont {Shaghoulian},\ and\
  \citenamefont {Tajdini}}]{Almheiri:2019qdq}%
  \BibitemOpen
  \bibfield  {author} {\bibinfo {author} {\bibfnamefont {A.}~\bibnamefont
  {Almheiri}}, \bibinfo {author} {\bibfnamefont {T.}~\bibnamefont {Hartman}},
  \bibinfo {author} {\bibfnamefont {J.}~\bibnamefont {Maldacena}}, \bibinfo
  {author} {\bibfnamefont {E.}~\bibnamefont {Shaghoulian}}, \ and\ \bibinfo
  {author} {\bibfnamefont {A.}~\bibnamefont {Tajdini}},\ }\href {\doibase
  10.1007/JHEP05(2020)013} {\bibfield  {journal} {\bibinfo  {journal} {JHEP}\
  }\textbf {\bibinfo {volume} {05}},\ \bibinfo {pages} {013} (\bibinfo {year}
  {2020}{\natexlab{a}})},\ \Eprint {http://arxiv.org/abs/1911.12333}
  {arXiv:1911.12333 [hep-th]} \BibitemShut {NoStop}%
\bibitem [{\citenamefont {Almheiri}\ \emph {et~al.}(2019)\citenamefont
  {Almheiri}, \citenamefont {Engelhardt}, \citenamefont {Marolf},\ and\
  \citenamefont {Maxfield}}]{Almheiri:2019psf}%
  \BibitemOpen
  \bibfield  {author} {\bibinfo {author} {\bibfnamefont {A.}~\bibnamefont
  {Almheiri}}, \bibinfo {author} {\bibfnamefont {N.}~\bibnamefont
  {Engelhardt}}, \bibinfo {author} {\bibfnamefont {D.}~\bibnamefont {Marolf}},
  \ and\ \bibinfo {author} {\bibfnamefont {H.}~\bibnamefont {Maxfield}},\
  }\href {\doibase 10.1007/JHEP12(2019)063} {\bibfield  {journal} {\bibinfo
  {journal} {JHEP}\ }\textbf {\bibinfo {volume} {12}},\ \bibinfo {pages} {063}
  (\bibinfo {year} {2019})},\ \Eprint {http://arxiv.org/abs/1905.08762}
  {arXiv:1905.08762 [hep-th]} \BibitemShut {NoStop}%
\bibitem [{\citenamefont {Almheiri}\ \emph {et~al.}(2021)\citenamefont
  {Almheiri}, \citenamefont {Hartman}, \citenamefont {Maldacena}, \citenamefont
  {Shaghoulian},\ and\ \citenamefont {Tajdini}}]{Almheiri:2020cfm}%
  \BibitemOpen
  \bibfield  {author} {\bibinfo {author} {\bibfnamefont {A.}~\bibnamefont
  {Almheiri}}, \bibinfo {author} {\bibfnamefont {T.}~\bibnamefont {Hartman}},
  \bibinfo {author} {\bibfnamefont {J.}~\bibnamefont {Maldacena}}, \bibinfo
  {author} {\bibfnamefont {E.}~\bibnamefont {Shaghoulian}}, \ and\ \bibinfo
  {author} {\bibfnamefont {A.}~\bibnamefont {Tajdini}},\ }\href {\doibase
  10.1103/RevModPhys.93.035002} {\bibfield  {journal} {\bibinfo  {journal}
  {Rev. Mod. Phys.}\ }\textbf {\bibinfo {volume} {93}},\ \bibinfo {pages}
  {035002} (\bibinfo {year} {2021})},\ \Eprint
  {http://arxiv.org/abs/2006.06872} {arXiv:2006.06872 [hep-th]} \BibitemShut
  {NoStop}%
\bibitem [{\citenamefont {Penington}\ \emph {et~al.}(2022)\citenamefont
  {Penington}, \citenamefont {Shenker}, \citenamefont {Stanford},\ and\
  \citenamefont {Yang}}]{Penington:2019kki}%
  \BibitemOpen
  \bibfield  {author} {\bibinfo {author} {\bibfnamefont {G.}~\bibnamefont
  {Penington}}, \bibinfo {author} {\bibfnamefont {S.~H.}\ \bibnamefont
  {Shenker}}, \bibinfo {author} {\bibfnamefont {D.}~\bibnamefont {Stanford}}, \
  and\ \bibinfo {author} {\bibfnamefont {Z.}~\bibnamefont {Yang}},\ }\href
  {\doibase 10.1007/JHEP03(2022)205} {\bibfield  {journal} {\bibinfo  {journal}
  {JHEP}\ }\textbf {\bibinfo {volume} {03}},\ \bibinfo {pages} {205} (\bibinfo
  {year} {2022})},\ \Eprint {http://arxiv.org/abs/1911.11977} {arXiv:1911.11977
  [hep-th]} \BibitemShut {NoStop}%
\bibitem [{\citenamefont {Penington}(2020)}]{Penington:2019npb}%
  \BibitemOpen
  \bibfield  {author} {\bibinfo {author} {\bibfnamefont {G.}~\bibnamefont
  {Penington}},\ }\href {\doibase 10.1007/JHEP09(2020)002} {\bibfield
  {journal} {\bibinfo  {journal} {JHEP}\ }\textbf {\bibinfo {volume} {09}},\
  \bibinfo {pages} {002} (\bibinfo {year} {2020})},\ \Eprint
  {http://arxiv.org/abs/1905.08255} {arXiv:1905.08255 [hep-th]} \BibitemShut
  {NoStop}%
\bibitem [{\citenamefont {Anderson}\ \emph {et~al.}(2020)\citenamefont
  {Anderson}, \citenamefont {Parrikar},\ and\ \citenamefont
  {Soni}}]{Anderson:2020vwi}%
  \BibitemOpen
  \bibfield  {author} {\bibinfo {author} {\bibfnamefont {L.}~\bibnamefont
  {Anderson}}, \bibinfo {author} {\bibfnamefont {O.}~\bibnamefont {Parrikar}},
  \ and\ \bibinfo {author} {\bibfnamefont {R.~M.}\ \bibnamefont {Soni}},\
  }\href {\doibase 10.1007/JHEP10(2021)226} {\bibfield  {journal} {\bibinfo
  {journal} {JHEP}\ }\textbf {\bibinfo {volume} {21}},\ \bibinfo {pages} {226}
  (\bibinfo {year} {2020})},\ \Eprint {http://arxiv.org/abs/2103.14746}
  {arXiv:2103.14746 [hep-th]} \BibitemShut {NoStop}%
\bibitem [{\citenamefont {Andronic}\ \emph {et~al.}(2018)\citenamefont
  {Andronic}, \citenamefont {Braun-Munzinger}, \citenamefont {Redlich},\ and\
  \citenamefont {Stachel}}]{Andronic:2017pug}%
  \BibitemOpen
  \bibfield  {author} {\bibinfo {author} {\bibfnamefont {A.}~\bibnamefont
  {Andronic}}, \bibinfo {author} {\bibfnamefont {P.}~\bibnamefont
  {Braun-Munzinger}}, \bibinfo {author} {\bibfnamefont {K.}~\bibnamefont
  {Redlich}}, \ and\ \bibinfo {author} {\bibfnamefont {J.}~\bibnamefont
  {Stachel}},\ }\href {\doibase 10.1038/s41586-018-0491-6} {\bibfield
  {journal} {\bibinfo  {journal} {Nature}\ }\textbf {\bibinfo {volume} {561}},\
  \bibinfo {pages} {321} (\bibinfo {year} {2018})},\ \Eprint
  {http://arxiv.org/abs/1710.09425} {arXiv:1710.09425 [nucl-th]} \BibitemShut
  {NoStop}%
\bibitem [{\citenamefont {Andronic}\ \emph {et~al.}(2021)\citenamefont
  {Andronic}, \citenamefont {Braun-Munzinger}, \citenamefont {Redlich},\ and\
  \citenamefont {Stachel}}]{Andronic:2021dkw}%
  \BibitemOpen
  \bibfield  {author} {\bibinfo {author} {\bibfnamefont {A.}~\bibnamefont
  {Andronic}}, \bibinfo {author} {\bibfnamefont {P.}~\bibnamefont
  {Braun-Munzinger}}, \bibinfo {author} {\bibfnamefont {K.}~\bibnamefont
  {Redlich}}, \ and\ \bibinfo {author} {\bibfnamefont {J.}~\bibnamefont
  {Stachel}},\ }in\ \href@noop {} {\emph {\bibinfo {booktitle} {{Criticality in
  QCD and the Hadron Resonance Gas}}}}\ (\bibinfo {year} {2021})\ \Eprint
  {http://arxiv.org/abs/2101.05747} {arXiv:2101.05747 [nucl-th]} \BibitemShut
  {NoStop}%
\bibitem [{\citenamefont {Borsanyi}\ \emph {et~al.}(2020)\citenamefont
  {Borsanyi}, \citenamefont {Fodor}, \citenamefont {Guenther}, \citenamefont
  {Kara}, \citenamefont {Katz}, \citenamefont {Parotto}, \citenamefont
  {Pasztor}, \citenamefont {Ratti},\ and\ \citenamefont
  {Szabo}}]{Borsanyi:2020fev}%
  \BibitemOpen
  \bibfield  {author} {\bibinfo {author} {\bibfnamefont {S.}~\bibnamefont
  {Borsanyi}}, \bibinfo {author} {\bibfnamefont {Z.}~\bibnamefont {Fodor}},
  \bibinfo {author} {\bibfnamefont {J.~N.}\ \bibnamefont {Guenther}}, \bibinfo
  {author} {\bibfnamefont {R.}~\bibnamefont {Kara}}, \bibinfo {author}
  {\bibfnamefont {S.~D.}\ \bibnamefont {Katz}}, \bibinfo {author}
  {\bibfnamefont {P.}~\bibnamefont {Parotto}}, \bibinfo {author} {\bibfnamefont
  {A.}~\bibnamefont {Pasztor}}, \bibinfo {author} {\bibfnamefont
  {C.}~\bibnamefont {Ratti}}, \ and\ \bibinfo {author} {\bibfnamefont {K.~K.}\
  \bibnamefont {Szabo}},\ }\href {\doibase 10.1103/PhysRevLett.125.052001}
  {\bibfield  {journal} {\bibinfo  {journal} {Phys. Rev. Lett.}\ }\textbf
  {\bibinfo {volume} {125}},\ \bibinfo {pages} {052001} (\bibinfo {year}
  {2020})},\ \Eprint {http://arxiv.org/abs/2002.02821} {arXiv:2002.02821
  [hep-lat]} \BibitemShut {NoStop}%
\bibitem [{\citenamefont {Bazavov}\ \emph {et~al.}(2019)\citenamefont {Bazavov}
  \emph {et~al.}}]{HotQCD:2018pds}%
  \BibitemOpen
  \bibfield  {author} {\bibinfo {author} {\bibfnamefont {A.}~\bibnamefont
  {Bazavov}} \emph {et~al.} (\bibinfo {collaboration} {HotQCD}),\ }\href
  {\doibase 10.1016/j.physletb.2019.05.013} {\bibfield  {journal} {\bibinfo
  {journal} {Phys. Lett. B}\ }\textbf {\bibinfo {volume} {795}},\ \bibinfo
  {pages} {15} (\bibinfo {year} {2019})},\ \Eprint
  {http://arxiv.org/abs/1812.08235} {arXiv:1812.08235 [hep-lat]} \BibitemShut
  {NoStop}%
\bibitem [{\citenamefont {Acharya}\ \emph
  {et~al.}(2022{\natexlab{a}})\citenamefont {Acharya} \emph
  {et~al.}}]{ALICE:2021puh}%
  \BibitemOpen
  \bibfield  {author} {\bibinfo {author} {\bibfnamefont {S.}~\bibnamefont
  {Acharya}} \emph {et~al.} (\bibinfo {collaboration} {ALICE}),\ }\href
  {\doibase 10.1103/PhysRevLett.128.252003} {\bibfield  {journal} {\bibinfo
  {journal} {Phys. Rev. Lett.}\ }\textbf {\bibinfo {volume} {128}},\ \bibinfo
  {pages} {252003} (\bibinfo {year} {2022}{\natexlab{a}})},\ \Eprint
  {http://arxiv.org/abs/2107.10627} {arXiv:2107.10627 [nucl-ex]} \BibitemShut
  {NoStop}%
\bibitem [{\citenamefont {M{\"u}ller}\ and\ \citenamefont
  {Sch{\"a}fer}(2017)}]{Muller:2017vnp}%
  \BibitemOpen
  \bibfield  {author} {\bibinfo {author} {\bibfnamefont {B.}~\bibnamefont
  {M{\"u}ller}}\ and\ \bibinfo {author} {\bibfnamefont {A.}~\bibnamefont
  {Sch{\"a}fer}},\ }\href@noop {} {\  (\bibinfo {year} {2017})},\ \Eprint
  {http://arxiv.org/abs/1712.03567} {arXiv:1712.03567 [nucl-th]} \BibitemShut
  {NoStop}%
\bibitem [{\citenamefont {Deutsch}(1991)}]{Deutsch:1991qu}%
  \BibitemOpen
  \bibfield  {author} {\bibinfo {author} {\bibfnamefont {J.~M.}\ \bibnamefont
  {Deutsch}},\ }\href@noop {} {\bibfield  {journal} {\bibinfo  {journal}
  {Physical Review A}\ }\textbf {\bibinfo {volume} {43}},\ \bibinfo {pages}
  {2046} (\bibinfo {year} {1991})}\BibitemShut {NoStop}%
\bibitem [{\citenamefont {Srednicki}(1994)}]{Srednicki:1994mfb}%
  \BibitemOpen
  \bibfield  {author} {\bibinfo {author} {\bibfnamefont {M.}~\bibnamefont
  {Srednicki}},\ }\href {\doibase 10.1103/PhysRevE.50.888} {\bibfield
  {journal} {\bibinfo  {journal} {Physical Review E}\ }\textbf {\bibinfo
  {volume} {50}},\ \bibinfo {pages} {888} (\bibinfo {year} {1994})},\ \Eprint
  {http://arxiv.org/abs/cond-mat/9403051} {arXiv:cond-mat/9403051} \BibitemShut
  {NoStop}%
\bibitem [{\citenamefont {D'Alessio}\ \emph {et~al.}(2016)\citenamefont
  {D'Alessio}, \citenamefont {Kafri}, \citenamefont {Polkovnikov},\ and\
  \citenamefont {Rigol}}]{DAlessio:2015qtq}%
  \BibitemOpen
  \bibfield  {author} {\bibinfo {author} {\bibfnamefont {L.}~\bibnamefont
  {D'Alessio}}, \bibinfo {author} {\bibfnamefont {Y.}~\bibnamefont {Kafri}},
  \bibinfo {author} {\bibfnamefont {A.}~\bibnamefont {Polkovnikov}}, \ and\
  \bibinfo {author} {\bibfnamefont {M.}~\bibnamefont {Rigol}},\ }\href
  {\doibase 10.1080/00018732.2016.1198134} {\bibfield  {journal} {\bibinfo
  {journal} {Adv. Phys.}\ }\textbf {\bibinfo {volume} {65}},\ \bibinfo {pages}
  {239} (\bibinfo {year} {2016})},\ \Eprint {http://arxiv.org/abs/1509.06411}
  {arXiv:1509.06411 [cond-mat.stat-mech]} \BibitemShut {NoStop}%
\bibitem [{\citenamefont {Mateos}\ \emph {et~al.}(2006)\citenamefont {Mateos},
  \citenamefont {Myers},\ and\ \citenamefont {Thomson}}]{Mateos:2006nu}%
  \BibitemOpen
  \bibfield  {author} {\bibinfo {author} {\bibfnamefont {D.}~\bibnamefont
  {Mateos}}, \bibinfo {author} {\bibfnamefont {R.~C.}\ \bibnamefont {Myers}}, \
  and\ \bibinfo {author} {\bibfnamefont {R.~M.}\ \bibnamefont {Thomson}},\
  }\href {\doibase 10.1103/PhysRevLett.97.091601} {\bibfield  {journal}
  {\bibinfo  {journal} {Phys. Rev. Lett.}\ }\textbf {\bibinfo {volume} {97}},\
  \bibinfo {pages} {091601} (\bibinfo {year} {2006})},\ \Eprint
  {http://arxiv.org/abs/hep-th/0605046} {arXiv:hep-th/0605046} \BibitemShut
  {NoStop}%
\bibitem [{\citenamefont {Bigazzi}\ \emph {et~al.}(2020)\citenamefont
  {Bigazzi}, \citenamefont {Caddeo}, \citenamefont {Cotrone},\ and\
  \citenamefont {Paredes}}]{Bigazzi:2020phm}%
  \BibitemOpen
  \bibfield  {author} {\bibinfo {author} {\bibfnamefont {F.}~\bibnamefont
  {Bigazzi}}, \bibinfo {author} {\bibfnamefont {A.}~\bibnamefont {Caddeo}},
  \bibinfo {author} {\bibfnamefont {A.~L.}\ \bibnamefont {Cotrone}}, \ and\
  \bibinfo {author} {\bibfnamefont {A.}~\bibnamefont {Paredes}},\ }\href
  {\doibase 10.1007/JHEP12(2020)200} {\bibfield  {journal} {\bibinfo  {journal}
  {JHEP}\ }\textbf {\bibinfo {volume} {12}},\ \bibinfo {pages} {200} (\bibinfo
  {year} {2020})},\ \Eprint {http://arxiv.org/abs/2008.02579} {arXiv:2008.02579
  [hep-th]} \BibitemShut {NoStop}%
\bibitem [{Note1()}]{Note1}%
  \BibitemOpen
  \bibinfo {note} {The acceptance of detectors usually covers a certain
  pseudorapidity window rather than a rapidity window. At high energy, however,
  by far the largest number of emitted particles are pions whose mass is
  smaller than their average transverse momentum, the effect of the difference
  between pseudorapidity and rapidity is small for most global
  observables.}\BibitemShut {Stop}%
\bibitem [{Note2()}]{Note2}%
  \BibitemOpen
  \bibinfo {note} {There are exceptions to this statement, e.g., in the
  measurement of the collective flow vector or in fully resolved jet
  measurements, but even these analyses ignore the full correlation information
  that is, in principle, contained in the recorded data.}\BibitemShut {Stop}%
\bibitem [{\citenamefont {Duan}\ \emph {et~al.}(2020)\citenamefont {Duan},
  \citenamefont {Akkaya}, \citenamefont {Kovner},\ and\ \citenamefont
  {Skokov}}]{Duan:2020jkz}%
  \BibitemOpen
  \bibfield  {author} {\bibinfo {author} {\bibfnamefont {H.}~\bibnamefont
  {Duan}}, \bibinfo {author} {\bibfnamefont {C.}~\bibnamefont {Akkaya}},
  \bibinfo {author} {\bibfnamefont {A.}~\bibnamefont {Kovner}}, \ and\ \bibinfo
  {author} {\bibfnamefont {V.~V.}\ \bibnamefont {Skokov}},\ }\href {\doibase
  10.1103/PhysRevD.101.036017} {\bibfield  {journal} {\bibinfo  {journal}
  {Phys. Rev. D}\ }\textbf {\bibinfo {volume} {101}},\ \bibinfo {pages}
  {036017} (\bibinfo {year} {2020})},\ \Eprint
  {http://arxiv.org/abs/2001.01726} {arXiv:2001.01726 [hep-ph]} \BibitemShut
  {NoStop}%
\bibitem [{\citenamefont {Pal}\ and\ \citenamefont {Pratt}(2004)}]{Pal:2003rz}%
  \BibitemOpen
  \bibfield  {author} {\bibinfo {author} {\bibfnamefont {S.}~\bibnamefont
  {Pal}}\ and\ \bibinfo {author} {\bibfnamefont {S.}~\bibnamefont {Pratt}},\
  }\href {\doibase 10.1016/j.physletb.2003.10.054} {\bibfield  {journal}
  {\bibinfo  {journal} {Phys. Lett. B}\ }\textbf {\bibinfo {volume} {578}},\
  \bibinfo {pages} {310} (\bibinfo {year} {2004})},\ \Eprint
  {http://arxiv.org/abs/nucl-th/0308077} {arXiv:nucl-th/0308077} \BibitemShut
  {NoStop}%
\bibitem [{\citenamefont {M{\"u}ller}\ and\ \citenamefont
  {Rajagopal}(2005)}]{Muller:2005en}%
  \BibitemOpen
  \bibfield  {author} {\bibinfo {author} {\bibfnamefont {B.}~\bibnamefont
  {M{\"u}ller}}\ and\ \bibinfo {author} {\bibfnamefont {K.}~\bibnamefont
  {Rajagopal}},\ }\href {\doibase 10.1140/epjc/s2005-02256-3} {\bibfield
  {journal} {\bibinfo  {journal} {Eur. Phys. J. C}\ }\textbf {\bibinfo {volume}
  {43}},\ \bibinfo {pages} {15} (\bibinfo {year} {2005})},\ \Eprint
  {http://arxiv.org/abs/hep-ph/0502174} {arXiv:hep-ph/0502174} \BibitemShut
  {NoStop}%
\bibitem [{\citenamefont {Hanus}\ \emph {et~al.}(2019)\citenamefont {Hanus},
  \citenamefont {Mazeliauskas},\ and\ \citenamefont {Reygers}}]{Hanus:2019fnc}%
  \BibitemOpen
  \bibfield  {author} {\bibinfo {author} {\bibfnamefont {P.}~\bibnamefont
  {Hanus}}, \bibinfo {author} {\bibfnamefont {A.}~\bibnamefont {Mazeliauskas}},
  \ and\ \bibinfo {author} {\bibfnamefont {K.}~\bibnamefont {Reygers}},\ }\href
  {\doibase 10.1103/PhysRevC.100.064903} {\bibfield  {journal} {\bibinfo
  {journal} {Phys. Rev. C}\ }\textbf {\bibinfo {volume} {100}},\ \bibinfo
  {pages} {064903} (\bibinfo {year} {2019})},\ \Eprint
  {http://arxiv.org/abs/1908.02792} {arXiv:1908.02792 [hep-ph]} \BibitemShut
  {NoStop}%
\bibitem [{\citenamefont {Wigner}(1967)}]{Wigner:1967ran}%
  \BibitemOpen
  \bibfield  {author} {\bibinfo {author} {\bibfnamefont {E.~P.}\ \bibnamefont
  {Wigner}},\ }\href@noop {} {\bibfield  {journal} {\bibinfo  {journal} {SIAM
  review}\ }\textbf {\bibinfo {volume} {9}},\ \bibinfo {pages} {1} (\bibinfo
  {year} {1967})}\BibitemShut {NoStop}%
\bibitem [{\citenamefont {Dymarsky}(2019)}]{Dymarsky:2018sef}%
  \BibitemOpen
  \bibfield  {author} {\bibinfo {author} {\bibfnamefont {A.}~\bibnamefont
  {Dymarsky}},\ }\href {\doibase 10.1103/PhysRevB.99.224302} {\bibfield
  {journal} {\bibinfo  {journal} {Phys. Rev. B}\ }\textbf {\bibinfo {volume}
  {99}},\ \bibinfo {pages} {224302} (\bibinfo {year} {2019})},\ \Eprint
  {http://arxiv.org/abs/1806.04187} {arXiv:1806.04187 [cond-mat.stat-mech]}
  \BibitemShut {NoStop}%
\bibitem [{\citenamefont {Richter}\ \emph {et~al.}(2020)\citenamefont
  {Richter}, \citenamefont {Dymarsky}, \citenamefont {Steinigeweg},\ and\
  \citenamefont {Gemmer}}]{Richter:2020bkf}%
  \BibitemOpen
  \bibfield  {author} {\bibinfo {author} {\bibfnamefont {J.}~\bibnamefont
  {Richter}}, \bibinfo {author} {\bibfnamefont {A.}~\bibnamefont {Dymarsky}},
  \bibinfo {author} {\bibfnamefont {R.}~\bibnamefont {Steinigeweg}}, \ and\
  \bibinfo {author} {\bibfnamefont {J.}~\bibnamefont {Gemmer}},\ }\href
  {\doibase 10.1103/PhysRevE.102.042127} {\bibfield  {journal} {\bibinfo
  {journal} {Phys. Rev. E}\ }\textbf {\bibinfo {volume} {102}},\ \bibinfo
  {pages} {042127} (\bibinfo {year} {2020})},\ \Eprint
  {http://arxiv.org/abs/2007.15070} {arXiv:2007.15070 [cond-mat.stat-mech]}
  \BibitemShut {NoStop}%
\bibitem [{\citenamefont {Bolte}\ \emph {et~al.}(2000)\citenamefont {Bolte},
  \citenamefont {M{\"u}ller},\ and\ \citenamefont
  {Sch{\"a}fer}}]{Bolte:1999th}%
  \BibitemOpen
  \bibfield  {author} {\bibinfo {author} {\bibfnamefont {J.}~\bibnamefont
  {Bolte}}, \bibinfo {author} {\bibfnamefont {B.}~\bibnamefont {M{\"u}ller}}, \
  and\ \bibinfo {author} {\bibfnamefont {A.}~\bibnamefont {Sch{\"a}fer}},\
  }\href {\doibase 10.1103/PhysRevD.61.054506} {\bibfield  {journal} {\bibinfo
  {journal} {Phys. Rev. D}\ }\textbf {\bibinfo {volume} {61}},\ \bibinfo
  {pages} {054506} (\bibinfo {year} {2000})},\ \Eprint
  {http://arxiv.org/abs/hep-lat/9906037} {arXiv:hep-lat/9906037} \BibitemShut
  {NoStop}%
\bibitem [{\citenamefont {M{\"u}ller}\ and\ \citenamefont
  {Sch{\"a}fer}(2012)}]{Muller:2011bb}%
  \BibitemOpen
  \bibfield  {author} {\bibinfo {author} {\bibfnamefont {B.}~\bibnamefont
  {M{\"u}ller}}\ and\ \bibinfo {author} {\bibfnamefont {A.}~\bibnamefont
  {Sch{\"a}fer}},\ }\href {\doibase 10.1103/PhysRevD.85.114030} {\bibfield
  {journal} {\bibinfo  {journal} {Phys. Rev. D}\ }\textbf {\bibinfo {volume}
  {85}},\ \bibinfo {pages} {114030} (\bibinfo {year} {2012})},\ \bibinfo {note}
  {[Erratum: Phys.Rev.D 96, 059903 (2017)]},\ \Eprint
  {http://arxiv.org/abs/1111.3347} {arXiv:1111.3347 [hep-ph]} \BibitemShut
  {NoStop}%
\bibitem [{\citenamefont {Lappi}(2006)}]{Lappi:2006hq}%
  \BibitemOpen
  \bibfield  {author} {\bibinfo {author} {\bibfnamefont {T.}~\bibnamefont
  {Lappi}},\ }\href {\doibase 10.1016/j.physletb.2006.10.017} {\bibfield
  {journal} {\bibinfo  {journal} {Phys. Lett. B}\ }\textbf {\bibinfo {volume}
  {643}},\ \bibinfo {pages} {11} (\bibinfo {year} {2006})},\ \Eprint
  {http://arxiv.org/abs/hep-ph/0606207} {arXiv:hep-ph/0606207} \BibitemShut
  {NoStop}%
\bibitem [{Note3()}]{Note3}%
  \BibitemOpen
  \bibinfo {note} {We refer to any transition between AdS space with a black
  brane and thermal AdS space as ``Hawking-Page type'' transition, independent
  of the mechanism that sets the temperature scale at which the transition
  occurs.}\BibitemShut {Stop}%
\bibitem [{\citenamefont {Maldacena}(1998{\natexlab{a}})}]{Maldacena:1997re}%
  \BibitemOpen
  \bibfield  {author} {\bibinfo {author} {\bibfnamefont {J.~M.}\ \bibnamefont
  {Maldacena}},\ }\href {\doibase 10.1023/A:1026654312961} {\bibfield
  {journal} {\bibinfo  {journal} {Adv. Theor. Math. Phys.}\ }\textbf {\bibinfo
  {volume} {2}},\ \bibinfo {pages} {231} (\bibinfo {year}
  {1998}{\natexlab{a}})},\ \Eprint {http://arxiv.org/abs/hep-th/9711200}
  {arXiv:hep-th/9711200} \BibitemShut {NoStop}%
\bibitem [{\citenamefont {Bayona}\ and\ \citenamefont
  {Braga}(2007)}]{Bayona:2005nq}%
  \BibitemOpen
  \bibfield  {author} {\bibinfo {author} {\bibfnamefont {C.~A.}\ \bibnamefont
  {Bayona}}\ and\ \bibinfo {author} {\bibfnamefont {N.~R.~F.}\ \bibnamefont
  {Braga}},\ }\href {\doibase 10.1007/s10714-007-0446-y} {\bibfield  {journal}
  {\bibinfo  {journal} {Gen. Rel. Grav.}\ }\textbf {\bibinfo {volume} {39}},\
  \bibinfo {pages} {1367} (\bibinfo {year} {2007})},\ \Eprint
  {http://arxiv.org/abs/hep-th/0512182} {arXiv:hep-th/0512182} \BibitemShut
  {NoStop}%
\bibitem [{\citenamefont {Kruczenski}\ \emph {et~al.}(2004)\citenamefont
  {Kruczenski}, \citenamefont {Mateos}, \citenamefont {Myers},\ and\
  \citenamefont {Winters}}]{Kruczenski:2003uq}%
  \BibitemOpen
  \bibfield  {author} {\bibinfo {author} {\bibfnamefont {M.}~\bibnamefont
  {Kruczenski}}, \bibinfo {author} {\bibfnamefont {D.}~\bibnamefont {Mateos}},
  \bibinfo {author} {\bibfnamefont {R.~C.}\ \bibnamefont {Myers}}, \ and\
  \bibinfo {author} {\bibfnamefont {D.~J.}\ \bibnamefont {Winters}},\ }\href
  {\doibase 10.1088/1126-6708/2004/05/041} {\bibfield  {journal} {\bibinfo
  {journal} {JHEP}\ }\textbf {\bibinfo {volume} {05}},\ \bibinfo {pages} {041}
  (\bibinfo {year} {2004})},\ \Eprint {http://arxiv.org/abs/hep-th/0311270}
  {arXiv:hep-th/0311270} \BibitemShut {NoStop}%
\bibitem [{\citenamefont {Waeber}\ and\ \citenamefont
  {Yaffe}(2022)}]{Waeber:2022tts}%
  \BibitemOpen
  \bibfield  {author} {\bibinfo {author} {\bibfnamefont {S.}~\bibnamefont
  {Waeber}}\ and\ \bibinfo {author} {\bibfnamefont {L.~G.}\ \bibnamefont
  {Yaffe}},\ }\href@noop {} {\  (\bibinfo {year} {2022})},\ \Eprint
  {http://arxiv.org/abs/2206.01819} {arXiv:2206.01819 [hep-th]} \BibitemShut
  {NoStop}%
\bibitem [{\citenamefont {Van~Raamsdonk}(2020)}]{VanRaamsdonk:2020ydg}%
  \BibitemOpen
  \bibfield  {author} {\bibinfo {author} {\bibfnamefont {M.}~\bibnamefont
  {Van~Raamsdonk}},\ }\href {\doibase 10.1126/science.aay9560} {\bibfield
  {journal} {\bibinfo  {journal} {Science}\ }\textbf {\bibinfo {volume}
  {370}},\ \bibinfo {pages} {198} (\bibinfo {year} {2020})}\BibitemShut
  {NoStop}%
\bibitem [{\citenamefont {Ding}\ \emph {et~al.}(2012)\citenamefont {Ding},
  \citenamefont {Francis}, \citenamefont {Kaczmarek}, \citenamefont {Karsch},
  \citenamefont {Satz},\ and\ \citenamefont {Soeldner}}]{Ding:2012sp}%
  \BibitemOpen
  \bibfield  {author} {\bibinfo {author} {\bibfnamefont {H.~T.}\ \bibnamefont
  {Ding}}, \bibinfo {author} {\bibfnamefont {A.}~\bibnamefont {Francis}},
  \bibinfo {author} {\bibfnamefont {O.}~\bibnamefont {Kaczmarek}}, \bibinfo
  {author} {\bibfnamefont {F.}~\bibnamefont {Karsch}}, \bibinfo {author}
  {\bibfnamefont {H.}~\bibnamefont {Satz}}, \ and\ \bibinfo {author}
  {\bibfnamefont {W.}~\bibnamefont {Soeldner}},\ }\href {\doibase
  10.1103/PhysRevD.86.014509} {\bibfield  {journal} {\bibinfo  {journal} {Phys.
  Rev. D}\ }\textbf {\bibinfo {volume} {86}},\ \bibinfo {pages} {014509}
  (\bibinfo {year} {2012})},\ \Eprint {http://arxiv.org/abs/1204.4945}
  {arXiv:1204.4945 [hep-lat]} \BibitemShut {NoStop}%
\bibitem [{\citenamefont {Liu}\ and\ \citenamefont {Suh}(2014)}]{Liu:2013iza}%
  \BibitemOpen
  \bibfield  {author} {\bibinfo {author} {\bibfnamefont {H.}~\bibnamefont
  {Liu}}\ and\ \bibinfo {author} {\bibfnamefont {S.~J.}\ \bibnamefont {Suh}},\
  }\href {\doibase 10.1103/PhysRevLett.112.011601} {\bibfield  {journal}
  {\bibinfo  {journal} {Phys. Rev. Lett.}\ }\textbf {\bibinfo {volume} {112}},\
  \bibinfo {pages} {011601} (\bibinfo {year} {2014})},\ \Eprint
  {http://arxiv.org/abs/1305.7244} {arXiv:1305.7244 [hep-th]} \BibitemShut
  {NoStop}%
\bibitem [{\citenamefont {Popescu}\ \emph {et~al.}(2006)\citenamefont
  {Popescu}, \citenamefont {Short},\ and\ \citenamefont
  {Winter}}]{Popescu:2006en}%
  \BibitemOpen
  \bibfield  {author} {\bibinfo {author} {\bibfnamefont {S.}~\bibnamefont
  {Popescu}}, \bibinfo {author} {\bibfnamefont {A.~J.}\ \bibnamefont {Short}},
  \ and\ \bibinfo {author} {\bibfnamefont {A.}~\bibnamefont {Winter}},\
  }\href@noop {} {\bibfield  {journal} {\bibinfo  {journal} {Nature Physics}\
  }\textbf {\bibinfo {volume} {2}},\ \bibinfo {pages} {754} (\bibinfo {year}
  {2006})}\BibitemShut {NoStop}%
\bibitem [{\citenamefont {Gong}\ \emph {et~al.}(2022)\citenamefont {Gong},
  \citenamefont {Parida}, \citenamefont {Tu},\ and\ \citenamefont
  {Venugopalan}}]{Gong:2021bcp}%
  \BibitemOpen
  \bibfield  {author} {\bibinfo {author} {\bibfnamefont {W.}~\bibnamefont
  {Gong}}, \bibinfo {author} {\bibfnamefont {G.}~\bibnamefont {Parida}},
  \bibinfo {author} {\bibfnamefont {Z.}~\bibnamefont {Tu}}, \ and\ \bibinfo
  {author} {\bibfnamefont {R.}~\bibnamefont {Venugopalan}},\ }\href {\doibase
  10.1103/PhysRevD.106.L031501} {\bibfield  {journal} {\bibinfo  {journal}
  {Phys. Rev. D}\ }\textbf {\bibinfo {volume} {106}},\ \bibinfo {pages}
  {L031501} (\bibinfo {year} {2022})},\ \Eprint
  {http://arxiv.org/abs/2107.13007} {arXiv:2107.13007 [hep-ph]} \BibitemShut
  {NoStop}%
\bibitem [{\citenamefont {Anselm}\ and\ \citenamefont
  {Ryskin}(1997)}]{Anselm:1996vm}%
  \BibitemOpen
  \bibfield  {author} {\bibinfo {author} {\bibfnamefont {A.~A.}\ \bibnamefont
  {Anselm}}\ and\ \bibinfo {author} {\bibfnamefont {M.~G.}\ \bibnamefont
  {Ryskin}},\ }\href {\doibase 10.1007/s002180050338} {\bibfield  {journal}
  {\bibinfo  {journal} {Z. Phys. A}\ }\textbf {\bibinfo {volume} {358}},\
  \bibinfo {pages} {353} (\bibinfo {year} {1997})},\ \Eprint
  {http://arxiv.org/abs/hep-ph/9606319} {arXiv:hep-ph/9606319} \BibitemShut
  {NoStop}%
\bibitem [{\citenamefont {Mohanty}\ and\ \citenamefont
  {Serreau}(2005)}]{Mohanty:2005mv}%
  \BibitemOpen
  \bibfield  {author} {\bibinfo {author} {\bibfnamefont {B.}~\bibnamefont
  {Mohanty}}\ and\ \bibinfo {author} {\bibfnamefont {J.}~\bibnamefont
  {Serreau}},\ }\href {\doibase 10.1016/j.physrep.2005.04.004} {\bibfield
  {journal} {\bibinfo  {journal} {Phys. Rept.}\ }\textbf {\bibinfo {volume}
  {414}},\ \bibinfo {pages} {263} (\bibinfo {year} {2005})},\ \Eprint
  {http://arxiv.org/abs/hep-ph/0504154} {arXiv:hep-ph/0504154} \BibitemShut
  {NoStop}%
\bibitem [{\citenamefont {Acharya}\ \emph
  {et~al.}(2022{\natexlab{b}})\citenamefont {Acharya} \emph
  {et~al.}}]{ALICE:2021fpb}%
  \BibitemOpen
  \bibfield  {author} {\bibinfo {author} {\bibfnamefont {S.}~\bibnamefont
  {Acharya}} \emph {et~al.} (\bibinfo {collaboration} {ALICE}),\ }\href
  {\doibase 10.1016/j.physletb.2022.137242} {\bibfield  {journal} {\bibinfo
  {journal} {Phys. Lett. B}\ }\textbf {\bibinfo {volume} {832}},\ \bibinfo
  {pages} {137242} (\bibinfo {year} {2022}{\natexlab{b}})},\ \Eprint
  {http://arxiv.org/abs/2112.09482} {arXiv:2112.09482 [nucl-ex]} \BibitemShut
  {NoStop}%
\bibitem [{\citenamefont {Steinhauer}(2016)}]{Steinhauer:2015saa}%
  \BibitemOpen
  \bibfield  {author} {\bibinfo {author} {\bibfnamefont {J.}~\bibnamefont
  {Steinhauer}},\ }\href {\doibase 10.1038/nphys3863} {\bibfield  {journal}
  {\bibinfo  {journal} {Nature Phys.}\ }\textbf {\bibinfo {volume} {12}},\
  \bibinfo {pages} {959} (\bibinfo {year} {2016})},\ \Eprint
  {http://arxiv.org/abs/1510.00621} {arXiv:1510.00621 [gr-qc]} \BibitemShut
  {NoStop}%
\bibitem [{\citenamefont {Kolobov}\ \emph {et~al.}(2021)\citenamefont
  {Kolobov}, \citenamefont {Golubkov}, \citenamefont {Mu\~noz~de Nova},\ and\
  \citenamefont {Steinhauer}}]{Kolobov:2019qfs}%
  \BibitemOpen
  \bibfield  {author} {\bibinfo {author} {\bibfnamefont {V.~I.}\ \bibnamefont
  {Kolobov}}, \bibinfo {author} {\bibfnamefont {K.}~\bibnamefont {Golubkov}},
  \bibinfo {author} {\bibfnamefont {J.~R.}\ \bibnamefont {Mu\~noz~de Nova}}, \
  and\ \bibinfo {author} {\bibfnamefont {J.}~\bibnamefont {Steinhauer}},\
  }\href {\doibase 10.1038/s41567-020-01076-0} {\bibfield  {journal} {\bibinfo
  {journal} {Nature Phys.}\ }\textbf {\bibinfo {volume} {17}},\ \bibinfo
  {pages} {362} (\bibinfo {year} {2021})},\ \Eprint
  {http://arxiv.org/abs/1910.09363} {arXiv:1910.09363 [gr-qc]} \BibitemShut
  {NoStop}%
\bibitem [{\citenamefont {Chesler}\ and\ \citenamefont
  {Yaffe}(2009)}]{Chesler:2008hg}%
  \BibitemOpen
  \bibfield  {author} {\bibinfo {author} {\bibfnamefont {P.~M.}\ \bibnamefont
  {Chesler}}\ and\ \bibinfo {author} {\bibfnamefont {L.~G.}\ \bibnamefont
  {Yaffe}},\ }\href {\doibase 10.1103/PhysRevLett.102.211601} {\bibfield
  {journal} {\bibinfo  {journal} {Phys. Rev. Lett.}\ }\textbf {\bibinfo
  {volume} {102}},\ \bibinfo {pages} {211601} (\bibinfo {year} {2009})},\
  \Eprint {http://arxiv.org/abs/0812.2053} {arXiv:0812.2053 [hep-th]}
  \BibitemShut {NoStop}%
\bibitem [{\citenamefont {Heller}\ \emph
  {et~al.}(2012{\natexlab{b}})\citenamefont {Heller}, \citenamefont {Mateos},
  \citenamefont {van~der Schee},\ and\ \citenamefont
  {Trancanelli}}]{Heller:2012km}%
  \BibitemOpen
  \bibfield  {author} {\bibinfo {author} {\bibfnamefont {M.~P.}\ \bibnamefont
  {Heller}}, \bibinfo {author} {\bibfnamefont {D.}~\bibnamefont {Mateos}},
  \bibinfo {author} {\bibfnamefont {W.}~\bibnamefont {van~der Schee}}, \ and\
  \bibinfo {author} {\bibfnamefont {D.}~\bibnamefont {Trancanelli}},\ }\href
  {\doibase 10.1103/PhysRevLett.108.191601} {\bibfield  {journal} {\bibinfo
  {journal} {Phys. Rev. Lett.}\ }\textbf {\bibinfo {volume} {108}},\ \bibinfo
  {pages} {191601} (\bibinfo {year} {2012}{\natexlab{b}})},\ \Eprint
  {http://arxiv.org/abs/1202.0981} {arXiv:1202.0981 [hep-th]} \BibitemShut
  {NoStop}%
\bibitem [{\citenamefont {Casalderrey-Solana}\ \emph
  {et~al.}(2014)\citenamefont {Casalderrey-Solana}, \citenamefont {Heller},
  \citenamefont {Mateos},\ and\ \citenamefont {van~der
  Schee}}]{Casalderrey-Solana:2013sxa}%
  \BibitemOpen
  \bibfield  {author} {\bibinfo {author} {\bibfnamefont {J.}~\bibnamefont
  {Casalderrey-Solana}}, \bibinfo {author} {\bibfnamefont {M.~P.}\ \bibnamefont
  {Heller}}, \bibinfo {author} {\bibfnamefont {D.}~\bibnamefont {Mateos}}, \
  and\ \bibinfo {author} {\bibfnamefont {W.}~\bibnamefont {van~der Schee}},\
  }\href {\doibase 10.1103/PhysRevLett.112.221602} {\bibfield  {journal}
  {\bibinfo  {journal} {Phys. Rev. Lett.}\ }\textbf {\bibinfo {volume} {112}},\
  \bibinfo {pages} {221602} (\bibinfo {year} {2014})},\ \Eprint
  {http://arxiv.org/abs/1312.2956} {arXiv:1312.2956 [hep-th]} \BibitemShut
  {NoStop}%
\bibitem [{\citenamefont {Chesler}\ \emph {et~al.}(2015)\citenamefont
  {Chesler}, \citenamefont {Kilbertus},\ and\ \citenamefont {van~der
  Schee}}]{Chesler:2015fpa}%
  \BibitemOpen
  \bibfield  {author} {\bibinfo {author} {\bibfnamefont {P.~M.}\ \bibnamefont
  {Chesler}}, \bibinfo {author} {\bibfnamefont {N.}~\bibnamefont {Kilbertus}},
  \ and\ \bibinfo {author} {\bibfnamefont {W.}~\bibnamefont {van~der Schee}},\
  }\href {\doibase 10.1007/JHEP11(2015)135} {\bibfield  {journal} {\bibinfo
  {journal} {JHEP}\ }\textbf {\bibinfo {volume} {11}},\ \bibinfo {pages} {135}
  (\bibinfo {year} {2015})},\ \Eprint {http://arxiv.org/abs/1507.02548}
  {arXiv:1507.02548 [hep-th]} \BibitemShut {NoStop}%
\bibitem [{\citenamefont {Ecker}\ \emph {et~al.}(2016)\citenamefont {Ecker},
  \citenamefont {Grumiller}, \citenamefont {Stanzer}, \citenamefont
  {Stricker},\ and\ \citenamefont {van~der Schee}}]{Ecker:2016thn}%
  \BibitemOpen
  \bibfield  {author} {\bibinfo {author} {\bibfnamefont {C.}~\bibnamefont
  {Ecker}}, \bibinfo {author} {\bibfnamefont {D.}~\bibnamefont {Grumiller}},
  \bibinfo {author} {\bibfnamefont {P.}~\bibnamefont {Stanzer}}, \bibinfo
  {author} {\bibfnamefont {S.~A.}\ \bibnamefont {Stricker}}, \ and\ \bibinfo
  {author} {\bibfnamefont {W.}~\bibnamefont {van~der Schee}},\ }\href {\doibase
  10.1007/JHEP11(2016)054} {\bibfield  {journal} {\bibinfo  {journal} {JHEP}\
  }\textbf {\bibinfo {volume} {11}},\ \bibinfo {pages} {054} (\bibinfo {year}
  {2016})},\ \Eprint {http://arxiv.org/abs/1609.03676} {arXiv:1609.03676
  [hep-th]} \BibitemShut {NoStop}%
\bibitem [{\citenamefont {Casalderrey-Solana}\ \emph
  {et~al.}(2016)\citenamefont {Casalderrey-Solana}, \citenamefont {Mateos},
  \citenamefont {van~der Schee},\ and\ \citenamefont
  {Triana}}]{Casalderrey-Solana:2016xfq}%
  \BibitemOpen
  \bibfield  {author} {\bibinfo {author} {\bibfnamefont {J.}~\bibnamefont
  {Casalderrey-Solana}}, \bibinfo {author} {\bibfnamefont {D.}~\bibnamefont
  {Mateos}}, \bibinfo {author} {\bibfnamefont {W.}~\bibnamefont {van~der
  Schee}}, \ and\ \bibinfo {author} {\bibfnamefont {M.}~\bibnamefont
  {Triana}},\ }\href {\doibase 10.1007/JHEP09(2016)108} {\bibfield  {journal}
  {\bibinfo  {journal} {JHEP}\ }\textbf {\bibinfo {volume} {09}},\ \bibinfo
  {pages} {108} (\bibinfo {year} {2016})},\ \Eprint
  {http://arxiv.org/abs/1607.05273} {arXiv:1607.05273 [hep-th]} \BibitemShut
  {NoStop}%
\bibitem [{\citenamefont {Endrodi}\ \emph {et~al.}(2018)\citenamefont
  {Endrodi}, \citenamefont {Kaminski}, \citenamefont {Sch{\"a}fer},
  \citenamefont {Wu},\ and\ \citenamefont {Yaffe}}]{Endrodi:2018ikq}%
  \BibitemOpen
  \bibfield  {author} {\bibinfo {author} {\bibfnamefont {G.}~\bibnamefont
  {Endrodi}}, \bibinfo {author} {\bibfnamefont {M.}~\bibnamefont {Kaminski}},
  \bibinfo {author} {\bibfnamefont {A.}~\bibnamefont {Sch{\"a}fer}}, \bibinfo
  {author} {\bibfnamefont {J.}~\bibnamefont {Wu}}, \ and\ \bibinfo {author}
  {\bibfnamefont {L.}~\bibnamefont {Yaffe}},\ }\href {\doibase
  10.1007/JHEP09(2018)070} {\bibfield  {journal} {\bibinfo  {journal} {JHEP}\
  }\textbf {\bibinfo {volume} {09}},\ \bibinfo {pages} {070} (\bibinfo {year}
  {2018})},\ \Eprint {http://arxiv.org/abs/1806.09632} {arXiv:1806.09632
  [hep-th]} \BibitemShut {NoStop}%
\bibitem [{\citenamefont {Waeber}\ \emph {et~al.}(2019)\citenamefont {Waeber},
  \citenamefont {Rabenstein}, \citenamefont {Sch{\"a}fer},\ and\ \citenamefont
  {Yaffe}}]{Waeber:2019nqd}%
  \BibitemOpen
  \bibfield  {author} {\bibinfo {author} {\bibfnamefont {S.}~\bibnamefont
  {Waeber}}, \bibinfo {author} {\bibfnamefont {A.}~\bibnamefont {Rabenstein}},
  \bibinfo {author} {\bibfnamefont {A.}~\bibnamefont {Sch{\"a}fer}}, \ and\
  \bibinfo {author} {\bibfnamefont {L.~G.}\ \bibnamefont {Yaffe}},\ }\href
  {\doibase 10.1007/JHEP08(2019)005} {\bibfield  {journal} {\bibinfo  {journal}
  {JHEP}\ }\textbf {\bibinfo {volume} {08}},\ \bibinfo {pages} {005} (\bibinfo
  {year} {2019})},\ \Eprint {http://arxiv.org/abs/1906.05086} {arXiv:1906.05086
  [hep-th]} \BibitemShut {NoStop}%
\bibitem [{\citenamefont {M{\"u}ller}\ \emph {et~al.}(2020)\citenamefont
  {M{\"u}ller}, \citenamefont {Rabenstein}, \citenamefont {Sch{\"a}fer},
  \citenamefont {Waeber},\ and\ \citenamefont {Yaffe}}]{Muller:2020ziz}%
  \BibitemOpen
  \bibfield  {author} {\bibinfo {author} {\bibfnamefont {B.}~\bibnamefont
  {M{\"u}ller}}, \bibinfo {author} {\bibfnamefont {A.}~\bibnamefont
  {Rabenstein}}, \bibinfo {author} {\bibfnamefont {A.}~\bibnamefont
  {Sch{\"a}fer}}, \bibinfo {author} {\bibfnamefont {S.}~\bibnamefont {Waeber}},
  \ and\ \bibinfo {author} {\bibfnamefont {L.~G.}\ \bibnamefont {Yaffe}},\
  }\href {\doibase 10.1103/PhysRevD.101.076008} {\bibfield  {journal} {\bibinfo
   {journal} {Phys. Rev. D}\ }\textbf {\bibinfo {volume} {101}},\ \bibinfo
  {pages} {076008} (\bibinfo {year} {2020})},\ \Eprint
  {http://arxiv.org/abs/2001.07161} {arXiv:2001.07161 [hep-ph]} \BibitemShut
  {NoStop}%
\bibitem [{\citenamefont {Gubser}\ \emph {et~al.}(1998)\citenamefont {Gubser},
  \citenamefont {Klebanov},\ and\ \citenamefont {Tseytlin}}]{Gubser:1998nz}%
  \BibitemOpen
  \bibfield  {author} {\bibinfo {author} {\bibfnamefont {S.~S.}\ \bibnamefont
  {Gubser}}, \bibinfo {author} {\bibfnamefont {I.~R.}\ \bibnamefont
  {Klebanov}}, \ and\ \bibinfo {author} {\bibfnamefont {A.~A.}\ \bibnamefont
  {Tseytlin}},\ }\href {\doibase 10.1016/S0550-3213(98)00514-8} {\bibfield
  {journal} {\bibinfo  {journal} {Nucl. Phys. B}\ }\textbf {\bibinfo {volume}
  {534}},\ \bibinfo {pages} {202} (\bibinfo {year} {1998})},\ \Eprint
  {http://arxiv.org/abs/hep-th/9805156} {arXiv:hep-th/9805156} \BibitemShut
  {NoStop}%
\bibitem [{\citenamefont {Buchel}(2008)}]{Buchel:2008sh}%
  \BibitemOpen
  \bibfield  {author} {\bibinfo {author} {\bibfnamefont {A.}~\bibnamefont
  {Buchel}},\ }\href {\doibase 10.1016/j.nuclphysb.2008.05.024} {\bibfield
  {journal} {\bibinfo  {journal} {Nucl. Phys. B}\ }\textbf {\bibinfo {volume}
  {803}},\ \bibinfo {pages} {166} (\bibinfo {year} {2008})},\ \Eprint
  {http://arxiv.org/abs/0805.2683} {arXiv:0805.2683 [hep-th]} \BibitemShut
  {NoStop}%
\bibitem [{\citenamefont {Waeber}\ and\ \citenamefont
  {Sch{\"a}fer}(2018)}]{Waeber:2018bea}%
  \BibitemOpen
  \bibfield  {author} {\bibinfo {author} {\bibfnamefont {S.}~\bibnamefont
  {Waeber}}\ and\ \bibinfo {author} {\bibfnamefont {A.}~\bibnamefont
  {Sch{\"a}fer}},\ }\href {\doibase 10.1007/JHEP07(2018)069} {\bibfield
  {journal} {\bibinfo  {journal} {JHEP}\ }\textbf {\bibinfo {volume} {07}},\
  \bibinfo {pages} {069} (\bibinfo {year} {2018})},\ \Eprint
  {http://arxiv.org/abs/1804.01912} {arXiv:1804.01912 [hep-th]} \BibitemShut
  {NoStop}%
\bibitem [{\citenamefont {Chesler}\ and\ \citenamefont {van~der
  Schee}(2015)}]{Chesler:2015lsa}%
  \BibitemOpen
  \bibfield  {author} {\bibinfo {author} {\bibfnamefont {P.~M.}\ \bibnamefont
  {Chesler}}\ and\ \bibinfo {author} {\bibfnamefont {W.}~\bibnamefont {van~der
  Schee}},\ }\href {\doibase 10.1142/S0218301315300118} {\bibfield  {journal}
  {\bibinfo  {journal} {Int. J. Mod. Phys. E}\ }\textbf {\bibinfo {volume}
  {24}},\ \bibinfo {pages} {1530011} (\bibinfo {year} {2015})},\ \Eprint
  {http://arxiv.org/abs/1501.04952} {arXiv:1501.04952 [nucl-th]} \BibitemShut
  {NoStop}%
\bibitem [{\citenamefont {van~der Schee}\ \emph {et~al.}(2013)\citenamefont
  {van~der Schee}, \citenamefont {Romatschke},\ and\ \citenamefont
  {Pratt}}]{vanderSchee:2013pia}%
  \BibitemOpen
  \bibfield  {author} {\bibinfo {author} {\bibfnamefont {W.}~\bibnamefont
  {van~der Schee}}, \bibinfo {author} {\bibfnamefont {P.}~\bibnamefont
  {Romatschke}}, \ and\ \bibinfo {author} {\bibfnamefont {S.}~\bibnamefont
  {Pratt}},\ }\href {\doibase 10.1103/PhysRevLett.111.222302} {\bibfield
  {journal} {\bibinfo  {journal} {Phys. Rev. Lett.}\ }\textbf {\bibinfo
  {volume} {111}},\ \bibinfo {pages} {222302} (\bibinfo {year} {2013})},\
  \Eprint {http://arxiv.org/abs/1307.2539} {arXiv:1307.2539 [nucl-th]}
  \BibitemShut {NoStop}%
\bibitem [{\citenamefont {Chesler}\ and\ \citenamefont
  {Yaffe}(2015)}]{Chesler:2015wra}%
  \BibitemOpen
  \bibfield  {author} {\bibinfo {author} {\bibfnamefont {P.~M.}\ \bibnamefont
  {Chesler}}\ and\ \bibinfo {author} {\bibfnamefont {L.~G.}\ \bibnamefont
  {Yaffe}},\ }\href {\doibase 10.1007/JHEP10(2015)070} {\bibfield  {journal}
  {\bibinfo  {journal} {JHEP}\ }\textbf {\bibinfo {volume} {10}},\ \bibinfo
  {pages} {070} (\bibinfo {year} {2015})},\ \Eprint
  {http://arxiv.org/abs/1501.04644} {arXiv:1501.04644 [hep-th]} \BibitemShut
  {NoStop}%
\bibitem [{\citenamefont {Schenke}(2021)}]{Schenke:2021mxx}%
  \BibitemOpen
  \bibfield  {author} {\bibinfo {author} {\bibfnamefont {B.}~\bibnamefont
  {Schenke}},\ }\href {\doibase 10.1088/1361-6633/ac14c9} {\bibfield  {journal}
  {\bibinfo  {journal} {Rept. Prog. Phys.}\ }\textbf {\bibinfo {volume} {84}},\
  \bibinfo {pages} {082301} (\bibinfo {year} {2021})},\ \Eprint
  {http://arxiv.org/abs/2102.11189} {arXiv:2102.11189 [nucl-th]} \BibitemShut
  {NoStop}%
\bibitem [{\citenamefont {Israel}(1976)}]{Israel:1976ur}%
  \BibitemOpen
  \bibfield  {author} {\bibinfo {author} {\bibfnamefont {W.}~\bibnamefont
  {Israel}},\ }\href {\doibase 10.1016/0375-9601(76)90178-X} {\bibfield
  {journal} {\bibinfo  {journal} {Phys. Lett. A}\ }\textbf {\bibinfo {volume}
  {57}},\ \bibinfo {pages} {107} (\bibinfo {year} {1976})}\BibitemShut
  {NoStop}%
\bibitem [{\citenamefont {Maldacena}(2003)}]{Maldacena:2001kr}%
  \BibitemOpen
  \bibfield  {author} {\bibinfo {author} {\bibfnamefont {J.~M.}\ \bibnamefont
  {Maldacena}},\ }\href {\doibase 10.1088/1126-6708/2003/04/021} {\bibfield
  {journal} {\bibinfo  {journal} {JHEP}\ }\textbf {\bibinfo {volume} {04}},\
  \bibinfo {pages} {021} (\bibinfo {year} {2003})},\ \Eprint
  {http://arxiv.org/abs/hep-th/0106112} {arXiv:hep-th/0106112} \BibitemShut
  {NoStop}%
\bibitem [{\citenamefont {Almheiri}\ \emph
  {et~al.}(2020{\natexlab{b}})\citenamefont {Almheiri}, \citenamefont
  {Mahajan}, \citenamefont {Maldacena},\ and\ \citenamefont
  {Zhao}}]{Almheiri:2019hni}%
  \BibitemOpen
  \bibfield  {author} {\bibinfo {author} {\bibfnamefont {A.}~\bibnamefont
  {Almheiri}}, \bibinfo {author} {\bibfnamefont {R.}~\bibnamefont {Mahajan}},
  \bibinfo {author} {\bibfnamefont {J.}~\bibnamefont {Maldacena}}, \ and\
  \bibinfo {author} {\bibfnamefont {Y.}~\bibnamefont {Zhao}},\ }\href {\doibase
  10.1007/JHEP03(2020)149} {\bibfield  {journal} {\bibinfo  {journal} {JHEP}\
  }\textbf {\bibinfo {volume} {03}},\ \bibinfo {pages} {149} (\bibinfo {year}
  {2020}{\natexlab{b}})},\ \Eprint {http://arxiv.org/abs/1908.10996}
  {arXiv:1908.10996 [hep-th]} \BibitemShut {NoStop}%
\bibitem [{\citenamefont {Van~Raamsdonk}(2017)}]{VanRaamsdonk:2016exw}%
  \BibitemOpen
  \bibfield  {author} {\bibinfo {author} {\bibfnamefont {M.}~\bibnamefont
  {Van~Raamsdonk}},\ }in\ \href {\doibase 10.1142/9789813149441_0005} {\emph
  {\bibinfo {booktitle} {{Theoretical Advanced Study Institute in Elementary
  Particle Physics}: {New Frontiers in Fields and Strings}}}}\ (\bibinfo {year}
  {2017})\ pp.\ \bibinfo {pages} {297--351},\ \Eprint
  {http://arxiv.org/abs/1609.00026} {arXiv:1609.00026 [hep-th]} \BibitemShut
  {NoStop}%
\bibitem [{\citenamefont {Van~Raamsdonk}(2021)}]{VanRaamsdonk:2020tlr}%
  \BibitemOpen
  \bibfield  {author} {\bibinfo {author} {\bibfnamefont {M.}~\bibnamefont
  {Van~Raamsdonk}},\ }\href {\doibase 10.1007/JHEP12(2021)156} {\bibfield
  {journal} {\bibinfo  {journal} {JHEP}\ }\textbf {\bibinfo {volume} {12}},\
  \bibinfo {pages} {156} (\bibinfo {year} {2021})},\ \Eprint
  {http://arxiv.org/abs/2008.02259} {arXiv:2008.02259 [hep-th]} \BibitemShut
  {NoStop}%
\bibitem [{\citenamefont {Raamsdonk}\ and\ \citenamefont
  {Waddell}(2021)}]{Raamsdonk:2020tin}%
  \BibitemOpen
  \bibfield  {author} {\bibinfo {author} {\bibfnamefont {M.~V.}\ \bibnamefont
  {Raamsdonk}}\ and\ \bibinfo {author} {\bibfnamefont {C.}~\bibnamefont
  {Waddell}},\ }\href {\doibase 10.1007/JHEP02(2021)222} {\bibfield  {journal}
  {\bibinfo  {journal} {JHEP}\ }\textbf {\bibinfo {volume} {02}},\ \bibinfo
  {pages} {222} (\bibinfo {year} {2021})},\ \Eprint
  {http://arxiv.org/abs/2010.14520} {arXiv:2010.14520 [hep-th]} \BibitemShut
  {NoStop}%
\bibitem [{\citenamefont {May}\ and\ \citenamefont
  {Van~Raamsdonk}(2021)}]{May:2020tch}%
  \BibitemOpen
  \bibfield  {author} {\bibinfo {author} {\bibfnamefont {A.}~\bibnamefont
  {May}}\ and\ \bibinfo {author} {\bibfnamefont {M.}~\bibnamefont
  {Van~Raamsdonk}},\ }\href {\doibase 10.1007/JHEP04(2021)185} {\bibfield
  {journal} {\bibinfo  {journal} {JHEP}\ }\textbf {\bibinfo {volume} {04}},\
  \bibinfo {pages} {185} (\bibinfo {year} {2021})},\ \Eprint
  {http://arxiv.org/abs/2011.14258} {arXiv:2011.14258 [hep-th]} \BibitemShut
  {NoStop}%
\bibitem [{Note4()}]{Note4}%
  \BibitemOpen
  \bibinfo {note} {Our concept inverts van Raamdonk's reasoning. His aim was to
  show that continuous space can emerge from the entanglement of isolated
  domains; we want to describe entanglement among isolated components of the
  final state geometrically.}\BibitemShut {Stop}%
\bibitem [{\citenamefont {Takayanagi}(2011)}]{Takayanagi:2011zk}%
  \BibitemOpen
  \bibfield  {author} {\bibinfo {author} {\bibfnamefont {T.}~\bibnamefont
  {Takayanagi}},\ }\href {\doibase 10.1103/PhysRevLett.107.101602} {\bibfield
  {journal} {\bibinfo  {journal} {Phys. Rev. Lett.}\ }\textbf {\bibinfo
  {volume} {107}},\ \bibinfo {pages} {101602} (\bibinfo {year} {2011})},\
  \Eprint {http://arxiv.org/abs/1105.5165} {arXiv:1105.5165 [hep-th]}
  \BibitemShut {NoStop}%
\bibitem [{\citenamefont {Gale}\ \emph {et~al.}(2021)\citenamefont {Gale},
  \citenamefont {Paquet}, \citenamefont {Schenke},\ and\ \citenamefont
  {Shen}}]{Gale:2020dum}%
  \BibitemOpen
  \bibfield  {author} {\bibinfo {author} {\bibfnamefont {C.}~\bibnamefont
  {Gale}}, \bibinfo {author} {\bibfnamefont {J.-F.}\ \bibnamefont {Paquet}},
  \bibinfo {author} {\bibfnamefont {B.}~\bibnamefont {Schenke}}, \ and\
  \bibinfo {author} {\bibfnamefont {C.}~\bibnamefont {Shen}},\ }\href {\doibase
  10.22323/1.387.0039} {\bibfield  {journal} {\bibinfo  {journal} {PoS}\
  }\textbf {\bibinfo {volume} {HardProbes2020}},\ \bibinfo {pages} {039}
  (\bibinfo {year} {2021})},\ \Eprint {http://arxiv.org/abs/2009.07841}
  {arXiv:2009.07841 [nucl-th]} \BibitemShut {NoStop}%
\bibitem [{Note5()}]{Note5}%
  \BibitemOpen
  \bibinfo {note} {As some regions at the fringe of the nuclear fireball may
  never become hot enough for deconfinement to occur, a small fraction of the
  final state hadrons may be produced directly without going through an
  intervening plasma phase. This phenomenon is described in so-called
  core-corona models \cite {Werner:2007bf}. The relative magnitude of the
  corona contribution to hadron production shrinks with increasing size of the
  collision region and increasing collision energy \cite
  {Petrovici:2017izo}.}\BibitemShut {Stop}%
\bibitem [{\citenamefont {Plumberg}\ and\ \citenamefont
  {Heinz}(2015)}]{Plumberg:2015eia}%
  \BibitemOpen
  \bibfield  {author} {\bibinfo {author} {\bibfnamefont {C.}~\bibnamefont
  {Plumberg}}\ and\ \bibinfo {author} {\bibfnamefont {U.}~\bibnamefont
  {Heinz}},\ }\href {\doibase 10.1103/PhysRevC.91.054905} {\bibfield  {journal}
  {\bibinfo  {journal} {Phys. Rev. C}\ }\textbf {\bibinfo {volume} {91}},\
  \bibinfo {pages} {054905} (\bibinfo {year} {2015})},\ \Eprint
  {http://arxiv.org/abs/1503.05605} {arXiv:1503.05605 [nucl-th]} \BibitemShut
  {NoStop}%
\bibitem [{\citenamefont {Gibbons}\ and\ \citenamefont
  {Hawking}(1977)}]{Gibbons:1976ue}%
  \BibitemOpen
  \bibfield  {author} {\bibinfo {author} {\bibfnamefont {G.~W.}\ \bibnamefont
  {Gibbons}}\ and\ \bibinfo {author} {\bibfnamefont {S.~W.}\ \bibnamefont
  {Hawking}},\ }\href {\doibase 10.1103/PhysRevD.15.2752} {\bibfield  {journal}
  {\bibinfo  {journal} {Phys. Rev. D}\ }\textbf {\bibinfo {volume} {15}},\
  \bibinfo {pages} {2752} (\bibinfo {year} {1977})}\BibitemShut {NoStop}%
\bibitem [{\citenamefont {Aharony}\ \emph {et~al.}(2004)\citenamefont
  {Aharony}, \citenamefont {Marsano}, \citenamefont {Minwalla}, \citenamefont
  {Papadodimas},\ and\ \citenamefont {Van~Raamsdonk}}]{Aharony:2003sx}%
  \BibitemOpen
  \bibfield  {author} {\bibinfo {author} {\bibfnamefont {O.}~\bibnamefont
  {Aharony}}, \bibinfo {author} {\bibfnamefont {J.}~\bibnamefont {Marsano}},
  \bibinfo {author} {\bibfnamefont {S.}~\bibnamefont {Minwalla}}, \bibinfo
  {author} {\bibfnamefont {K.}~\bibnamefont {Papadodimas}}, \ and\ \bibinfo
  {author} {\bibfnamefont {M.}~\bibnamefont {Van~Raamsdonk}},\ }\href {\doibase
  10.4310/ATMP.2004.v8.n4.a1} {\bibfield  {journal} {\bibinfo  {journal} {Adv.
  Theor. Math. Phys.}\ }\textbf {\bibinfo {volume} {8}},\ \bibinfo {pages}
  {603} (\bibinfo {year} {2004})},\ \Eprint
  {http://arxiv.org/abs/hep-th/0310285} {arXiv:hep-th/0310285} \BibitemShut
  {NoStop}%
\bibitem [{\citenamefont {Maldacena}(1998{\natexlab{b}})}]{Maldacena:1998im}%
  \BibitemOpen
  \bibfield  {author} {\bibinfo {author} {\bibfnamefont {J.~M.}\ \bibnamefont
  {Maldacena}},\ }\href {\doibase 10.1103/PhysRevLett.80.4859} {\bibfield
  {journal} {\bibinfo  {journal} {Phys. Rev. Lett.}\ }\textbf {\bibinfo
  {volume} {80}},\ \bibinfo {pages} {4859} (\bibinfo {year}
  {1998}{\natexlab{b}})},\ \Eprint {http://arxiv.org/abs/hep-th/9803002}
  {arXiv:hep-th/9803002} \BibitemShut {NoStop}%
\bibitem [{\citenamefont {Rey}\ \emph {et~al.}(1998)\citenamefont {Rey},
  \citenamefont {Theisen},\ and\ \citenamefont {Yee}}]{Rey:1998bq}%
  \BibitemOpen
  \bibfield  {author} {\bibinfo {author} {\bibfnamefont {S.-J.}\ \bibnamefont
  {Rey}}, \bibinfo {author} {\bibfnamefont {S.}~\bibnamefont {Theisen}}, \ and\
  \bibinfo {author} {\bibfnamefont {J.-T.}\ \bibnamefont {Yee}},\ }\href
  {\doibase 10.1016/S0550-3213(98)00471-4} {\bibfield  {journal} {\bibinfo
  {journal} {Nucl. Phys. B}\ }\textbf {\bibinfo {volume} {527}},\ \bibinfo
  {pages} {171} (\bibinfo {year} {1998})},\ \Eprint
  {http://arxiv.org/abs/hep-th/9803135} {arXiv:hep-th/9803135} \BibitemShut
  {NoStop}%
\bibitem [{\citenamefont {Brandhuber}\ \emph {et~al.}(1998)\citenamefont
  {Brandhuber}, \citenamefont {Itzhaki}, \citenamefont {Sonnenschein},\ and\
  \citenamefont {Yankielowicz}}]{Brandhuber:1998bs}%
  \BibitemOpen
  \bibfield  {author} {\bibinfo {author} {\bibfnamefont {A.}~\bibnamefont
  {Brandhuber}}, \bibinfo {author} {\bibfnamefont {N.}~\bibnamefont {Itzhaki}},
  \bibinfo {author} {\bibfnamefont {J.}~\bibnamefont {Sonnenschein}}, \ and\
  \bibinfo {author} {\bibfnamefont {S.}~\bibnamefont {Yankielowicz}},\ }\href
  {\doibase 10.1016/S0370-2693(98)00730-8} {\bibfield  {journal} {\bibinfo
  {journal} {Phys. Lett. B}\ }\textbf {\bibinfo {volume} {434}},\ \bibinfo
  {pages} {36} (\bibinfo {year} {1998})},\ \Eprint
  {http://arxiv.org/abs/hep-th/9803137} {arXiv:hep-th/9803137} \BibitemShut
  {NoStop}%
\bibitem [{\citenamefont {Liu}\ \emph {et~al.}(2007)\citenamefont {Liu},
  \citenamefont {Rajagopal},\ and\ \citenamefont {Wiedemann}}]{Liu:2006nn}%
  \BibitemOpen
  \bibfield  {author} {\bibinfo {author} {\bibfnamefont {H.}~\bibnamefont
  {Liu}}, \bibinfo {author} {\bibfnamefont {K.}~\bibnamefont {Rajagopal}}, \
  and\ \bibinfo {author} {\bibfnamefont {U.~A.}\ \bibnamefont {Wiedemann}},\
  }\href {\doibase 10.1103/PhysRevLett.98.182301} {\bibfield  {journal}
  {\bibinfo  {journal} {Phys. Rev. Lett.}\ }\textbf {\bibinfo {volume} {98}},\
  \bibinfo {pages} {182301} (\bibinfo {year} {2007})},\ \Eprint
  {http://arxiv.org/abs/hep-ph/0607062} {arXiv:hep-ph/0607062} \BibitemShut
  {NoStop}%
\bibitem [{\citenamefont {Eune}\ \emph {et~al.}(2013)\citenamefont {Eune},
  \citenamefont {Kim},\ and\ \citenamefont {Yi}}]{Eune:2013qs}%
  \BibitemOpen
  \bibfield  {author} {\bibinfo {author} {\bibfnamefont {M.}~\bibnamefont
  {Eune}}, \bibinfo {author} {\bibfnamefont {W.}~\bibnamefont {Kim}}, \ and\
  \bibinfo {author} {\bibfnamefont {S.-H.}\ \bibnamefont {Yi}},\ }\href
  {\doibase 10.1007/JHEP03(2013)020} {\bibfield  {journal} {\bibinfo  {journal}
  {JHEP}\ }\textbf {\bibinfo {volume} {03}},\ \bibinfo {pages} {020} (\bibinfo
  {year} {2013})},\ \Eprint {http://arxiv.org/abs/1301.0395} {arXiv:1301.0395
  [gr-qc]} \BibitemShut {NoStop}%
\bibitem [{\citenamefont {Panero}(2009)}]{Panero:2009tv}%
  \BibitemOpen
  \bibfield  {author} {\bibinfo {author} {\bibfnamefont {M.}~\bibnamefont
  {Panero}},\ }\href {\doibase 10.1103/PhysRevLett.103.232001} {\bibfield
  {journal} {\bibinfo  {journal} {Phys. Rev. Lett.}\ }\textbf {\bibinfo
  {volume} {103}},\ \bibinfo {pages} {232001} (\bibinfo {year} {2009})},\
  \Eprint {http://arxiv.org/abs/0907.3719} {arXiv:0907.3719 [hep-lat]}
  \BibitemShut {NoStop}%
\bibitem [{\citenamefont {Attems}\ \emph {et~al.}(2018)\citenamefont {Attems},
  \citenamefont {Bea}, \citenamefont {Casalderrey-Solana}, \citenamefont
  {Mateos}, \citenamefont {Triana},\ and\ \citenamefont
  {Zilh\~ao}}]{Attems:2018gou}%
  \BibitemOpen
  \bibfield  {author} {\bibinfo {author} {\bibfnamefont {M.}~\bibnamefont
  {Attems}}, \bibinfo {author} {\bibfnamefont {Y.}~\bibnamefont {Bea}},
  \bibinfo {author} {\bibfnamefont {J.}~\bibnamefont {Casalderrey-Solana}},
  \bibinfo {author} {\bibfnamefont {D.}~\bibnamefont {Mateos}}, \bibinfo
  {author} {\bibfnamefont {M.}~\bibnamefont {Triana}}, \ and\ \bibinfo {author}
  {\bibfnamefont {M.}~\bibnamefont {Zilh\~ao}},\ }\href {\doibase
  10.1103/PhysRevLett.121.261601} {\bibfield  {journal} {\bibinfo  {journal}
  {Phys. Rev. Lett.}\ }\textbf {\bibinfo {volume} {121}},\ \bibinfo {pages}
  {261601} (\bibinfo {year} {2018})},\ \Eprint
  {http://arxiv.org/abs/1807.05175} {arXiv:1807.05175 [hep-th]} \BibitemShut
  {NoStop}%
\bibitem [{\citenamefont {Janik}\ \emph {et~al.}(2021)\citenamefont {Janik},
  \citenamefont {J{\"a}rvinen},\ and\ \citenamefont
  {Sonnenschein}}]{Janik:2021jbq}%
  \BibitemOpen
  \bibfield  {author} {\bibinfo {author} {\bibfnamefont {R.~A.}\ \bibnamefont
  {Janik}}, \bibinfo {author} {\bibfnamefont {M.}~\bibnamefont {J{\"a}rvinen}},
  \ and\ \bibinfo {author} {\bibfnamefont {J.}~\bibnamefont {Sonnenschein}},\
  }\href {\doibase 10.1007/JHEP09(2021)129} {\bibfield  {journal} {\bibinfo
  {journal} {JHEP}\ }\textbf {\bibinfo {volume} {09}},\ \bibinfo {pages} {129}
  (\bibinfo {year} {2021})},\ \Eprint {http://arxiv.org/abs/2106.02642}
  {arXiv:2106.02642 [hep-th]} \BibitemShut {NoStop}%
\bibitem [{\citenamefont {Gregory}\ and\ \citenamefont
  {Laflamme}(1993)}]{Gregory:1993vy}%
  \BibitemOpen
  \bibfield  {author} {\bibinfo {author} {\bibfnamefont {R.}~\bibnamefont
  {Gregory}}\ and\ \bibinfo {author} {\bibfnamefont {R.}~\bibnamefont
  {Laflamme}},\ }\href {\doibase 10.1103/PhysRevLett.70.2837} {\bibfield
  {journal} {\bibinfo  {journal} {Phys. Rev. Lett.}\ }\textbf {\bibinfo
  {volume} {70}},\ \bibinfo {pages} {2837} (\bibinfo {year} {1993})},\ \Eprint
  {http://arxiv.org/abs/hep-th/9301052} {arXiv:hep-th/9301052} \BibitemShut
  {NoStop}%
\bibitem [{\citenamefont {Hubeny}\ and\ \citenamefont
  {Rangamani}(2002)}]{Hubeny:2002xn}%
  \BibitemOpen
  \bibfield  {author} {\bibinfo {author} {\bibfnamefont {V.~E.}\ \bibnamefont
  {Hubeny}}\ and\ \bibinfo {author} {\bibfnamefont {M.}~\bibnamefont
  {Rangamani}},\ }\href {\doibase 10.1088/1126-6708/2002/05/027} {\bibfield
  {journal} {\bibinfo  {journal} {JHEP}\ }\textbf {\bibinfo {volume} {05}},\
  \bibinfo {pages} {027} (\bibinfo {year} {2002})},\ \Eprint
  {http://arxiv.org/abs/hep-th/0202189} {arXiv:hep-th/0202189} \BibitemShut
  {NoStop}%
\bibitem [{\citenamefont {Buchel}\ and\ \citenamefont
  {Lehner}(2015)}]{Buchel:2015gxa}%
  \BibitemOpen
  \bibfield  {author} {\bibinfo {author} {\bibfnamefont {A.}~\bibnamefont
  {Buchel}}\ and\ \bibinfo {author} {\bibfnamefont {L.}~\bibnamefont
  {Lehner}},\ }\href {\doibase 10.1088/0264-9381/32/14/145003} {\bibfield
  {journal} {\bibinfo  {journal} {Class. Quant. Grav.}\ }\textbf {\bibinfo
  {volume} {32}},\ \bibinfo {pages} {145003} (\bibinfo {year} {2015})},\
  \Eprint {http://arxiv.org/abs/1502.01574} {arXiv:1502.01574 [hep-th]}
  \BibitemShut {NoStop}%
\bibitem [{\citenamefont {Dias}\ \emph {et~al.}(2016)\citenamefont {Dias},
  \citenamefont {Santos},\ and\ \citenamefont {Way}}]{Dias:2016eto}%
  \BibitemOpen
  \bibfield  {author} {\bibinfo {author} {\bibfnamefont {O.~J.~C.}\
  \bibnamefont {Dias}}, \bibinfo {author} {\bibfnamefont {J.~E.}\ \bibnamefont
  {Santos}}, \ and\ \bibinfo {author} {\bibfnamefont {B.}~\bibnamefont {Way}},\
  }\href {\doibase 10.1103/PhysRevLett.117.151101} {\bibfield  {journal}
  {\bibinfo  {journal} {Phys. Rev. Lett.}\ }\textbf {\bibinfo {volume} {117}},\
  \bibinfo {pages} {151101} (\bibinfo {year} {2016})},\ \Eprint
  {http://arxiv.org/abs/1605.04911} {arXiv:1605.04911 [hep-th]} \BibitemShut
  {NoStop}%
\bibitem [{\citenamefont {Yaffe}(2018)}]{Yaffe:2017axl}%
  \BibitemOpen
  \bibfield  {author} {\bibinfo {author} {\bibfnamefont {L.~G.}\ \bibnamefont
  {Yaffe}},\ }\href {\doibase 10.1103/PhysRevD.97.026010} {\bibfield  {journal}
  {\bibinfo  {journal} {Phys. Rev. D}\ }\textbf {\bibinfo {volume} {97}},\
  \bibinfo {pages} {026010} (\bibinfo {year} {2018})},\ \Eprint
  {http://arxiv.org/abs/1710.06455} {arXiv:1710.06455 [hep-th]} \BibitemShut
  {NoStop}%
\bibitem [{\citenamefont {Fries}\ \emph {et~al.}(2008)\citenamefont {Fries},
  \citenamefont {Greco},\ and\ \citenamefont {Sorensen}}]{Fries:2008hs}%
  \BibitemOpen
  \bibfield  {author} {\bibinfo {author} {\bibfnamefont {R.~J.}\ \bibnamefont
  {Fries}}, \bibinfo {author} {\bibfnamefont {V.}~\bibnamefont {Greco}}, \ and\
  \bibinfo {author} {\bibfnamefont {P.}~\bibnamefont {Sorensen}},\ }\href
  {\doibase 10.1146/annurev.nucl.58.110707.171134} {\bibfield  {journal}
  {\bibinfo  {journal} {Ann. Rev. Nucl. Part. Sci.}\ }\textbf {\bibinfo
  {volume} {58}},\ \bibinfo {pages} {177} (\bibinfo {year} {2008})},\ \Eprint
  {http://arxiv.org/abs/0807.4939} {arXiv:0807.4939 [nucl-th]} \BibitemShut
  {NoStop}%
\bibitem [{\citenamefont {M{\"u}ller}\ \emph {et~al.}(2005)\citenamefont
  {M{\"u}ller}, \citenamefont {Fries},\ and\ \citenamefont
  {Bass}}]{Muller:2005pv}%
  \BibitemOpen
  \bibfield  {author} {\bibinfo {author} {\bibfnamefont {B.}~\bibnamefont
  {M{\"u}ller}}, \bibinfo {author} {\bibfnamefont {R.~J.}\ \bibnamefont
  {Fries}}, \ and\ \bibinfo {author} {\bibfnamefont {S.~A.}\ \bibnamefont
  {Bass}},\ }\href {\doibase 10.1016/j.physletb.2005.05.025} {\bibfield
  {journal} {\bibinfo  {journal} {Phys. Lett. B}\ }\textbf {\bibinfo {volume}
  {618}},\ \bibinfo {pages} {77} (\bibinfo {year} {2005})},\ \Eprint
  {http://arxiv.org/abs/nucl-th/0503003} {arXiv:nucl-th/0503003} \BibitemShut
  {NoStop}%
\bibitem [{Note6()}]{Note6}%
  \BibitemOpen
  \bibinfo {note} {The ground state of a colliding nucleus is unique, and any
  interaction of the nucleus with the accelerator structure is completely
  negligible on the scale of nuclear excitations. Finally, although the two
  nuclei can be Coulomb excited on their approach to each other, the excitation
  is coherent and does not change the fact that the nuclear quantum state is
  pure.}\BibitemShut {Stop}%
\bibitem [{Note7()}]{Note7}%
  \BibitemOpen
  \bibinfo {note} {We remind the reader that use the term ``AdS edge'' to avoid
  confusion of the asymptotic region of AdS space with the boundary of the QCD
  fireball.}\BibitemShut {Stop}%
\bibitem [{\citenamefont {Wootters}\ and\ \citenamefont
  {Zurek}(1982)}]{Wootters:1982zz}%
  \BibitemOpen
  \bibfield  {author} {\bibinfo {author} {\bibfnamefont {W.~K.}\ \bibnamefont
  {Wootters}}\ and\ \bibinfo {author} {\bibfnamefont {W.~H.}\ \bibnamefont
  {Zurek}},\ }\href {\doibase 10.1038/299802a0} {\bibfield  {journal} {\bibinfo
   {journal} {Nature}\ }\textbf {\bibinfo {volume} {299}},\ \bibinfo {pages}
  {802} (\bibinfo {year} {1982})}\BibitemShut {NoStop}%
\bibitem [{\citenamefont {Wootters}(1998)}]{Wootters:1997id}%
  \BibitemOpen
  \bibfield  {author} {\bibinfo {author} {\bibfnamefont {W.~K.}\ \bibnamefont
  {Wootters}},\ }\href {\doibase 10.1103/PhysRevLett.80.2245} {\bibfield
  {journal} {\bibinfo  {journal} {Phys. Rev. Lett.}\ }\textbf {\bibinfo
  {volume} {80}},\ \bibinfo {pages} {2245} (\bibinfo {year} {1998})},\ \Eprint
  {http://arxiv.org/abs/quant-ph/9709029} {arXiv:quant-ph/9709029} \BibitemShut
  {NoStop}%
\bibitem [{\citenamefont {Coffman}\ \emph {et~al.}(2000)\citenamefont
  {Coffman}, \citenamefont {Kundu},\ and\ \citenamefont
  {Wootters}}]{Coffman:1999jd}%
  \BibitemOpen
  \bibfield  {author} {\bibinfo {author} {\bibfnamefont {V.}~\bibnamefont
  {Coffman}}, \bibinfo {author} {\bibfnamefont {J.}~\bibnamefont {Kundu}}, \
  and\ \bibinfo {author} {\bibfnamefont {W.~K.}\ \bibnamefont {Wootters}},\
  }\href {\doibase 10.1103/PhysRevA.61.052306} {\bibfield  {journal} {\bibinfo
  {journal} {Phys. Rev. A}\ }\textbf {\bibinfo {volume} {61}},\ \bibinfo
  {pages} {052306} (\bibinfo {year} {2000})},\ \Eprint
  {http://arxiv.org/abs/quant-ph/9907047} {arXiv:quant-ph/9907047} \BibitemShut
  {NoStop}%
\bibitem [{\citenamefont {Osborne}\ and\ \citenamefont
  {Verstraete}(2006)}]{Osborne2006xx}%
  \BibitemOpen
  \bibfield  {author} {\bibinfo {author} {\bibfnamefont {T.~J.}\ \bibnamefont
  {Osborne}}\ and\ \bibinfo {author} {\bibfnamefont {F.}~\bibnamefont
  {Verstraete}},\ }\href@noop {} {\bibfield  {journal} {\bibinfo  {journal}
  {Physical Review Letters}\ }\textbf {\bibinfo {volume} {96}},\ \bibinfo
  {pages} {220503} (\bibinfo {year} {2006})}\BibitemShut {NoStop}%
\bibitem [{\citenamefont {Yang}\ and\ \citenamefont
  {Fries}(2022)}]{Yang:2022yxa}%
  \BibitemOpen
  \bibfield  {author} {\bibinfo {author} {\bibfnamefont {Z.}~\bibnamefont
  {Yang}}\ and\ \bibinfo {author} {\bibfnamefont {R.~J.}\ \bibnamefont
  {Fries}},\ }\href {\doibase 10.1103/PhysRevC.105.014910} {\bibfield
  {journal} {\bibinfo  {journal} {Phys. Rev. C}\ }\textbf {\bibinfo {volume}
  {105}},\ \bibinfo {pages} {014910} (\bibinfo {year} {2022})}\BibitemShut
  {NoStop}%
\bibitem [{\citenamefont {Schuckert}\ and\ \citenamefont
  {Knap}(2020)}]{Schuckert:2020qeo}%
  \BibitemOpen
  \bibfield  {author} {\bibinfo {author} {\bibfnamefont {A.}~\bibnamefont
  {Schuckert}}\ and\ \bibinfo {author} {\bibfnamefont {M.}~\bibnamefont
  {Knap}},\ }\href {\doibase 10.1103/PhysRevResearch.2.043315} {\bibfield
  {journal} {\bibinfo  {journal} {Phys. Rev. Research}\ }\textbf {\bibinfo
  {volume} {2}},\ \bibinfo {pages} {043315} (\bibinfo {year}
  {2020})}\BibitemShut {NoStop}%
\bibitem [{\citenamefont {Berbenni-Bitsch}\ \emph {et~al.}(1998)\citenamefont
  {Berbenni-Bitsch}, \citenamefont {Meyer}, \citenamefont {Sch{\"a}fer},
  \citenamefont {Verbaarschot},\ and\ \citenamefont
  {Wettig}}]{Berbenni-Bitsch:1997zmi}%
  \BibitemOpen
  \bibfield  {author} {\bibinfo {author} {\bibfnamefont {M.~E.}\ \bibnamefont
  {Berbenni-Bitsch}}, \bibinfo {author} {\bibfnamefont {S.}~\bibnamefont
  {Meyer}}, \bibinfo {author} {\bibfnamefont {A.}~\bibnamefont {Sch{\"a}fer}},
  \bibinfo {author} {\bibfnamefont {J.~J.~M.}\ \bibnamefont {Verbaarschot}}, \
  and\ \bibinfo {author} {\bibfnamefont {T.}~\bibnamefont {Wettig}},\ }\href
  {\doibase 10.1103/PhysRevLett.80.1146} {\bibfield  {journal} {\bibinfo
  {journal} {Phys. Rev. Lett.}\ }\textbf {\bibinfo {volume} {80}},\ \bibinfo
  {pages} {1146} (\bibinfo {year} {1998})},\ \Eprint
  {http://arxiv.org/abs/hep-lat/9704018} {arXiv:hep-lat/9704018} \BibitemShut
  {NoStop}%
\bibitem [{\citenamefont {Verbaarschot}\ and\ \citenamefont
  {Wettig}(2000)}]{Verbaarschot:2000dy}%
  \BibitemOpen
  \bibfield  {author} {\bibinfo {author} {\bibfnamefont {J.~J.~M.}\
  \bibnamefont {Verbaarschot}}\ and\ \bibinfo {author} {\bibfnamefont
  {T.}~\bibnamefont {Wettig}},\ }\href {\doibase 10.1146/annurev.nucl.50.1.343}
  {\bibfield  {journal} {\bibinfo  {journal} {Ann. Rev. Nucl. Part. Sci.}\
  }\textbf {\bibinfo {volume} {50}},\ \bibinfo {pages} {343} (\bibinfo {year}
  {2000})},\ \Eprint {http://arxiv.org/abs/hep-ph/0003017}
  {arXiv:hep-ph/0003017} \BibitemShut {NoStop}%
\bibitem [{\citenamefont {Harrow}\ and\ \citenamefont
  {Huang}(2022)}]{Harrow:2022znr}%
  \BibitemOpen
  \bibfield  {author} {\bibinfo {author} {\bibfnamefont {A.~W.}\ \bibnamefont
  {Harrow}}\ and\ \bibinfo {author} {\bibfnamefont {Y.}~\bibnamefont {Huang}},\
  }\href@noop {} {\  (\bibinfo {year} {2022})},\ \Eprint
  {http://arxiv.org/abs/2209.09826} {arXiv:2209.09826 [cond-mat.stat-mech]}
  \BibitemShut {NoStop}%
\bibitem [{\citenamefont {Majidy}\ \emph {et~al.}(2022)\citenamefont {Majidy},
  \citenamefont {Lasek}, \citenamefont {Huse},\ and\ \citenamefont
  {Halpern}}]{Majidy:2022kzx}%
  \BibitemOpen
  \bibfield  {author} {\bibinfo {author} {\bibfnamefont {S.}~\bibnamefont
  {Majidy}}, \bibinfo {author} {\bibfnamefont {A.}~\bibnamefont {Lasek}},
  \bibinfo {author} {\bibfnamefont {D.~A.}\ \bibnamefont {Huse}}, \ and\
  \bibinfo {author} {\bibfnamefont {N.~Y.}\ \bibnamefont {Halpern}},\
  }\href@noop {} {\  (\bibinfo {year} {2022})},\ \Eprint
  {http://arxiv.org/abs/2209.14303} {arXiv:2209.14303 [quant-ph]} \BibitemShut
  {NoStop}%
\bibitem [{\citenamefont {Turner}\ \emph {et~al.}(2018)\citenamefont {Turner},
  \citenamefont {Michailidis}, \citenamefont {Abanin}, \citenamefont {Serbyn},\
  and\ \citenamefont {Papi{\'c}}}]{Turner2018:weak}%
  \BibitemOpen
  \bibfield  {author} {\bibinfo {author} {\bibfnamefont {C.~J.}\ \bibnamefont
  {Turner}}, \bibinfo {author} {\bibfnamefont {A.~A.}\ \bibnamefont
  {Michailidis}}, \bibinfo {author} {\bibfnamefont {D.~A.}\ \bibnamefont
  {Abanin}}, \bibinfo {author} {\bibfnamefont {M.}~\bibnamefont {Serbyn}}, \
  and\ \bibinfo {author} {\bibfnamefont {Z.}~\bibnamefont {Papi{\'c}}},\
  }\href@noop {} {\bibfield  {journal} {\bibinfo  {journal} {Nature Physics}\
  }\textbf {\bibinfo {volume} {14}},\ \bibinfo {pages} {745} (\bibinfo {year}
  {2018})}\BibitemShut {NoStop}%
\bibitem [{\citenamefont {Michailidis}\ \emph {et~al.}(2020)\citenamefont
  {Michailidis}, \citenamefont {Turner}, \citenamefont {Papi\'c}, \citenamefont
  {Abanin},\ and\ \citenamefont {Serbyn}}]{Michailidis:2019tgg}%
  \BibitemOpen
  \bibfield  {author} {\bibinfo {author} {\bibfnamefont {A.~A.}\ \bibnamefont
  {Michailidis}}, \bibinfo {author} {\bibfnamefont {C.~J.}\ \bibnamefont
  {Turner}}, \bibinfo {author} {\bibfnamefont {Z.}~\bibnamefont {Papi\'c}},
  \bibinfo {author} {\bibfnamefont {D.~A.}\ \bibnamefont {Abanin}}, \ and\
  \bibinfo {author} {\bibfnamefont {M.}~\bibnamefont {Serbyn}},\ }\href
  {\doibase 10.1103/PhysRevX.10.011055} {\bibfield  {journal} {\bibinfo
  {journal} {Phys. Rev. X}\ }\textbf {\bibinfo {volume} {10}},\ \bibinfo
  {pages} {011055} (\bibinfo {year} {2020})},\ \Eprint
  {http://arxiv.org/abs/1905.08564} {arXiv:1905.08564 [quant-ph]} \BibitemShut
  {NoStop}%
\bibitem [{\citenamefont {Hanada}\ and\ \citenamefont
  {Watanabe}(2022)}]{Hanada:2022wcq}%
  \BibitemOpen
  \bibfield  {author} {\bibinfo {author} {\bibfnamefont {M.}~\bibnamefont
  {Hanada}}\ and\ \bibinfo {author} {\bibfnamefont {H.}~\bibnamefont
  {Watanabe}},\ }\href@noop {} {\  (\bibinfo {year} {2022})},\ \Eprint
  {http://arxiv.org/abs/2210.11216} {arXiv:2210.11216 [hep-th]} \BibitemShut
  {NoStop}%
\bibitem [{\citenamefont {Pateloudis}\ \emph {et~al.}(2022)\citenamefont
  {Pateloudis}, \citenamefont {Bergner}, \citenamefont {Hanada}, \citenamefont
  {Rinaldi}, \citenamefont {Sch\"afer}, \citenamefont {Vranas}, \citenamefont
  {Watanabe},\ and\ \citenamefont {Bodendorfer}}]{Pateloudis:2022ijr}%
  \BibitemOpen
  \bibfield  {author} {\bibinfo {author} {\bibfnamefont {S.}~\bibnamefont
  {Pateloudis}}, \bibinfo {author} {\bibfnamefont {G.}~\bibnamefont {Bergner}},
  \bibinfo {author} {\bibfnamefont {M.}~\bibnamefont {Hanada}}, \bibinfo
  {author} {\bibfnamefont {E.}~\bibnamefont {Rinaldi}}, \bibinfo {author}
  {\bibfnamefont {A.}~\bibnamefont {Sch\"afer}}, \bibinfo {author}
  {\bibfnamefont {P.}~\bibnamefont {Vranas}}, \bibinfo {author} {\bibfnamefont
  {H.}~\bibnamefont {Watanabe}}, \ and\ \bibinfo {author} {\bibfnamefont
  {N.}~\bibnamefont {Bodendorfer}},\ }\href@noop {} {\  (\bibinfo {year}
  {2022})},\ \Eprint {http://arxiv.org/abs/2210.04881} {arXiv:2210.04881
  [hep-th]} \BibitemShut {NoStop}%
\bibitem [{\citenamefont {Buividovich}(2022)}]{Buividovich:2022udc}%
  \BibitemOpen
  \bibfield  {author} {\bibinfo {author} {\bibfnamefont {P.}~\bibnamefont
  {Buividovich}},\ }in\ \href@noop {} {\emph {\bibinfo {booktitle} {{39th
  International Symposium on Lattice Field Theory}}}}\ (\bibinfo {year}
  {2022})\ \Eprint {http://arxiv.org/abs/2210.05288} {arXiv:2210.05288
  [hep-lat]} \BibitemShut {NoStop}%
\bibitem [{\citenamefont {Buividovich}\ \emph {et~al.}(2019)\citenamefont
  {Buividovich}, \citenamefont {Hanada},\ and\ \citenamefont
  {Sch\"afer}}]{Buividovich:2018scl}%
  \BibitemOpen
  \bibfield  {author} {\bibinfo {author} {\bibfnamefont {P.~V.}\ \bibnamefont
  {Buividovich}}, \bibinfo {author} {\bibfnamefont {M.}~\bibnamefont {Hanada}},
  \ and\ \bibinfo {author} {\bibfnamefont {A.}~\bibnamefont {Sch\"afer}},\
  }\href {\doibase 10.1103/PhysRevD.99.046011} {\bibfield  {journal} {\bibinfo
  {journal} {Phys. Rev. D}\ }\textbf {\bibinfo {volume} {99}},\ \bibinfo
  {pages} {046011} (\bibinfo {year} {2019})},\ \Eprint
  {http://arxiv.org/abs/1810.03378} {arXiv:1810.03378 [hep-th]} \BibitemShut
  {NoStop}%
\bibitem [{\citenamefont {G{\"u}rsoy}\ \emph
  {et~al.}(2009{\natexlab{b}})\citenamefont {G{\"u}rsoy}, \citenamefont
  {Kiritsis}, \citenamefont {Mazzanti},\ and\ \citenamefont
  {Nitti}}]{Gursoy:2009jd}%
  \BibitemOpen
  \bibfield  {author} {\bibinfo {author} {\bibfnamefont {U.}~\bibnamefont
  {G{\"u}rsoy}}, \bibinfo {author} {\bibfnamefont {E.}~\bibnamefont
  {Kiritsis}}, \bibinfo {author} {\bibfnamefont {L.}~\bibnamefont {Mazzanti}},
  \ and\ \bibinfo {author} {\bibfnamefont {F.}~\bibnamefont {Nitti}},\ }\href
  {\doibase 10.1016/j.nuclphysb.2009.05.017} {\bibfield  {journal} {\bibinfo
  {journal} {Nucl. Phys. B}\ }\textbf {\bibinfo {volume} {820}},\ \bibinfo
  {pages} {148} (\bibinfo {year} {2009}{\natexlab{b}})},\ \Eprint
  {http://arxiv.org/abs/0903.2859} {arXiv:0903.2859 [hep-th]} \BibitemShut
  {NoStop}%
\bibitem [{\citenamefont {Megias}\ \emph {et~al.}(2011)\citenamefont {Megias},
  \citenamefont {Pirner},\ and\ \citenamefont {Veschgini}}]{Megias:2010ku}%
  \BibitemOpen
  \bibfield  {author} {\bibinfo {author} {\bibfnamefont {E.}~\bibnamefont
  {Megias}}, \bibinfo {author} {\bibfnamefont {H.~J.}\ \bibnamefont {Pirner}},
  \ and\ \bibinfo {author} {\bibfnamefont {K.}~\bibnamefont {Veschgini}},\
  }\href {\doibase 10.1103/PhysRevD.83.056003} {\bibfield  {journal} {\bibinfo
  {journal} {Phys. Rev. D}\ }\textbf {\bibinfo {volume} {83}},\ \bibinfo
  {pages} {056003} (\bibinfo {year} {2011})},\ \Eprint
  {http://arxiv.org/abs/1009.2953} {arXiv:1009.2953 [hep-ph]} \BibitemShut
  {NoStop}%
\bibitem [{\citenamefont {G{\"u}rsoy}\ \emph {et~al.}(2013)\citenamefont
  {G{\"u}rsoy}, \citenamefont {Lin},\ and\ \citenamefont
  {Shuryak}}]{Gursoy:2013zxa}%
  \BibitemOpen
  \bibfield  {author} {\bibinfo {author} {\bibfnamefont {U.}~\bibnamefont
  {G{\"u}rsoy}}, \bibinfo {author} {\bibfnamefont {S.}~\bibnamefont {Lin}}, \
  and\ \bibinfo {author} {\bibfnamefont {E.}~\bibnamefont {Shuryak}},\ }\href
  {\doibase 10.1103/PhysRevD.88.105021} {\bibfield  {journal} {\bibinfo
  {journal} {Phys. Rev. D}\ }\textbf {\bibinfo {volume} {88}},\ \bibinfo
  {pages} {105021} (\bibinfo {year} {2013})},\ \Eprint
  {http://arxiv.org/abs/1309.0789} {arXiv:1309.0789 [hep-th]} \BibitemShut
  {NoStop}%
\bibitem [{\citenamefont {Attems}\ \emph {et~al.}(2020)\citenamefont {Attems},
  \citenamefont {Bea}, \citenamefont {Casalderrey-Solana}, \citenamefont
  {Mateos},\ and\ \citenamefont {Zilh\~ao}}]{Attems:2019yqn}%
  \BibitemOpen
  \bibfield  {author} {\bibinfo {author} {\bibfnamefont {M.}~\bibnamefont
  {Attems}}, \bibinfo {author} {\bibfnamefont {Y.}~\bibnamefont {Bea}},
  \bibinfo {author} {\bibfnamefont {J.}~\bibnamefont {Casalderrey-Solana}},
  \bibinfo {author} {\bibfnamefont {D.}~\bibnamefont {Mateos}}, \ and\ \bibinfo
  {author} {\bibfnamefont {M.}~\bibnamefont {Zilh\~ao}},\ }\href {\doibase
  10.1007/JHEP01(2020)106} {\bibfield  {journal} {\bibinfo  {journal} {JHEP}\
  }\textbf {\bibinfo {volume} {01}},\ \bibinfo {pages} {106} (\bibinfo {year}
  {2020})},\ \Eprint {http://arxiv.org/abs/1905.12544} {arXiv:1905.12544
  [hep-th]} \BibitemShut {NoStop}%
\bibitem [{\citenamefont {Ecker}\ \emph {et~al.}(2021)\citenamefont {Ecker},
  \citenamefont {Erdmenger},\ and\ \citenamefont {van~der
  Schee}}]{Ecker:2021ukv}%
  \BibitemOpen
  \bibfield  {author} {\bibinfo {author} {\bibfnamefont {C.}~\bibnamefont
  {Ecker}}, \bibinfo {author} {\bibfnamefont {J.}~\bibnamefont {Erdmenger}}, \
  and\ \bibinfo {author} {\bibfnamefont {W.}~\bibnamefont {van~der Schee}},\
  }\href {\doibase 10.21468/SciPostPhys.11.3.047} {\bibfield  {journal}
  {\bibinfo  {journal} {SciPost Phys.}\ }\textbf {\bibinfo {volume} {11}},\
  \bibinfo {pages} {047} (\bibinfo {year} {2021})},\ \Eprint
  {http://arxiv.org/abs/2103.10435} {arXiv:2103.10435 [hep-th]} \BibitemShut
  {NoStop}%
\bibitem [{\citenamefont {Attems}(2021)}]{Attems:2020qkg}%
  \BibitemOpen
  \bibfield  {author} {\bibinfo {author} {\bibfnamefont {M.}~\bibnamefont
  {Attems}},\ }\href {\doibase 10.1007/JHEP08(2021)155} {\bibfield  {journal}
  {\bibinfo  {journal} {JHEP}\ }\textbf {\bibinfo {volume} {08}},\ \bibinfo
  {pages} {155} (\bibinfo {year} {2021})},\ \Eprint
  {http://arxiv.org/abs/2012.15687} {arXiv:2012.15687 [hep-th]} \BibitemShut
  {NoStop}%
\bibitem [{\citenamefont {Shimaji}\ \emph {et~al.}(2019)\citenamefont
  {Shimaji}, \citenamefont {Takayanagi},\ and\ \citenamefont
  {Wei}}]{Shimaji:2018czt}%
  \BibitemOpen
  \bibfield  {author} {\bibinfo {author} {\bibfnamefont {T.}~\bibnamefont
  {Shimaji}}, \bibinfo {author} {\bibfnamefont {T.}~\bibnamefont {Takayanagi}},
  \ and\ \bibinfo {author} {\bibfnamefont {Z.}~\bibnamefont {Wei}},\ }\href
  {\doibase 10.1007/JHEP03(2019)165} {\bibfield  {journal} {\bibinfo  {journal}
  {JHEP}\ }\textbf {\bibinfo {volume} {03}},\ \bibinfo {pages} {165} (\bibinfo
  {year} {2019})},\ \Eprint {http://arxiv.org/abs/1812.01176} {arXiv:1812.01176
  [hep-th]} \BibitemShut {NoStop}%
\bibitem [{\citenamefont {Caputa}\ \emph {et~al.}(2019)\citenamefont {Caputa},
  \citenamefont {Numasawa}, \citenamefont {Shimaji}, \citenamefont
  {Takayanagi},\ and\ \citenamefont {Wei}}]{Caputa:2019avh}%
  \BibitemOpen
  \bibfield  {author} {\bibinfo {author} {\bibfnamefont {P.}~\bibnamefont
  {Caputa}}, \bibinfo {author} {\bibfnamefont {T.}~\bibnamefont {Numasawa}},
  \bibinfo {author} {\bibfnamefont {T.}~\bibnamefont {Shimaji}}, \bibinfo
  {author} {\bibfnamefont {T.}~\bibnamefont {Takayanagi}}, \ and\ \bibinfo
  {author} {\bibfnamefont {Z.}~\bibnamefont {Wei}},\ }\href {\doibase
  10.1007/JHEP09(2019)018} {\bibfield  {journal} {\bibinfo  {journal} {JHEP}\
  }\textbf {\bibinfo {volume} {09}},\ \bibinfo {pages} {018} (\bibinfo {year}
  {2019})},\ \Eprint {http://arxiv.org/abs/1905.08265} {arXiv:1905.08265
  [hep-th]} \BibitemShut {NoStop}%
\bibitem [{\citenamefont {Bellwied}\ \emph {et~al.}(2021)\citenamefont
  {Bellwied}, \citenamefont {Borsanyi}, \citenamefont {Fodor}, \citenamefont
  {Guenther}, \citenamefont {Katz}, \citenamefont {Parotto}, \citenamefont
  {Pasztor}, \citenamefont {Pesznyak}, \citenamefont {Ratti},\ and\
  \citenamefont {Szabo}}]{Bellwied:2021nrt}%
  \BibitemOpen
  \bibfield  {author} {\bibinfo {author} {\bibfnamefont {R.}~\bibnamefont
  {Bellwied}}, \bibinfo {author} {\bibfnamefont {S.}~\bibnamefont {Borsanyi}},
  \bibinfo {author} {\bibfnamefont {Z.}~\bibnamefont {Fodor}}, \bibinfo
  {author} {\bibfnamefont {J.~N.}\ \bibnamefont {Guenther}}, \bibinfo {author}
  {\bibfnamefont {S.~D.}\ \bibnamefont {Katz}}, \bibinfo {author}
  {\bibfnamefont {P.}~\bibnamefont {Parotto}}, \bibinfo {author} {\bibfnamefont
  {A.}~\bibnamefont {Pasztor}}, \bibinfo {author} {\bibfnamefont
  {D.}~\bibnamefont {Pesznyak}}, \bibinfo {author} {\bibfnamefont
  {C.}~\bibnamefont {Ratti}}, \ and\ \bibinfo {author} {\bibfnamefont {K.~K.}\
  \bibnamefont {Szabo}},\ }\href {\doibase 10.1103/PhysRevD.104.094508}
  {\bibfield  {journal} {\bibinfo  {journal} {Phys. Rev. D}\ }\textbf {\bibinfo
  {volume} {104}},\ \bibinfo {pages} {094508} (\bibinfo {year} {2021})},\
  \Eprint {http://arxiv.org/abs/2102.06625} {arXiv:2102.06625 [hep-lat]}
  \BibitemShut {NoStop}%
\bibitem [{\citenamefont {Werner}(2007)}]{Werner:2007bf}%
  \BibitemOpen
  \bibfield  {author} {\bibinfo {author} {\bibfnamefont {K.}~\bibnamefont
  {Werner}},\ }\href {\doibase 10.1103/PhysRevLett.98.152301} {\bibfield
  {journal} {\bibinfo  {journal} {Phys. Rev. Lett.}\ }\textbf {\bibinfo
  {volume} {98}},\ \bibinfo {pages} {152301} (\bibinfo {year} {2007})},\
  \Eprint {http://arxiv.org/abs/0704.1270} {arXiv:0704.1270 [nucl-th]}
  \BibitemShut {NoStop}%
\bibitem [{\citenamefont {Petrovici}\ \emph {et~al.}(2017)\citenamefont
  {Petrovici}, \citenamefont {Berceanu}, \citenamefont {Pop}, \citenamefont
  {T\^arzil\u{a}},\ and\ \citenamefont {Andrei}}]{Petrovici:2017izo}%
  \BibitemOpen
  \bibfield  {author} {\bibinfo {author} {\bibfnamefont {M.}~\bibnamefont
  {Petrovici}}, \bibinfo {author} {\bibfnamefont {I.}~\bibnamefont {Berceanu}},
  \bibinfo {author} {\bibfnamefont {A.}~\bibnamefont {Pop}}, \bibinfo {author}
  {\bibfnamefont {M.}~\bibnamefont {T\^arzil\u{a}}}, \ and\ \bibinfo {author}
  {\bibfnamefont {C.}~\bibnamefont {Andrei}},\ }\href {\doibase
  10.1103/PhysRevC.96.014908} {\bibfield  {journal} {\bibinfo  {journal} {Phys.
  Rev. C}\ }\textbf {\bibinfo {volume} {96}},\ \bibinfo {pages} {014908}
  (\bibinfo {year} {2017})},\ \Eprint {http://arxiv.org/abs/1703.05805}
  {arXiv:1703.05805 [nucl-th]} \BibitemShut {NoStop}%
\end{thebibliography}%

\end{document}